\documentclass[fleqn,usenatbib]{mnras}

\usepackage{newtxtext,newtxmath}

\usepackage{ae,aecompl,placeins}
\usepackage[T1]{fontenc}

\usepackage{hyperref}
\usepackage{graphicx}	
\usepackage{amsmath}	
\usepackage{times,txfonts,mathptmx,textcomp,threeparttable,makecell}
\usepackage[dvipsnames]{xcolor}

\newcommand{\Plus}{\texttt{+}}
\newcommand{\Minus}{\texttt{--}}
\newcommand{\Equals}{\texttt{=}}
\newcommand{\Frac}{\texttt{/}}

\title[CCSN Host galaxies from ASAS-SN]{Core-collapse, superluminous, and gamma-ray burst supernova host galaxy populations at low redshift: the importance of dwarf and starbursting galaxies}

\author[K. Taggart et al.]{
K. Taggart, $^{1,2}$\thanks{E-mail: kltaggar@ucsc.edu}
D. A. Perley$^{1}$
\\
$^{1}$ Astrophysics Research Institute, Liverpool John Moores University, 146 Brownlow Hill, Liverpool L3 5RF\\
$^{2}$ Department of Astronomy and Astrophysics, University of California, Santa Cruz, CA 95064, USA \\
}

\date{Accepted 2021 January 18. Received 2021 January 18; in original form 2020 July 4}
\pubyear{2021}

\begin{document}
\label{firstpage}
\pagerange{\pageref{firstpage}--\pageref{lastpage}}
\maketitle

\begin{abstract}
We present a comprehensive study of an unbiased sample of 150 nearby (median redshift, $z$ = 0.014) core-collapse supernova (CCSN) host galaxies drawn from the All-Sky Automated Survey for Supernovae (ASAS-SN) for direct comparison to the nearest long-duration gamma-ray burst (LGRB) and superluminous supernova (SLSN) hosts. We use public imaging surveys to gather multi-wavelength photometry for all CCSN host galaxies and fit their spectral energy distributions (SEDs) to derive stellar masses and integrated star formation rates. CCSNe populate galaxies across a wide range of stellar masses, from blue and compact dwarf galaxies to large spiral galaxies. We find 33$^{\Plus 4}_{\Minus 4}$ per cent of CCSNe are in dwarf galaxies (M$_*$ $<$ 10$^9$ M$_{\odot}$) and 2$^{\Plus 2}_{\Minus 1}$ per cent are in dwarf starburst galaxies (sSFR > 10$^{-8}$ yr$^{-1}$). We reanalyse low-redshift SLSN and LGRB hosts from the literature (out to $z<0.3$) in a homogeneous way and compare against the CCSN host sample. The relative SLSN to CCSN supernova rate is increased in low-mass galaxies and at high specific star-formation rates. These parameters are strongly covariant and we cannot break the degeneracy between them with our current sample, although there is some evidence that both factors may play a role. Larger unbiased samples of CCSNe from projects such as ZTF and LSST will be needed to determine whether host-galaxy mass (a proxy for metallicity) or specific star-formation rate (a proxy for star-formation intensity and potential IMF variation) is more fundamental in driving the preference for SLSNe and LGRBs in unusual galaxy environments.
\end{abstract}




\section{Introduction}\label{sec:introduction}

Massive stars ($>$8 M$_{\odot}$) evolve rapidly, and after a short life (up to a few tens of million years), they die in violent core-collapse supernova (CCSN) explosions. CCSNe have a profound influence on their environment: they produce heavy elements and deposit large amounts of energy into their environments, driving feedback and chemical evolution in galaxies \citep[e.g.,][]{Chevalier1977}. In addition, because of the short progenitor lifetime, the volumetric CCSN rate is a direct tracer of star-formation. Thus, CCSNe can be used to quantify the contribution to cosmic star-formation from distinct galaxy sub-classes and to pinpoint rare individual star-forming galaxies, especially at low stellar mass, where galaxy catalogues are incomplete \citep[e.g.,][]{Sedgwick2019}. 

Candidate CCSN progenitors are diverse, as are the explosion properties they produce. Observations of CCSN explosions and their progenitors provide a means to test theories of stellar evolution and the explosion channels of very massive stars. However, despite the importance of CCSNe to many areas of astrophysics, mapping a star's evolution (accounting for complicating factors such as metallicity, binarity, and rotation) from its beginning to end is a complex problem.

Observationally, CCSNe are classified into types I and II based on the presence (II) or absence (I) of hydrogen emission lines in their spectra at maximum light \citep{Filippenko1997}. Some CCSN progenitors lose part/all of their hydrogen stellar envelope prior to their explosion due to stellar winds \citep{Maeder2000} or binary mass transfer \citep{Podsiadlowsk1992} and are observed as a helium-rich (Ib and IIb) or helium-poor (Ic) stripped-envelope SNe \citep{Smartt2009}. In recent years, due to a new generation of all-sky surveys and ever-increasing observational capabilities, many new types of stellar explosion have emerged beyond this classical picture. One example is the class of superluminous supernovae (SLSNe) which are also classified into types I and II, but whose extreme luminosities exceed ordinary CCSNe by a factor of 10--100 (\citealt{Quimby2011,2012Sci...337..927G}; see \citealt{Moriya2018, Gal-Yam2019} for more recent reviews) and likely require an additional power source.

SLSNe-II are most likely powered by SN interaction with a dense circumstellar shell of hydrogen created by an ultra-massive progenitor star before the explosion \citep{Chevalier2011, Ginzburg2012, Moriya2013} or episodic mass-loss in a pulsational pair-instability explosion \citep[PPISNe;][]{Woosley2007, Chatzopoulos2012}. However, the mechanism that powers SLSN-I is still puzzling. In theory, an extremely massive stellar core \citep{Moriya2010, Young2010} could produce enough $^{56}$Ni to power a SLSN via radioactive decay, but mass-loss during a star's lifetime makes it difficult to retain such a massive core. Several other theoretical mechanisms have been proposed to explain SLSN-I, including interaction with non-hydrogen circumstellar-material \citep{Chatzopoulos2012,Sorokina2016,Vreeswijk2017}, a Pair-Instability SN \citep[PISN,][]{Barkat1967,1967Rakavy} from a very massive \textit{and} metal-poor star \citep[$\sim$0.2 Z$_\odot$;][]{Yusof2013} or an engine-driven scenario (similar to that invoked for long-duration gamma-ray bursts) which would provide a long-lived energy source behind the SN ejecta \citep[e.g.,][]{2010Kasen,Metzger2015}. 

Long-duration gamma-ray bursts (LGRBs) are brief, but extremely luminous flashes of high-energy radiation associated with the formation of a relativistic jet from a `central engine' (a fast-spinning neutron star or black hole) at the centre of a collapsing and rapidly rotating massive stellar core. While most LGRBs occur at high redshifts, events that occur sufficiently nearby are typically observed in association with CCSNe \citep{Galama1998, Hjorth2003, Woosley2006}; these SNe are universally luminous, helium-poor stripped-envelope SNe with broad spectral features (Ic-BL) indicating large ejecta velocities \citep{Cano2017review}.

However, despite this association,  the nature of LGRB progenitors is uncertain, including whether the progenitor is a single star \citep{Yoon2006} or a binary system \citep{Cantiello2007} and it is not yet firmly established whether all LGRBs occur in association with SN Ic-BL, and vice versa. Two LGRBs from 2006 have no reported SN association to deep limits \citep{Fynbo+2006, GalYam2006, Gehrels2006, DellaValle2006}, although it has been suggested that some SN-less LGRBs are not associated with the death of massive stars, but may be compact binary mergers with unusually long duration \citep[e.g.,][]{Ofek2007,Kann2011}. In addition, most known SN Ic-BL are found in optical surveys with no observed association with a LGRB.  Some of these may represent LGRBs observed off-axis, but they could also represent events in which the jet fails to break out of the star or is not produced to begin with.


The physical powering mechanisms and progenitors of SLSNe and LGRBs are still under debate. However, it is highly unlikely that pre-explosion imaging will ever uncover the progenitor properties of SLSN or LGRBs due to a combination of their low volumetric rate \citep[$\sim$1 in 1,000 CCSNe;][]{Quimby2013, Prajs2017} and their high-redshift nature: the closest SLSN discovered to date is at a distance of $\sim$110 Mpc \citep[SN 2018bsz;][]{Anderson2018} and the closest LGRB-SN is at $\sim$40 Mpc \citep[SN 1998bw;][]{Galama1998}. This motivates the use of indirect methods to probe SLSN and LGRB progenitor properties and to constrain their poorly understood explosion mechanisms. One method is to analyse the properties of the galaxies they inhabit, to search for trends in morphology, colour, chemical composition, and star-formation, which can be tied to the SN progenitor models themselves. For example, a PISN likely requires a low-metallicity, star-forming environment to produce a star with sufficient initial mass and to avoid losing its mass in line-driven winds. Single-star progenitor mechanisms for central-engine models of LGRBs also likely require a low metallicity, since line-driven winds would otherwise quickly sap the progenitor of its rotational energy. More exotically, some models postulate that LGRBs and/or SLSNe may arise as the result of runaway collisions in young and dense star clusters \citep{2013ApJ...779..114V}. In this scenario, one may expect to find SLSNe more frequently in galaxies undergoing an exceptionally high rate of star formation, even after accounting for the fact that any CCSN is proportionally more likely to occur in a galaxy with a high SFR.

There is ample evidence that LGRB and SLSN-I host galaxies differ from the bulk of the star-forming galaxy population. For example, both LGRBs and SLSNe-I seem to occur preferentially in faint, low-mass galaxies with irregular structure \citep{Neill2011, Lunnan2014, Angus2016, Fruchter2006}. \citet{Japelj2016} found the $B$-band luminosity, stellar mass, SFR and sSFR of SLSNe-I and LGRBs are statistically similar between a redshift range of $0.3<z<0.7$ and \citet{Lunnan2014} found that SLSN-I host galaxies at $0.1<z<1.6$ (discovered in the PS medium deep survey) are statistically indistinguishable from LGRB host galaxies. There is also good evidence in particular that metallicity affects SLSN and LGRB production: high-metallicity environments rarely produce LGRBs \citep{Kruhler2015, Vergani2015, Japelj2016a, Perleyb2016, Palmerio2019} or SLSNe \citep{2016ApJ...817....7P, Schulze2018,Chen2017}\footnote{However, this is not the entire picture since over the past few years, as the statistical sizes of nearby SLSN and LGRBs have increased, there have been a handful of cases of large spiral galaxies with high-metallicities hosting SLSN-I \citep[MLS121104, PTF10uhf, SN 2017egm,][]{Lunnan2014,2016Perley, Dong2017} and nearby LGRBs \citep[e.g.,][]{Izzo2019}.}.

However, population studies with larger sample sizes show that there may also be some subtle differences between the SLSNe and LGRBs populations themselves. For example, the median half-light radius of LGRB host galaxies is $\sim$1700 pc \citep{Lyman2017}, and for SLSNe it is $\sim$900 pc \citep{Lunnan2015}. In addition,\citet{Lunnan2014} bolstered the PS medium deep survey SLSNe-I with SLSNe-I from the literature (typically at lower redshift) and found that SLSNe-I were statistically distinct from LGRBs, with a fainter $B-$band luminosity and lower stellar mass. \citet{Leloudas2015} suggested that on average, SLSNe-I explode in lower mass and higher sSFR than the hosts of LGRBs ($0.1<z<1.6$). These findings were further supported by \citet{Schulze2018} who used the largest sample size of LGRBs and SLSNe (in comparison to previous studies) and found that the $B-$band luminosity, stellar mass and sSFR of SLSNe-I and LGRBs are statistically distinct over a redshift range of $0.3<z<1$. 

In contrast, CCSNe have typically been found in massive spiral galaxies. In part, this was a reflection of the fact that CCSN samples (prior to untargeted all-sky surveys) were found via targeted surveys of pre-selected nearby galaxies. Therefore CCSNe were always found in massive, nearby galaxies (most of which were massive spirals), but about half of high-redshift ($0.28<z<1.2$) CCSNe found blindly in deep surveys (covering small) fields of view also explode in spiral galaxies \citep{Svensson2010}, in contrast to only $\sim$10 per cent of LGRB hosts. 

 \citet{Graur2017a, Graur2017b} found that the relative rate of Ib/c stripped-envelope SNe versus non-stripped CCSNe declines in low-mass ($<$10$^{10}$ M$_\odot$) galaxies; they are underrepresented by a factor of $\sim$3. In addition, \citet{Graur2017a, Graur2017b} also note that there appears to be a strong metallicity bias, with the relative rate of Ib/c to II SNe increasing with metallicity. However, this is not interpreted as evidence for the single-star scenario: the single-star stellar evolution models underpredict the observed absolute numbers of SE-SN, therefore the binary scenario could be important and there could be multiple channels at play. In addition, the binary scenario can also show a strong metallicity dependency, although binary star channels are much more uncertain than the single-star channel.

 Nevertheless, there is some disagreement in the literature; \citet{Arcavi2010} found that while the relative proportion of Ic SNe versus non-stripped CCSNe decreased in low-mass galaxies, the relative rates of all other stripped-envelope SNe (Ic-BL, Ib, IIb) versus non-stripped CCSNe increased in low-mass, low-metallicity galaxies, which may be a result of a reduced metallicity-driven mass loss causing some massive stars that would have exploded as a Ic SN in a metal-rich galaxy to retain some H and He and explode as a Ib/IIb event instead. There are also differences in the environments of stripped-envelope CCSNe themselves. Ordinary Ic CCSNe are found in more metal-rich galaxies with lower sSFRs than their more energetic Ic-BL cousins  (with and without LGRB associations)  which may suggest that Ic-BL harbour LGRB jets from a compact central engine, which in turn requires a low-metallicity environment, whereas ordinary Ic SNe may not require such an environment \citep{Japelj2018, Modjaz2019}. 

Additionally, there are also some indications that metallicity alone may not fully explain the unusual properties of the host galaxies of SLSNe and LGRBs. In particular, many SLSN-I hosts show very high specific star-formation rates (sSFR=SFR/M$_*$) as well as low metallicities, evidenced by their very high equivalent widths \citep{Leloudas2015}: as many as $\sim$50 per cent of SLSNe-I are found in extreme emission line galaxies \citep[EELGs;][]{Leloudas2015}. While sometimes attributed to a very young progenitor that simply explodes earlier than other types of SNe \citep{Leloudas2015, Thone2015}, it could also point towards an \emph{intrinsic} preference in starbursting galaxies that favours the production of SLSNe, such as a top-heavy IMF \citep[e.g.,][]{Dabringhausen2009} or the collisional model of \cite{2013ApJ...779..114V}. 

A complicating factor is that all key galaxy observational parameters we may want to use to diagnose the nature of the progenitor (e.g., stellar mass, metallicity and sSFR) correlate across the star-forming galaxy population \citep[e.g.,][]{Tremonti2004, Salim2007}. For example, a low-mass and low-metallicity galaxy tends to have a star-formation history with short bursts of concentrated star-formation and therefore is more likely to be observed as a starburst than a high-mass and high-metallicity galaxy. Thus, it is still unclear to what extent the environmental properties of SLSNe and LGRBs (low-mass, low metallicity and high sSFR) reflect their specific physical influences (progenitor and explosion mechanism). 

In order to disentangle the role of metallicity and SFR and to determine if both properties are equally important in governing SLSN and LGRB production, we need an unbiased and representative sample of star-forming galaxies to provide testable predictions for where we might expect SLSNe and LGRBs to occur under various hypotheses about their formation preferences. Ideally, the sample of star-forming galaxies should be selected in the same manner as a SLSN or a LGRB--via the explosion of a massive star as detected in a time-domain imaging survey--to minimize the large methodological differences between selecting via SNe versus selecting via galaxy counts in flux-limited surveys. In other words, we require a high-quality sample of `ordinary' CCSNe. 

This sample must have several properties. First, it must enclose a sufficiently large volume to be representative of the average distribution of galaxies, since large-scale structure can potentially bias the galaxy population seen within smaller volumes. Second, the SNe must be discovered in an unbiased way (not via galaxy-targeted surveys). Third, the sample must be able to securely distinguish CCSNe from Ia SNe for all transients, ideally via spectroscopy. Finally, it must have multi-wavelength galaxy data from UV to NIR in order to derive physical parameters for the hosts. Few existing SN samples have these properties, and until recently, none of these samples have been at low redshift where detailed host studies are most practical. Examples of other large, untargeted SN samples include SDSS \citep{Frieman2008, Sako2008} and SNLS \citep{Bazin2009} but these surveys are not spectroscopically complete, and this leads to ambiguities in the classifications.

In this paper, we address this need by compiling a large, unbiased, representative sample of CCSN host galaxies (which we assume sample the explosions of `typical' massive stars, unlike SLSNe and LGRBs). We provide photometry of this sample with integrated UV-through-NIR SEDs and stellar masses and star-formation rates derived from these measurements. We investigate star-formation within the CCSN host galaxy sample and compare to a sample of SLSN and LGRBs. 

The paper is organized as follows: Section~\ref{sec:samples} describes how the transient host galaxies are selected to form our CCSN, SLSN and LGRB samples. In Section~\ref{sec:photometry}, we describe our photometry method and show all other archival photometry which has been used in this paper. In Section~\ref{sec:physparams}, we present the methodology used to measure the star-formation rates and stellar masses of each host galaxy based on UV through NIR colours. In Section~\ref{sec:results} we show our results and in Section~\ref{sec:conclusions}, we summarize our findings and present our conclusions. Throughout this paper we adopt $\Lambda$CDM cosmology, with $\Omega_{0}$ = 0.27, $\Omega_\Lambda$ = 0.73 and H$_0$ = 70kms$^{-1}$Mpc$^{-1}$ \citep{Komatsu2011}.

\begin{figure*}
	\includegraphics[width=\textwidth]{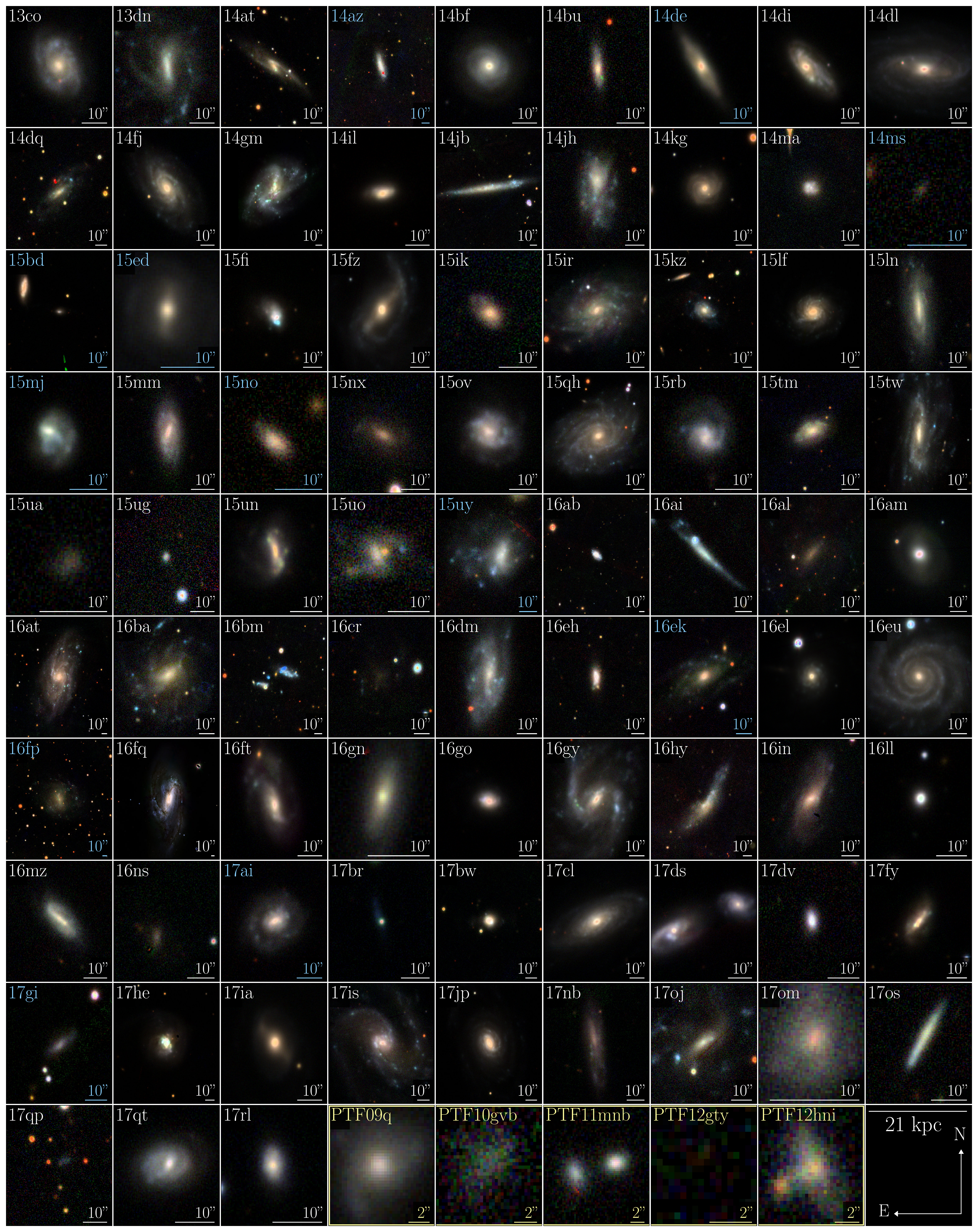}
	\vspace{-0.5cm}
	 \caption{Mosaic showing RGB ($gri$ PS1) colour composite images of our ASAS-SN CCSN host galaxy sample. Images labelled in white text are Type II CCSNe (excluding IIb) and images in blue are stripped-envelope SNe of type Ib/c or IIb. Each image has a constant physical size scale of 21 kpc in diameter at the redshift of the host galaxy and an angular scale of 10 arcsec is shown on each individual cutout. The image of low surface-brightness SN host 16ns is after the subtraction of a bright ($m_v \sim 17$) foreground star. The 
	 SLSN candidates that were discovered in archival PTF data are also included in the last row of the figure in yellow text. The same physical size as the CCSN is used, but with a scale bar of 2 arcsec due to their higher redshift nature.} 
   \label{fig:ccsn_rgb}
\end{figure*}

\begin{figure*}
	\includegraphics[width=\textwidth]{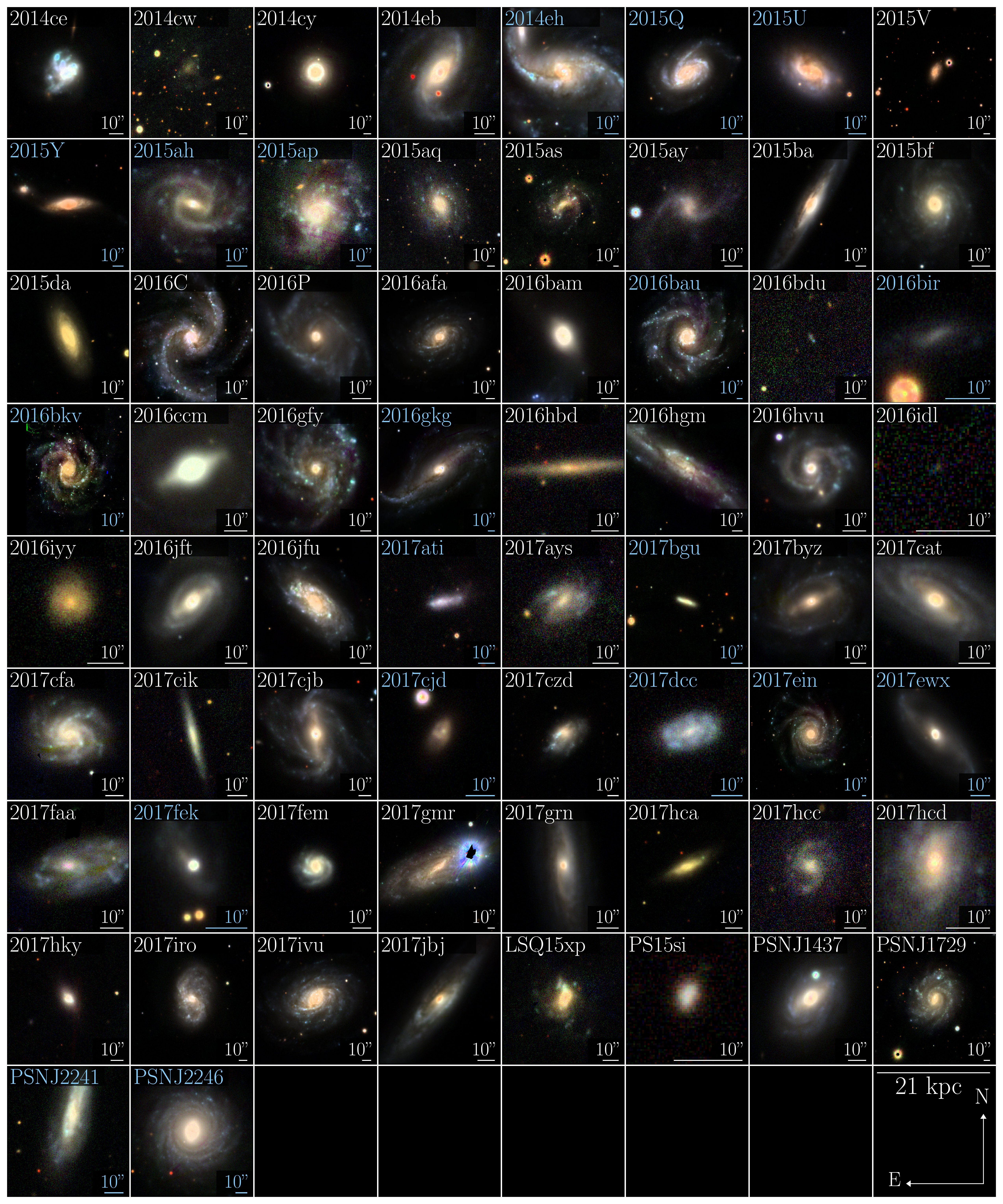}
	\vspace{-0.5cm}
	 \caption{Mosaic showing RGB ($gri$ PS1) colour composite images of hosts of additional CCSN recovered by ASAS-SN. Images labelled in white text are Type II CCSNe (excluding IIb) and images in blue are stripped-envelope SNe of type Ib/c or IIb. Each image has a constant physical size scale of 21 kpc in diameter at the redshift of the host galaxy and an angular scale of 10 arcsec is shown on each individual cutout.} 
   \label{fig:ccsn_rgb2}
\end{figure*}

\section{Host galaxy samples}\label{sec:samples}

\subsection{Core collapse supernovae}
A variety of galaxy-untargeted SN catalogues exist, including the Dark Energy Survey \citep{Flaugher2005}, Catalina Real-Time Survey \citep{Drake2009}, the Palomar Transient Factory \citep{Law2009}, SuperNova Legacy Survey \citep{Bazin2009}, Pan-STARRS \citep{Kaiser2002}, La Silla Quest \citep{Hadjiyska2012}, the Gaia transient survey \citep{Hodgkin2013}, SkyMapper \citep{Keller2007}, SDSS Supernova Survey \citep{Frieman2008} and the All-Sky Automated Survey for Supernovae \citep[ASAS-SN;][]{asas-sn}. We drew our CCSN sample from ASAS-SN since it is is shallow (m$_{V,\rm limit}$ $\sim$17 mag) but is all-sky, so the SNe it finds are bright and generally very nearby.  This means that excellent photometric galaxy information exists in public catalogues and that almost all SNe are bright enough (even with small telescopes) for the global SN community to follow-up, spectroscopically classify and derive a redshift estimate. Therefore the ASAS-SN sample is spectroscopically complete for SNe with peak \textit{V}-band light curve magnitudes $m_{V}<15.8$ and is roughly 50 per cent complete at $m_{V}$=$\sim$ 17 \citep{Holoien2017a}. This was important since we required an unambiguous sample of CCSN selected host galaxies and a reliable SN redshift estimate for our host analysis. 

We compiled all spectroscopically confirmed CCSNe discovered by ASAS-SN \citep[2013--2014, 2015, 2016, 2017;][]{Holoien2017a,Holoien2017b,Holoien2017c,Holoien2019}, and adopted any SN classifications and redshift estimates that were updated since the initial classification was made. We also included any SNe that were not discovered by ASAS-SN, but were `recovered' in their data and therefore do not have an ASAS-SN name designation. We refer to these SNe in the paper text using the designated IAU name, or the discovery group name (6 SNe) when there is no IAU name to our knowledge. For the sake of brevity, we shortened any possible supernova (PSN) object names to the first 8 digits.

There were some ambiguous classifications which we removed from the sample. We removed two claimed SLSNe: ASAS-SN 15lh was classified as a SLSN-I \citep{Dong257}, but was omitted since it was unclear whether this event was a SLSN or a tidal disruption event \citep{Leloudas2016b,Margutti2017} and ASAS-SN 17jz was re-classified as a SLSN-II, but its classification is ambiguous; it could be a very luminous SN-II \citep{Xhakaj2017} or alternatively it could be an AGN \citep{Arcavi2017}. In addition, we removed SN 2015bh since the classification was ambiguous. Despite having a dataset spanning a 21 year time period, it was unclear whether SN 2015bh was the terminal explosion of the star resulting in a CCSN or if it was a precursor LBV hyper-eruption \citep{Elias-Rosa2016,Thone2017}.

We limited our sample to a declination greater than -30\textdegree\ because uniform, public, deep optical survey data is not available across the entire southern hemisphere.  Two supernovae (SN 2016afa and 2017ivu) have the same host (NGC 5962) and this galaxy is included twice in the host galaxy analysis. We also imposed a galactic latitude cut (|$b$|$>$ 15\textdegree) in order to eliminate the galaxies where stellar crowding significantly affects the photometry and thus remove SN 2015an, 2015W, 2016bpq, 2016G, 2017eaw, 2017gpn, ASAS-SN 17ny, 17kr, and PSNJ1828 from the sample. In addition, we imposed a minimum distance cut out to 10 Mpc, meaning that one supernova (AT 2014ge) was removed from our sample. Primarily we made this cut since performing consistent photometry for very extended galaxies within this volume using the same methods used for more distant galaxies is difficult. Also making this cut avoided sample overlap with the comparison sample used in this analysis (the Local Volume Legacy (LVL) survey) which is volume-complete to within $\sim$11 Mpc.

Our sample is comprised of 150 SNe discovered from 2013 to the end of 2017. The redshift distribution covers the range 0.00198--0.08, with a median value of 0.014. A table with details of these galaxies can be found in Appendix ~\ref{tab:asassntable}. A mosaic showing our ASAS-SN CCSN host galaxy sample is provided in Fig.~\ref{fig:ccsn_rgb}--~\ref{fig:ccsn_rgb2}. We used methods detailed by \citet{Lupton2004} to convert PS1 $gri$  images into a colour composite image. Each cutout has a constant physical size scale in the rest frame of the SN host of 21 kpc on each side and a scale bar showing an angular size of 10 arcsec is shown on each cutout. 

\begin{table}
\centering
\caption{Division of transient types within our samples.}
\begin{threeparttable}
\label{tab:ccsn_table}
\begin{tabular}{p{0.50\linewidth}p{0.10\linewidth}}
\hline
Transient Type & Number \\
\hline
\hline
CCSN-II &  72 \\
CCSN-IIP &  26 \\
CCSN-IIb & 10 \\ 
CCSN-Ib/Ic & 19  \\
CCSN-IIn/Ibn & 21  \\
CCSN-Ic-BL & 2 \\
\hline 
CCSN Total	& \textbf{150} \\
\hline \hline
SLSN-I & 29 \\
Possible SLSN-I & 3 \\
SLSN-II & 21 \\
\hline
SLSN Total	& \textbf{53} \\
\hline \hline
LGRB-SN & 12	\\
SN-less LGRB & 5	\\
\hline
LGRB Total	& \textbf{17} \\
\hline \hline
\end{tabular}
\end{threeparttable}
\end{table}

\subsection{Superluminous supernovae}

We collated our initial SLSN sample based on archival SLSNe in the literature. We included SLSN hosts from \citet{Neill2011}\footnote{We did not include SN1995av and SN1997cy since their classifications are unclear: SN1997cy could be a SN Ia or IIn and  SN1995av may have been associated with a LGRB.}, SUSHIES \citep{Schulze2018} and PTF \citep{2016Perley}. In addition, we included five candidates identified by \cite{Quimby2018} following their reanalysis of archival PTF spectra: two likely SLSNe-I (PTF12gty and PTF12hni) and three possible SLSNe-I (PTF09q, PTF10gvb and PTF11mnb) at slightly lower luminosities (M$\,>-21$ mag) than the PTF sample of SLSN host galaxies by \cite{2016Perley}. These SLSN candidates and their properties are summarized in Table~\ref{tab:newptftable}. Rest frame \textit{g}-band magnitudes for PTF12gty and PTF12hni were taken from \citet{DeCia2018} and PTF09q, PTF10gvb and PTF11mnb were taken from \cite{Quimby2018}. Thumbnail images of each host are shown in the bottom row of Fig.~\ref{fig:ccsn_rgb}; the physical scale is the same as for the CCSN hosts, with a yellow scale bar of 2 arcsec. 

We restricted our analysis to SLSNe with a redshift of $z<0.3$ for two main reasons. Firstly, including distant SLSNe could have caused incompleteness in the sample due to the increased difficulty in spectroscopically confirming members of this class without a bright associated host galaxy. Secondly, we wanted to reduce cosmic evolution effects when comparing to the $z\sim0.014$ CCSN sample. After making this cut and excluding PTF09q, PTF10gvb and PTF11mnb, our final statistical sample consisted of 29 SLSNe-I and 21 SLSNe-II in total. 

\begin{table}
\centering
\caption{New PTF SLSN-I candidates from archival PTF search.}
\label{tab:newptftable}
\begin{threeparttable}
\begin{tabular}{llllll} 
\hline
PTF ID & M$_{\rm peak}$ & $\alpha$(2000) & $\delta$(2000) & $z$ & E($B$-$V$)\\
\hline
09q*     &  $\sim$--20  & 12:24:50.11 & +08:25:58.8  & 0.09   & 0.021 \\
10gvb*   &  --19.6 [1]  & 12:15:32.28 & +40:18:09.5  & 0.098  & 0.022 \\ 
11mnb*   &  --18.9 [1]  & 00:34:13.25 & +02:48:31.4  & 0.0603 & 0.016 \\
12gty    &  --20.1 [2]  & 16:01:15.23 & +21:23:17.4  & 0.1768 & 0.061 \\
12hni    &  --19.9 [2]  & 22:31:55.86 & --06:47:49.0 & 0.1056 & 0.054 \\
\hline
\end{tabular}
\begin{tablenotes}[flushleft]
\item \textbf{Notes.} Possible SLSNe-I from \cite{Quimby2018} are indicated by a *; host analysis is done, but not included the SLSN statistical analysis due to uncertainty about the nature of the classification. PTF09q is reclassified as a SN Ia in \citep{Modjaz2019}. 
\item \textbf{References}:  [1] \cite{Quimby2018}, [2] \cite{DeCia2018}.
\end{tablenotes}
\end{threeparttable}
\end{table}

\subsection{LGRBs}
Our LGRB sample consists of all $z<0.3$ LGRBs discovered prior to the end of 2017 with an associated optical counterpart: a supernova, an optical afterglow, or both. The requirement for an optical afterglow was imposed to better match the optical selection of SNe used for comparison and to ensure a high degree of confidence in the host-galaxy association: while many additional low-$z$ LGRBs have been reported based on X-ray associations alone, it is difficult to rule out the possibility that these are higher-$z$ events seen in coincidence with a foreground galaxy. This sample was comprised of 17 LGRBs; 12 of which had confirmed SN associations and 5 without any reported SN (see Table~\ref{tab:grbsnassociations}). 

Of the 5 LGRBs without reported SNe, two were highly-publicised events from 2006 (LGRBs 060505 and 060614) for which a SN was ruled out to deep limits\citep{Fynbo+2006,GalYam2006,DellaValle2006,Gehrels2006}. These appear to have genuinely different progenitors (such as compact binary mergers) and/or explosion mechanisms from ordinary SN-associated long-duration GRBs, a possibility that makes scrutiny of their host properties particularly relevant. The remaining events, LGRBs 050826, 080517 and 111225A, have relatively poor constraints on the extinction column towards the LGRB and/or on the presence of a SN peaking 1--3 weeks after the event (e.g., \citealt{Stanway2015}).

 \begin{table}
\caption{Table of LGRB sources with and without SN associations. SN names and discovery reports are referenced and photometric (P) or spectroscopic (S) reports are indicated.}
\begin{threeparttable}
\renewcommand\TPTminimum{\linewidth}
\makebox[\linewidth]{
\label{tab:grbsnassociations}
\centering
\begin{tabular}{lll}
\hline
LGRB & SN name &  SN Reference  \\
\hline
980425  & 1998bw  & [1,S]  \\
020903  & SN$^\dagger$       & [1,S]  \\
030329A & 2003dh  & [1,S]  \\
031203  & 2003lw  & [1,S]  \\
050826  & --      &  --     \\
060218  & 2006aj  & [1,S] 	\\
060505  & --      &  -- 	\\
060614  & --      &  --		\\
080517  & --      &  --		\\
100316D & 2010bh  & [1,S]   \\
111225A & --	  &  --		\\
120422A & 2012bz  & [1,S] 	\\
130702A & 2013dx  & [1,S] 	\\
150518A & SN$^\dagger$       & [1,P] 	\\
150818A & SN$^\dagger$       & [1,S]	\\
161219B & 2016jca & [2--6,S] 	\\
171205A  & 2017iuk & [7--13,S] 	\\
\hline
\end{tabular}}
\begin{tablenotes}[flushleft]
\item \textbf{Notes.} $^\dagger$In these cases, the LGRBs do have associated SNe but there is no known SN name designation on TNS. 
\item \textbf{References:} [1] Refer to Table 4. in \cite{Cano2017review}, [2] \cite{deUgartePostigo2016}, [3]\cite{Volnova2017}, [4] \cite{2017chen}, [5] \cite{Ashall2017}, [6] \cite{Cano2017grb}, [7] \cite{deUgartePostigo2017}, [8] \cite{Cobb2017}, [9] \cite{Prentice2017}, [10] \cite{DElia2018}, [11] \cite{Wang2018}, [12]  \cite{Suzuki2019}, [13] \cite{Izzo2019}.
 \end{tablenotes}
\end{threeparttable}
\end{table}

\section{Photometry}\label{sec:photometry}

\subsection{CCSN host multi-wavelength data}\label{sec:ccsndata}

The galaxies in our CCSN sample are nearby ($z<0.08$), so most were detectable in all-sky multi-wavelength surveys. Therefore our primary image and source catalogues were public surveys. We used images from: the \textit{Galaxy Evolution Explorer} \citep[\textit{GALEX};][]{Martin2005}, the Panoramic Survey Telescope and Rapid Response System \citep[PS1;][]{2010SPIE.7733E..0EK}, the Sloan Digital Sky Survey \citep[SDSS;][]{York2000} and the Two Micron All-Sky Survey \citep[2MASS;][]{2MASS2012}.

Our aim was to derive consistent mass and star-formation estimates for our host galaxy sample, thus we matched aperture sizes across the optical and NIR wavelengths. This was particularly important for nearby and massive galaxies, since the aperture size can significantly increase or decrease the flux measurements. In addition, the automated pipeline of \textit{GALEX}, 2MASS and \textit{WISE} often incorrectly deblends galaxies with a large angular diameter on the sky and does not capture the low surface-brightness parts of the galaxy. If available, we used SDSS \textit{ugriz} and \textit{GALEX} FUV and NUV photometry from the NASA Sloan Atlas (NSA) \citep[NSA;][]{Blanton2011}. The NSA is a unified catalogue of galaxies out to $z\sim$ 0.05, optimised for nearby extended objects since the flux measurements are derived from reprocessed SDSS images with a better background subtraction \citep{Blanton2011}. We used the elliptical petrosian aperture photometry, with an elliptical aperture radius defined by the shape of the light profile of the galaxy as in \citet{Blanton2011} and \citet{Yasuda2001}. The NSA flux measurements were available for about half of the northern hemisphere sample. Otherwise, we performed the photometry using optical images downloaded from Pan-STARRS DR1 (PS1) \citep{pan-starrs1,pan-starrs2} and SDSS \textit{u}-band if available.

We used the 2MASS extended source catalogue to obtain NIR brightness measurements in the \textit{J}, \textit{H} and \textit{K$_{s}$} filters \citep{2MASS2012}. If the galaxy was in the NSA, we redid the 2MASS photometry with the same axis ratio and aperture orientation and use the curve of growth technique to adjust the size of the aperture. If the galaxy was not within the NSA, we checked whether the 2MASS extended aperture (which fits an ellipse to the 20 mag arcsec$^{-2}$ isophote in the \textit{K$_{s}$} band and uses a curve of growth technique to capture low surface-brightness flux of the galaxy) was adequate. In the cases of galaxies with small angular size, the aperture was usually adequate, but in the case of high-mass, extended galaxies the aperture often missed a substantial fraction of the low surface-brightness flux in the outskirts of the galaxy, thus we redid the 2MASS photometry for these sources. 

\subsection{Procedure for CCSN hosts}\label{sec:photomprocedure}

We performed aperture photometry using the python program \textsc{Photutils}\footnote{https://github.com/astropy/photutils/tree/v0.3}. We used an elliptical aperture and a curve-of-growth technique. We placed the elliptical aperture at gradually increasing radii, measuring the flux in each aperture until the curve-of-growth levelled, to the order of a few per cent, meaning the aperture was sufficiently large enough to include all of the host galaxy flux. We derived the uncertainties on these photometric measurements by using the galaxy aperture to determine the brightness of the background sky. We placed the galaxy aperture 30 times within the image on `blank' patches of the sky, making sure there was no overlap between apertures and used the standard deviation of these measurements to derive the uncertainty. 

In some cases, the galaxy was sufficiently massive and nearby that it covered a large angular diameter on the sky: placing 30 apertures of this size on blank patches of the sky was not feasible in these instances (the aperture region will always contain field sources), and in many cases the image itself was simply not large enough to place the aperture in 30 non-overlapping locations. In these cases, we removed the sources from the image and calculated the uncertainty based on the standard deviation of the sky background.

For image calibration we used catalogues of stars (PS1 Object Catalogue, 2MASS Point Source Catalogue and the SDSS Imaging Catalogue) to calculate the zero point for each image. Instrumental magnitudes were calibrated directly to the AB system with photometry from PS1 and all other magnitudes were converted into the AB system \citep{Oke1983}. In addition, we corrected all photometry for Galactic foreground extinction \citep{Schlafly2011}.\footnote{SN 2003ma pierces through the Large Magellanic Cloud. Hence the Galactic extinction of E($B$--$V$) = 0.348 mag is the lower limit of what we would expect in this direction \citep{Rest20112003ma}.}

\subsection{Galaxies requiring special attention}

Some host galaxies in our sample required extra care when performing photometry and when fitting SED models. These galaxies were either diffuse, low surface-brightness galaxies, galaxies which showed signs of interaction with nearby galaxies, galaxies contaminated with foreground stars (or other objects), or galaxies where there was some prior indication of an AGN. Below we briefly describe these cases below. 

\subsubsection{Interacting galaxies}

A significant number of host galaxies (in both the CCSN and extreme-SN samples) showed evidence of physical companions, some of which appeared to be in the process of interacting or merging. Since our general philosophy was to mimic the photometry steps and subtraction we would do if the ASAS-SN galaxies were observed at $z\sim0.2$ (for comparison to the LGRB and SLSN samples), we treated the merger as one system if it was in the advanced merger stages and would not be resolved at $z\sim0.2$. Whereas if the galaxy could be resolved at $z\sim0.2$, we measured the photometry of the single galaxy. 

\textit{ASAS-SN 14de}: This galaxy was possibly undergoing an interaction or merger. This system would barely be detectable as two individual galaxies if it was discovered at a similar redshift ($z\sim0.2$) to the SLSN or LGRB sample, therefore we quoted two different measurements for photometry: one of the entire system and one of the single galaxy from which the SN originated.

\textit{SN 2015Y}:
This SN exploded in NGC 2735 at $z$ = 0.00817, which is interacting with MCG+04-22-003 at $z\,$=$\,0.00827$. We did not include MCG+04-22-003 in the flux measurement. 

\textit{ASAS-SN 16bm}: This host galaxy did not have a catalogued redshift. However, it was possibly undergoing an interaction or merger since the SN redshift $z$ = 0.007 was similar to the redshift of a companion galaxy at $z$ = 0.00686. The galaxies are 35 arcsec apart, but if the system was at $z\sim0.2$ their centres would be separated only by 1 arcsecond. Thus, this system would barely be detectable as two individual galaxies if it was discovered at a similar redshift to the SLSN/LGRB sample ($z\sim0.2$). We quoted two different measurements for photometry: one of the entire system and one of the single galaxy from which the SN originated. We used the photometry of the system for the SED fit. 

\textit{ASAS-SN 17ds}: This host galaxy appeared to have a companion in the PS1 imaging. However, an SDSS spectrum confirmed that the redshift of this galaxy was $z$ = 0.046, compared with the host galaxy which has a redshift $z$ = 0.022. 

\textit{PTF12hni}: There was a small, red object to east of the host galaxy (see panel 5 in Fig.~\ref{fig:ccsn_rgb}). An archival KeckII/DEIMOS spectrum from 2017 July 13 confirmed that this red object was at $z$ = 0.185 and was not associated with the host galaxy with redshift $z$ = 0.1056. For this reason, we were careful not to include this object in the photometry aperture.

\textit{PTF11mnb}: The host appeared to have a companion galaxy (see bottom right panel in Fig.~\ref{fig:ccsn_rgb}). Thus, the galaxy on the west of the image was removed since the low surface-brightness flux of the galaxy overlaps. We used the program \textsc{galfit} \, \citep{Peng2002} to model and subtract any contaminating objects from the image and then used the procedure outlined in Section ~\ref{sec:photomprocedure} to perform aperture photometry on the galaxy.

\subsubsection{Unclear host galaxy}

\indent \textit{SN 2016bam:} This SN was originally reported to TNS as being hosted by the elliptical galaxy NGC 2444, which is interacting with NGC 2445. The supernova exploded between these galaxies, so even at low-redshift, this was a difficult case to judge which was the true host. At the typical redshift of SLSNe it would also be tricky. However, we made the decision to attribute this supernova to NGC 2445 (the southern object) instead of NGC 2444 because it is a star-forming galaxy and the supernova position is near (3.54 arcsec away from) an \textsc{H\,ii} region associated with NGC 2445. \\

\indent \textit{SN 2017ati:} was originally reported to TNS as a hostless supernova. However, when we looked at a larger image of the field, the SN was located between two galaxies. The SN was 36 arcsec from one galaxy nucleus and 76 arcsec from the other galaxy. This remote location is unusual for a CCSN, but these galaxies may possibly be interacting and plausibly there could be a faint (unseen) bridge of star-formation between these galaxies. The redshift of the SN is consistent with the nearest galaxy (KUG 0946+674), but no spectra exist to confirm whether both galaxies are at the same redshift. This placed the supernova $\sim$10 kpc (36 arcsec) away from the galaxy nucleus. Although the remote location of the supernova defied any prescriptive attempt to assign a host galaxy, in our analysis we assigned the SN to the nearest galaxy since this would be how we would treat this SN if it were at a typical SLSN redshift.

\subsubsection{Foreground star contamination}

\indent \textit{ASAS-SN 14dq, SN 2014cw, SN 2016bir and SN 2017fek}: These hosts were large and extended objects low surface-brightness hosts. Flux from foreground stars in these images were subtracted from these hosts. 
    
\textit{SN 2014eh}: This host galaxy has a small background galaxy and a few foreground stars covering the host. We removed the flux from these stars in the images.

\textit{SN 2015V, SN 2015ay, SN 2016P and 2016ccm}: These host galaxies all have bright stars (between 12--16 mag) nearby.  Therefore in each case, the aperture was chosen carefully so that the stellar flux was not included in the flux measurements. 

\textit{2017gmr}: There was a very bright, saturated star (HD 16152, m$_V\sim$ 7.1) covering a large area ($\sim$50 per cent) of the host. The stellar flux was removed. However the host flux measurement was very uncertain.

\textit{ASAS-SN 16al}: There was a very bright star (BD-12 4185, m$_V\sim$ 9.8) in the nearby field, causing large variations in the sky background. In addition, this object was aligned with many foreground stars which contributed to around 50 per cent of the light from the galaxy aperture. We modelled and subtracted these stars from images, but accurate photometry of the galaxy remained difficult. Thus we estimated the uncertainty in the removal of the foreground stars and incorporated an extra photometric error of 0.1 magnitudes into the photometry measurements.

\textit{ASAS-SN 16ns}: This system had a foreground star (m $\sim$ 17 mag) which masked a large percentage of the galaxy flux due to the small and low surface-brightness nature of the galaxy. We removed this star, but the subtraction residuals remained at approximately $\sim$10 per cent of the object flux in the \textit{i} and \textit{z} bands. Photometric uncertainties were increased accordingly.

\textit{ASAS-SN 17oj}:
We removed foreground stars from this image. This was a low surface-brightness galaxy, so a large aperture was used to incorporate the flux in the outskirts of the galaxy.

\textit{SN 2017fek}:
We removed multiple foreground stars from this image before we performed aperture photometry.

\subsubsection{Active Galactic Nuclei}

We checked if any of the host galaxies in our sample had an observable AGN present. First, we inspected the SDSS spectra where available (55/150 galaxies) to check for an AGN flag. Three host galaxies were flagged as an AGN in the SDSS spectra: ASAS-SN 14de (SN Ic), SN 2016afa/2017ivu (SN II/IIP) and PSNJ1437 (SN II).

The line ratio \hbox{[N~\textsc{ii}]}6583/\hbox{H\,$\alpha$} was used to identify the presence of an AGN \citep{Baldwin1981,Carter2001}. If $\log$(\hbox{[N~\textsc{ii}]}6583/\hbox{H$\alpha$})$>$ --0.25, we assumed the spectrum could be dominated by an AGN. According to this metric, only ASAS-SN 14de (SN Ic) hosts an (observable) AGN ($\log$(\hbox{[N~\textsc{ii}]}6583/\hbox{H\,$\alpha$}=$-0.32$); strong [O~\textsc{iii}] emission confirmed it as a Seyfert II galaxy.  While visual inspection of the host galaxy suggested that the AGN is unlikely to contribute significantly to the optical flux measured in SDSS/PS1, it could contribute more significantly to the IR flux, which could in turn affect the SED derived parameters including ages of the stellar populations, star-formation rates and also dust attenuation in the host galaxy. Hence, for 14de we excluded NIR photometry for the SED fit.

Since we did not have spectra for every galaxy in our sample, we also inspected the images of each host (see Fig.~\ref{fig:ccsn_rgb}) to check for a clear nuclear point source. Almost all galaxies were well resolved and few showed evidence for any sort of central point source (much less a photometrically-dominant AGN). However, the following sources in Fig.~\ref{fig:ccsn_rgb} did seem to have a red point source located at the centre of the host which could be either a galaxy bulge or an AGN:  14de (a Seyfert galaxy), 14di, 14dl, 14kg, 16am, 16go, 17br and 17cl. The following sources in Fig.~\ref{fig:ccsn_rgb2} also seemed to have a red point source located at the centre of the host which could be either a galaxy bulge or an AGN: SN 2014cy, 2014eb, and 2015bf. However, in all cases, given the huge and bright galaxies an AGN could not contribute much ($<20$ per cent) to the integrated flux in any band relevant to our SED fitting procedure.
We also checked the ALLWISE colours ($W1$--$W2$ and $W3$--$W2$) of the host galaxies as another diagnostic to test whether an AGN was present in the host galaxies \citep[see Fig. 12 of ][]{Wright2010}. Aside from 14de, we found that two galaxies (15fi and 14ma) had WISE colours suggestive of a possible AGN. ASAS-SN 15fi (Mrk 0884) had an SDSS spectrum with a line ratio of $\log$(\hbox{[N~{\sc ii}]}6583 /\hbox{H$\alpha$})=--1.13, therefore we estimated the maximum contribution to be $\sim$15 per cent. We also obtained a spectrum of ASAS-SN 14ma in Taggart et al (in prep) from the WHT and we found a line ratio of $\log$(\hbox{[N~{\sc ii}]}6583 /\hbox{H$\alpha$})=-0.83, indicating that AGN contribution was minimal.

\subsection{Literature Photometry}\label{sec:archivalphotometry}

Photometry of the SLSN and LGRB hosts was gathered primarily from the published literature. For clarity, all sources are listed in Tables ~\ref{tab:slsn}--~\ref{tab:grb} and are available in a machine readable form. If the uncertainties were not given in the photometry from the literature, it was assumed that they were negligible and we therefore assign an uncertainty of 0.01 mag when performing the SED modelling. 

We omitted photometric data points from the literature if they were inconsistent with the other photometric points at nearby wavelengths at high significance, if there was suspected contamination from the transient given the time that the data were taken, or (in cases where contamination with other galaxies is possible) if it was unclear whether the authors took deblending into account in their host photometry. 

\subsection{New LGRB host photometry}\label{sec:newphotometry}

We supplemented the LGRB photometry from the literature with new photometry from a variety of sources, detailed below. A summary of our photometry is presented in Table \ref{tab:grb}.

\subsubsection{\textit{Spitzer}/ IRAC}
Most of the LGRB hosts in our sample were observed using the Infrared Array Camera (IRAC; \citealt{2004ApJS..154...10F}) on the \textit{Spitzer Space Telescope} \citep{2004ApJS..154....1W} as part of the extended \textit{Swift/Spitzer} Host Galaxy Legacy Survey (SHOALS; \citealt{2016ApJ...817....7P}). These observations were generally carried out in channel 1 (3.6 $\mu$m) only, although LGRB\,060218 was also observed in channel 2 (4.5 $\mu$m). We used the PBCD images from the \textit{Spitzer} Heritage Archive and photometric techniques detailed in \citet{Perleyb2016}, including subtraction of all neighbouring objects that might contaminate the aperture or sky background. Data from some archival programs were also reanalysed using a consistent methodology. In most cases this was straightforward. In the case of LGRB\,020903, isolating the host galaxy was challenging due to the presence of a dense group of merging galaxies with complicated light profiles in the foreground. The dwarf host of LGRB\,130702A is part of a smaller and more distant galaxy group \citep{Kelly2013}. The companion spiral is approximately 6 magnitudes brighter and offset by 6.5 arcsec; subtraction of its halo also left some residuals in the sky background. As a result, in both these cases the uncertainty on the host flux is relatively large.

\subsubsection{Keck / MOSFIRE}
LGRB\,130702A was observed in imaging mode using the Multi-Object Spectrograph for Infrared Exploration (MOSFIRE;  \citealt{2010SPIE.7735E..1EM,2012SPIE.8446E..0JM}) at Keck Observatory on the night of 2014 Jun 16 in the $J$ and $K_s$ filters. We reduced these data using a custom pipeline. The resolution of these images (and of archival optical data) are sufficient that there are no issues with background contamination from the nearby galaxies. Aperture photometry was performed in a standard fashion using nearby 2MASS standards.

\subsubsection{Palomar / WIRC}
LGRB 120422A was observed with the Wide-Field Infrared Camera (WIRC; \citealt{2003SPIE.4841..451W}) on the Palomar 200-inch Hale telescope on the night of 2013 Feb 17 in the $J$ and $K_s$ filters. We reduced these data using our custom pipeline, which included cleaning of noise signatures associated with the replacement-detector. Aperture photometry was performed in a standard fashion using nearby 2MASS standards.

\subsubsection{Palomar / P60}
LGRB 150818A was observed extensively with the CCD imager on the Palomar 60-inch robotic telescope \citep{2006PASP..118.1396C} as part of a campaign to follow-up the supernova associated with this event (Sanchez-Ramirez et al., in prep.). A series of late-time reference images in $g r i z$ filters were taken on 2016 February 14 for the purposes of galaxy subtraction against the earlier supernova imaging; we employed these here to measure the host flux in these bands.

\subsubsection{Keck / LRIS}
LGRB 150518A was observed in imaging mode with LRIS \citep{Oke1995} in the \textit{u}-band filter on 2016 June 07. The observations were reduced with LPipe \citep{2019Perley} and aperture photometry of the host galaxy was measured relative to SDSS secondary standards in the field.

\subsubsection{Magellan / FourStar}
LGRB 150518A was observed in \textit{J}-band with the near-infrared (NIR) camera FourStar \citep{Persson2013a} at the 6.5-m Magellan/Baade Telescope (Las Campanas Observatory, Chile) on 2016 March 27 as a part of the programme CN2016A-108. The observation sequence consisted of 39 dithered images with individual integration time of 32 s. These data were reduced with the software package \textsc{theli} version 2.10.0 \citep{Erben2005a, Schirmer2013a}.

\begin{table}
\centering
\caption{New LGRB host galaxy photometry.}
\begin{threeparttable}
\label{tab:grb}
\centering
\begin{tabular}{llclc}
\hline
LGRB & Filter & AB Mag & Instrument & Date  \\
\hline
020903 	& 3.6$\mu$m & 22.30 $\pm$ 0.30 	& \textit{Spitzer}/IRAC & 2006-06-07 \\
030329A & 3.6$\mu$m & 23.71 $\pm$ 0.11 	& \textit{Spitzer}/IRAC & 2017-03-31 \\
031203	& 3.6$\mu$m & 18.19 $\pm$ 0.01	& \textit{Spitzer}/IRAC & 2005-11-29  \\
060218 	& 3.6$\mu$m & 20.77 $\pm$ 0.02 	& \textit{Spitzer}/IRAC & 2012-11-07 \\
	 	& 4.5$\mu$m & 21.06 $\pm$ 0.05  & \textit{Spitzer}/IRAC & 2012-11-07 \\
060614	& 3.6$\mu$m & 22.96 $\pm$ 0.10 	& \textit{Spitzer}/IRAC & 2012-11-25 \\
080517  & $J$       & 16.90 $\pm$ 0.14	& 2MASS & - \\
        & $H$       & 17.12 $\pm$ 0.24	& 2MASS & - \\
        & $K_s$     & 16.87 $\pm$ 0.21  & 2MASS & - \\
111225A & 3.6$\mu$m & 24.00 $\pm$ 0.30 	& \textit{Spitzer}/IRAC & 2016-12-05 \\
120422A & 3.6$\mu$m & 21.12 $\pm$ 0.03	& \textit{Spitzer}/IRAC & 2017-02-21 \\
 		& $J$       & 20.34 $\pm$ 0.09	& P200/WIRC   & 2013-02-17 \\
 		& $K_s$     & 20.35 $\pm$ 0.17 	& P200/WIRC  & 2013-02-17 \\
130702A & $J$       & 22.63 $\pm$ 0.17 	& Keck/MOSFIRE & 2014-06-16 \\
		& $K$       & 21.41 $\pm$ 0.45 	& Keck/MOSFIRE & 2014-06-16 \\
		& 3.6$\mu$m & 23.80 $\pm$ 0.30  & \textit{Spitzer}/IRAC & 2016-11-05 \\
150518A & $u'$      & 22.78 $\pm$ 0.03  & KeckI/LRIS & 2016-06-07	\\
        & $g'$      & 22.07 $\pm$ 0.14  & PS1 & - \\
        & $r'$      & 21.43 $\pm$ 0.08  & PS1 & - \\
        & $i'$      & 21.25 $\pm$ 0.13  & PS1 & - \\
        & $z'$      & 20.65 $\pm$ 0.11  & PS1 & - \\
        & $y'$      & 20.80 $\pm$ 0.34  & PS1 & - \\
        & $J$       & 19.78 $\pm$ 0.03  & Magellan/FourStar & 2016-03-27 \\		
150818A	& $g'$      & 22.30 $\pm$ 0.16  & P60 & 2016-02-14 \\
		& $r'$      & 22.10 $\pm$ 0.20	& P60 & 2016-02-14 \\
		& $i'$      & 21.70 $\pm$ 0.20	& P60 & 2016-02-14 \\
		& $z'$      & $>$ 21.30 		& P60 & 2016-02-14 \\
        & 3.6$\mu$m & 21.89	$\pm$ 0.05  & \textit{Spitzer}/IRAC & 2017-02-03 \\
161219B & 3.6$\mu$m & 20.70 $\pm$ 0.02  & \textit{Spitzer}/IRAC & 2018-01-04 \\
\hline
\end{tabular}
\begin{tablenotes}[flushleft]
\item \textbf{Notes.} Photometry is not corrected for Galactic foreground extinction. Upper limits are 2-$\sigma$. All photometry is available online in a machine-readable form.
\end{tablenotes}
\end{threeparttable}
\end{table}

\begin{table}
\centering
\caption{New PTF SLSN host photometry.}
\begin{threeparttable}
\label{tab:ptfslsn}
\centering
\begin{tabular}{lccl}
\hline
PTF ID & Filter & AB Mag & Instrument   \\
\hline
PTF09q   & $u'$ &  18.20 $\pm$ 0.08 & SDSS \\ 	
         & $g'$ &  17.13 $\pm$ 0.05 & PS1  \\
         & $r'$ &  16.54 $\pm$ 0.04 & PS1  \\
       	 & $i'$ &  16.14 $\pm$ 0.03 & PS1  \\
         & $z'$ &  15.98 $\pm$ 0.03 & PS1  \\
         & $y'$ &  15.74 $\pm$ 0.06 & PS1  \\
PTF10gvb & $u'$ &  21.10 $\pm$ 0.22 & SDSS \\
         & $g'$ &  20.14 $\pm$ 0.07 & PS1  \\ 
         & $r'$ &  19.85 $\pm$ 0.07 & PS1  \\
         & $i'$ &  19.70 $\pm$ 0.09 & PS1  \\
         & $z'$ &  19.38 $\pm$ 0.12 & PS1  \\
         & $y'$ &  19.89 $\pm$ 0.32 & PS1  \\
PTF11mnb & $u'$ &  20.42 $\pm$ 0.08 & SDSS \\
         & $g'$ &  19.42 $\pm$ 0.02 & PS1  \\
         & $r'$ &  19.27 $\pm$ 0.02 & PS1  \\
         & $i'$ &  18.96 $\pm$ 0.02 & PS1  \\
         & $z'$ &  18.88 $\pm$ 0.03 & PS1  \\
       	 & $y'$ &  18.91 $\pm$ 0.07 & PS1  \\
PTF12gty & $u'$ &  $>$ 21.62        & SDSS \\
         & $g'$ &  $>$ 24.23  		& PS1  \\
         & $r'$ &  $>$ 24.27 		& PS1  \\
         & $i'$ &  23.78 $\pm$ 0.24 & PS1  \\
         & $z'$ &  22.53 $\pm$ 0.21 & PS1  \\
         & $y'$ &  $>$ 24.28		& PS1  \\
PTF12hni & $u'$ &  20.16 $\pm$ 0.20 & SDSS \\
         & $g'$ &  19.19 $\pm$ 0.01 & PS1  \\
         & $r'$ &  18.94 $\pm$ 0.03 & PS1  \\
         & $i'$ &  18.86 $\pm$ 0.02 & PS1  \\
         & $z'$ &  18.56 $\pm$ 0.04 & PS1  \\
         & $y'$ &  18.50 $\pm$ 0.10 & PS1  \\
\hline
\end{tabular}
\begin{tablenotes}[flushleft]
\item \textbf{Notes.} Photometry are not corrected for Galactic foreground extinction. Upper limits are 2-$\sigma$. All photometry is available online in a machine-readable form.
\end{tablenotes}
\end{threeparttable}
\end{table}

\begin{figure*}
\includegraphics[width=\textwidth]{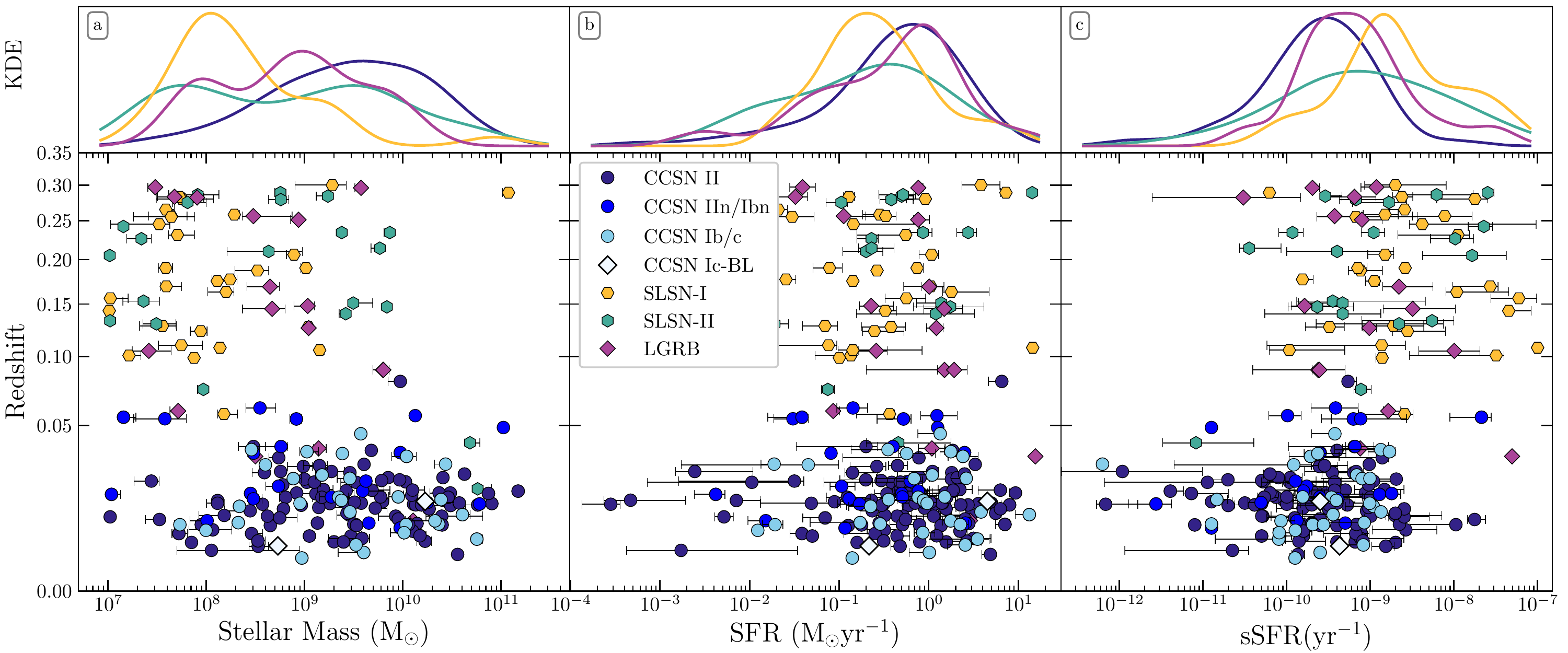}
\caption{Distribution of the physical properties plotted against redshift for each host galaxy sample. Panel (a) shows the stellar mass, (b) the star formation rate, and (c) the specific star formation rate all plotted against redshift using a square root scale. Each upper panel is a Gaussian kernel density estimation of each physical property. For the kernel density estimation all subtypes of CCSNe are grouped together and plotted in dark blue. Redshift evolution is not corrected for in the physical parameters.}
\label{fig:panel_redshift}
\end{figure*}

\subsection{CCSN distances}

We did not have our own spectroscopy for each CCSN host galaxy. Thus, we obtained distances to each galaxy from redshift measurements as published in the NASA Extra-galactic Database (NED; \url{https://ned.ipac.caltech.edu/}) where available (114/150 galaxies).

Since the CCSNe in the sample primarily exploded in very low-redshift galaxies (median luminosity distance $\sim$70 Mpc and all galaxies $<400$ Mpc), they had non-negligible peculiar velocities relative to the motion due to the isotropic expansion of the Universe as described by the Hubble Flow. The fractional distance errors from peculiar velocities could have has implications for the analysis of our hosts. Thus, we corrected for peculiar velocity using the velocity field model in \cite{Mould2000}. This model accounted for peculiar velocities due to the Virgo Cluster, the Great Attractor and the Shapley Supercluster and was typically a 6--8 per cent correction. 

If a catalogued redshift was not available for the host galaxy (34/150), we adopted the redshift of the supernova, since the SN redshift is a good estimator of the host galaxy redshift \citep{Fremling2020}. We estimated the uncertainty based on data from the Bright Transient Survey \citep{Fremling2020}. In this study the authors concluded that the standard deviation of the derived supernova redshift versus the host galaxy redshift was 0.005, therefore we adopted this uncertainty estimate in the distance. 

\section{Physical Parameters}\label{sec:physparams}

\subsection{Spectral Energy Distribution Fitting}\label{subsec:sedfitting}

To quantify the stellar parameters of the host galaxies, including stellar mass and star-formation rate, we modelled the spectral energy distribution (SED) of each host galaxy using UV through NIR photometry. We used the code \textsc{Le PHARE}\footnote{http://www.cfht.hawaii.edu/~arnouts/LEPHARE/lephare.html} \citep{2006A&A...457..841I} which used single-age stellar population synthesis model templates of \citet{bc03} summed according to a single-burst of exponentially declining star-formation. We assumed a Chabrier initial mass function \citep{ChabrierIMF2003} and a stellar metallicity set between 0.2--1.0 Z$_{\odot}$. The contribution of emission lines to the modelled spectra was based on the \cite{Kennicutt1998} relations between SFR and UV luminosity. The contribution of H$\alpha$ and [O~\textsc{ii}] lines to the photometry was included for galaxies with dust free colour bluer than (NUV--\textit{r})$_{\textrm{ABS}}$ $\leq$ 4 and the intensity of the emission lines was scaled according to the intrinsic UV luminosity of the galaxy. Dust attenuation in the galaxy was applied to the SED models using the \citet{Calzetti2000} extinction law for starburst galaxies. If spectroscopy of the host galaxy was available and showed little evidence for nebular emission, we fitted a continuum driven SED model.

\begin{figure*}
\includegraphics[width=\textwidth]{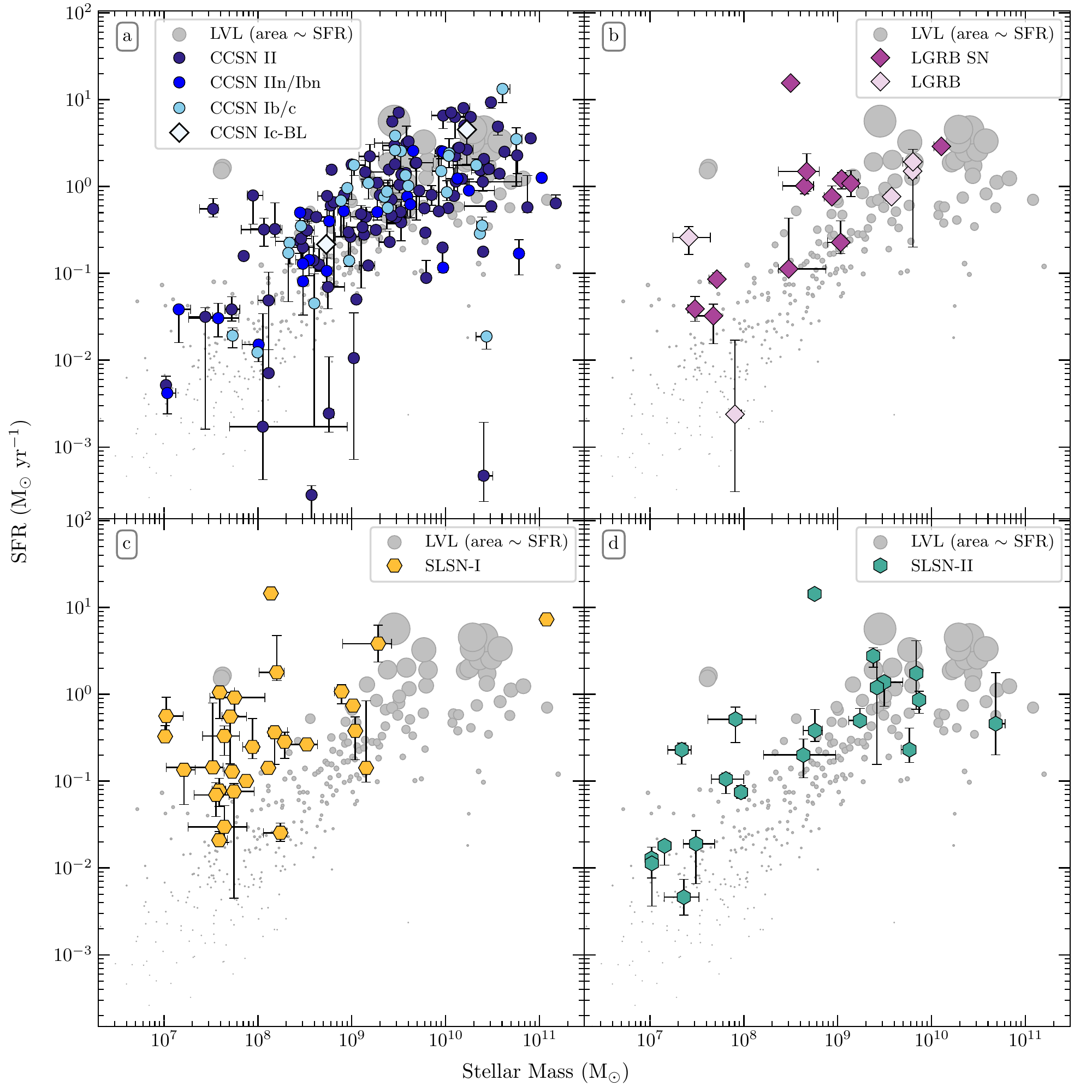}
\caption{Star formation rate vs. stellar mass for each host galaxy class. SFR has not been corrected for redshift evolution. Grey points are the LVL survey galaxies with their sizes scaled in proportion to SFR to show the probability of producing a SN per unit time. Panel (a) shows the unbiased CCSN sample divided into subtypes. Panel (b) shows the LGRB sample in purple; the darker shade indicates where the LGRB was associated with a SN or optical afterglow. Panel (c) shows the SLSN-I sample. Panel (d) shows the SLSN-II sample.}
\label{fig:panel_sfr}
\vspace{5cm}
\end{figure*}
\begin{figure*}
\includegraphics[width=\textwidth]{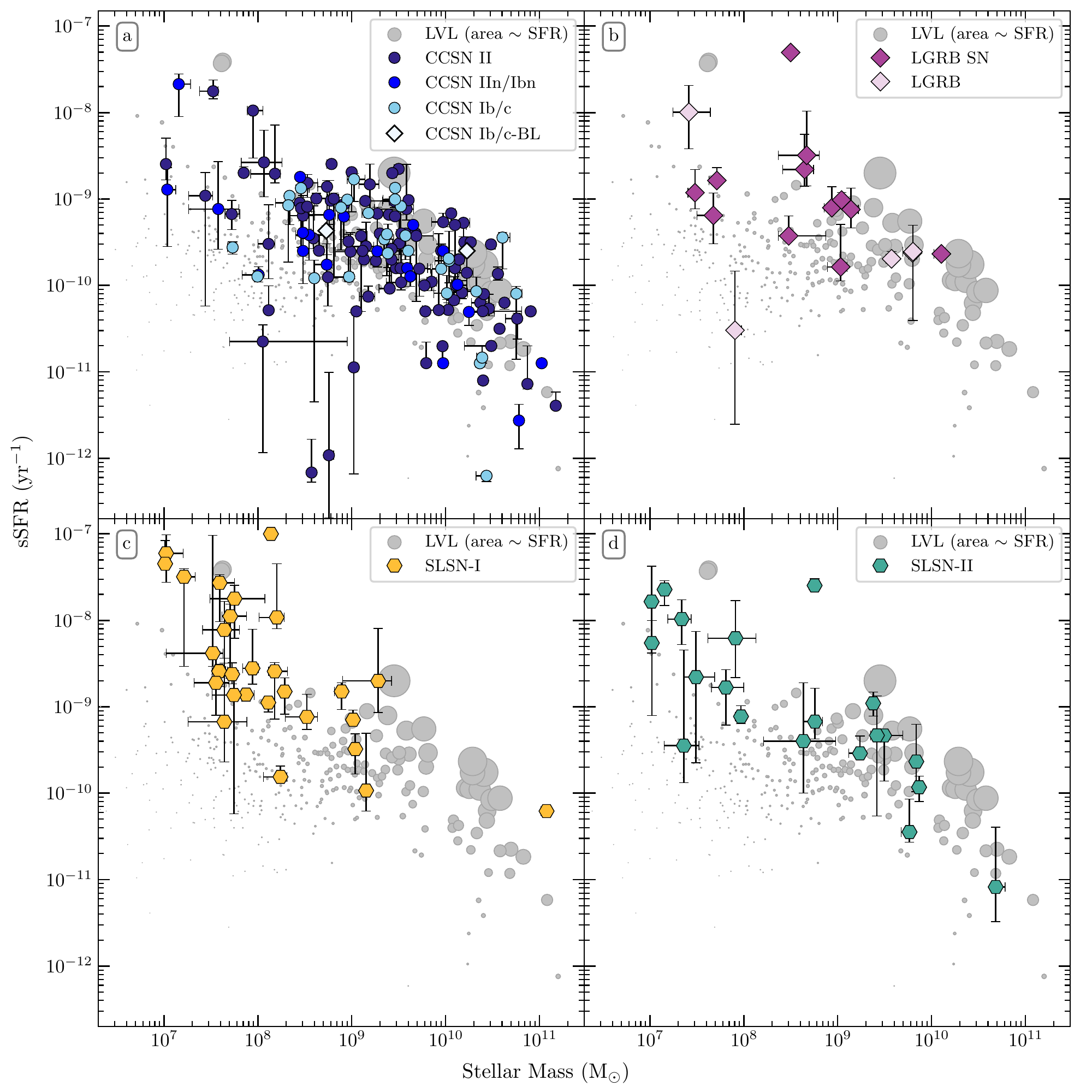}
\caption{Specific star formation rate vs. stellar mass.  The symbols and colours are the same as in Fig.~\ref{fig:panel_sfr}. As in previous figure, SFRs have not been corrected for redshift evolution. SLSN-I show a strong preference for galaxies with high sSFR and/or low stellar mass, (top left of panel c), whereas CCSN are broadly consistent with the distribution of LVL galaxies (panel a). SLSN-II and LGRB hosts (panels a and d respectively) also seem to show a preference towards galaxies with high sSFR and/or low stellar mass compared to CCSNe. There are very few SLSN-II and LGRB hosts with low sSFR and high-mass, but this trend is clearly not as strong as for SLSNe-I.}
\label{fig:panel_specsfr}
\vspace{5cm}
\end{figure*}

To calculate the uncertainties involved in deriving the mass and star-formation rate parameters, we performed a simple Monte Carlo analysis. We chose a random number from a Gaussian distribution in flux space with standard deviation equal to the photometric uncertainty on the derived magnitude for each filter and for each host. We sampled from the distribution 1000 times and then ran the SED fit on each set of `noisy' photometry and used the 16-to-84th percentile of each parameter as an estimate of its uncertainty. If the reduced chi-squared $ \gg 1$ (before the Monte Carlo sampling) and the SED photometry was well-sampled in the UV, optical and IR, we applied the additional uncertainty to the photometry. We applied the uncertainty equally across all photometric points, before the Monte Carlo sampling, in order to more appropriately fit these data until the reduced chi-squared was approximately one, and then we re-ran the Monte Carlo sampling. 

A polycyclic aromatic hydrocarbon \citep[PAH;][]{Leger1984} emission feature is present within the \textit{WISE}/\textit{W1} and \textit{Spitzer}/3.6$\mu$m bands at $z<0.2$. In most galaxies this emission is insignificant compared to the stellar continuum. However, in low-mass galaxies with extreme star-formation, this non-stellar feature can significantly contribute to the flux in the mid-IR. \textsc{Le PHARE} does not account for this emission feature. Thus we investigated if there was any evidence that the 3.6 $\mu$m feature may have affected the flux in this band, given our photometry. The only case where this might have been significant was for the host of LGRB 031203. However, \citet{Watson2011} studied the mid-infrared spectrum and did not find any evidence for PAH emission in the host of LGRB 031203.

\begin{table*}
\label{tab:statprop}
\caption{Statistical properties of galaxy samples. 10th, 50th(median) and 90th percentiles are given for each physical parameter. 1-$\sigma$ uncertainties are given on the median derived parameters. Star-formation rates are not corrected for redshift evolution. Ic-BL are not included as an individual subtype (only as part of the statistic for all subtypes) in this table since there are only two objects in this category.}
\centering
\begin{tabular}{|l|c|c|c|c|c|c|c|c|c|c|c|c|c|c|c|c|}
\hline
\multicolumn{2}{|c|}{}&\multicolumn{3}{c|}{\thead{$z$}} & \multicolumn{3}{c|}{\thead{log$_{10}$M$_{*}$ (M$_\odot$)}} & \multicolumn{3}{c|}{\thead{log$_{10}$SFR (M$_\odot$yr$^{-1}$)}} & \multicolumn{3}{c|}{\thead{log$_{10}$sSFR (yr$^{-1}$)}}&\multicolumn{3}{c|}{\thead{log$_{10}$\big($\Delta$S\big)}}\\
\hline
Transient & N & 10 & 50 & 90 & 10 & 50 & 90 & 10 & 50 & 90 & 10 & 50 & 90  & 10 & 50 & 90 \\
\hline
CCSN (all subtypes) & 150 & 0.005 & 0.014 & 0.033 & 8.1 & 9.5(0.1) & 10.4 & -1.4 & -0.2(0.1)  & 0.5 & -10.7 & -9.6(0.1)  & -8.9 & -0.8 & 0.1(0.1)  & 0.7 \\
CCSN II & 98 & 0.005 & 0.014 & 0.025 & 8.2 & 9.5(0.1) & 10.4 & -1.3 & -0.2(0.1)  & 0.6 & -10.7 & -9.6(0.1)  & -8.8 & -0.8 & 0.1(0.1)  & 0.7 \\
CCSN IIb/Ib/Ic & 29 & 0.004 & 0.014 & 0.035 & 8.3 & 9.5(0.2) & 10.4 & -1.4 & -0.1(0.1)  & 0.4 & -10.2 & -9.6(0.1)  & -8.9 & -0.7 & 0.2(0.1)  & 0.7 \\
CCSN IIn/Ibn & 21 & 0.009 & 0.020 & 0.054 & 7.6 & 8.9(0.4) & 10.2 & -1.5 & -0.4(0.3)  & 0.1 & -10.9 & -9.6(0.2)  & -8.9 & -0.7 & -0.1(0.1)  & 0.3 \\
\hline
SLSN-I       & 29 & 0.105 & 0.177 & 0.281 & 7.5 & 7.9(0.2) & 9.1 & -1.2 & -0.5(0.2)  & 0.3 & -9.6 & -8.6(0.1)  & -7.5 & -0.3 & 0.3(0.1)  & 1.3 \\
SLSN-II      & 21 & 0.074 & 0.210 & 0.284 & 7.2 & 8.8(0.5) & 9.9 & -1.9 & -0.6(0.3)  & 0.2 & -10.4 & -9.2(0.3)  & -7.8 & -0.8 & -0.1(0.1)  & 0.8 \\
LGRB SN      & 12 & 0.033 & 0.146 & 0.280 & 7.7 & 8.7(0.2) & 9.1 & -1.4 & -0.1(0.3)  & 0.4 & -9.6 & -9.1(0.1)  & -8.5 & -0.2 & 0.3(0.2)  & 0.8 \\
SN-less LGRB & 5  & 0.089 & 0.105 & 0.290 & 7.6 & 9.6(0.9) & 9.8 & -1.8 & -0.1(0.6)  & 0.2 & -10.2 & -9.6(0.4)  & -8.6 & -1.0 & 0.1(0.5)  & 0.7 \\ 
\hline
\end{tabular}
\end{table*}

\subsection{Redshift evolution correction}\label{subsec:redshiftcorrection}


The overall SFR density of the Universe, and of individual galaxies, rises rapidly with increasing redshift \citep[e.g.,][]{lilly1996}, making it likely that the rare, luminous SNe that are typically found at higher redshifts than common, less luminous SNe will tend to be found in galaxies with higher star-formation rates simply on account of the effects of cosmic evolution. While we restricted all of our samples to relatively low-redshift ($z<0.3$) since our our ultimate goal was to compare them against each other. Fig.~\ref{fig:panel_redshift} clearly shows that there are still redshift differences between our samples--in particular, between the CCSNe (nearly all at $z\sim0$) and the more extreme superluminous and LGRB supernovae (typically at $z\sim0.2$).

To make a direct comparison between our samples and to avoid systematic errors introduced by cosmic evolution, we corrected for redshift evolution in SFR by empirically re-scaling all star formation rates to $z$ = 0. We did this by measuring the ratio between the expected SFR for a $z$ = 0 galaxy on the main-sequence (for a given host galaxy stellar mass) versus the expected SFR for this galaxy at the redshift of the host SFR$_{ \rm MS\left(M,0\right)}$/SFR$_{ \rm MS\left(M,z\right)}$. We used this ratio to scale the measured SFR and sSFR down to $z$ = 0 as in Eq. \ref{eq:redshiftcorrectioninital}.

\begin{equation}
\rm SFR_{corrected} \, \Equals \, \rm SFR_{measured}  \frac{SFR_{MS\left(M,0\right)}}{SFR_{MS \left(M,z\right)}} \label{eq:redshiftcorrectioninital}
\end{equation}

We parametrized the main-sequence as a power-law, as in Eq. \ref{eq:redshiftcorrection}.

\begin{equation}
\rm SFR_{\rm MS} \, \Equals \, SFR_{0}\left(M_{*} \Frac 10^{10} \, M_{\odot}\right)^{\alpha}  \label{eq:redshiftcorrection}
\end{equation}

Parameter ($\alpha$) is the slope of the galaxy main-sequence and (SFR$_{0}$) describes the normalisation at a stellar mass of 10$^{10}$M$_\odot$, which varies as a function of redshift. Parameters were derived from observational data in \citet{Salim2007} ($z \sim$ 0.1) and \citet{Noeske2007} ($z \sim$ 0.36). The approximate values are (SFR$_{0}$ $\Frac$ M$_{\odot}$ yr$^{-1}$, $\alpha$) = (1.48,0.65) for the galaxy main-sequence at $z \sim$ 0.1 and (2.3,0.67) for $z \sim$ 0.36. We interpolated these parameters ($\alpha$ and SFR$_{0}$) over the redshift range of our sample in order to calculate the SFR of a main sequence galaxy (for a certain stellar mass) at every host redshift and at redshift zero.

We applied this correction to the sSFR and SFR for statistical comparison between the host galaxy populations of CCSNe, SLSNe and LGRBs. Once these corrections were applied, we found the median SFR was reduced by 0.02 dex for CCSNe, 0.42 dex for SLSNe-I, 0.15 dex for SLSNe-II and 0.20 dex for LGRBs. The SFR and sSFR parameters have not been corrected, unless specifically indicated in the text and figure caption. We provide the derived physical parameters from SED fits without applying this SFR correction in Tables~\ref{tab:asassntable}--~\ref{tab:slsnphottable}.

\subsection{Sequence-offset parameter}


As an alternative to applying a redshift evolution correction to the SFR to deal with cosmic evolution, we defined a metric of star-formation intensity, the `sequence-offset' parameter ($\Delta $S). This parameter, given by Eq.~\ref{eq:seqoff}, measures the ratio between the actual, SED-measured star formation rate of a galaxy in our sample (SFR$_{\rm host}$) vs. the predicted SFR (SFR$_{\rm MS}$) for a galaxy on the star-forming galaxy main-sequence (at the same redshift with the same stellar mass), based on the parametrization in Eq.~\ref{eq:redshiftcorrection}.

\begin{equation}
\rm \Delta \, S \, \Equals \, SFR_{host}\left( M_{*},z\right) \Frac \, SFR_{\rm MS}\left( M_{*},z\right)
 \label{eq:seqoff}
\end{equation}

\section{Results}\label{sec:results}

\begin{figure}
\caption{Cumulative distributions of the different galaxy samples with colours the same as in previous figures. We empirically re-scale all star formation rates to $z$ = 0 for all host galaxy samples (CCSNe, SLSNe-I, SLSNe-II and LGRBs) using the procedure in \ref{subsec:redshiftcorrection}. The LVL galaxies (in grey) are weighted here by SFR (step size) to create a galaxy population that traces star-formation. Panel (a): cumulative distributions of all galaxy populations by mass. Panel (b): cumulative distributions of all galaxy populations by star-formation rate. Panel (c) \& Panel (d) show measures of star-formation intensity via sSFR and sequence offsets from star-formation rate compared with the galaxy main sequence at that redshift. CCSN and the weighted LVL are similar, although not identical.}
\includegraphics[width=\columnwidth]{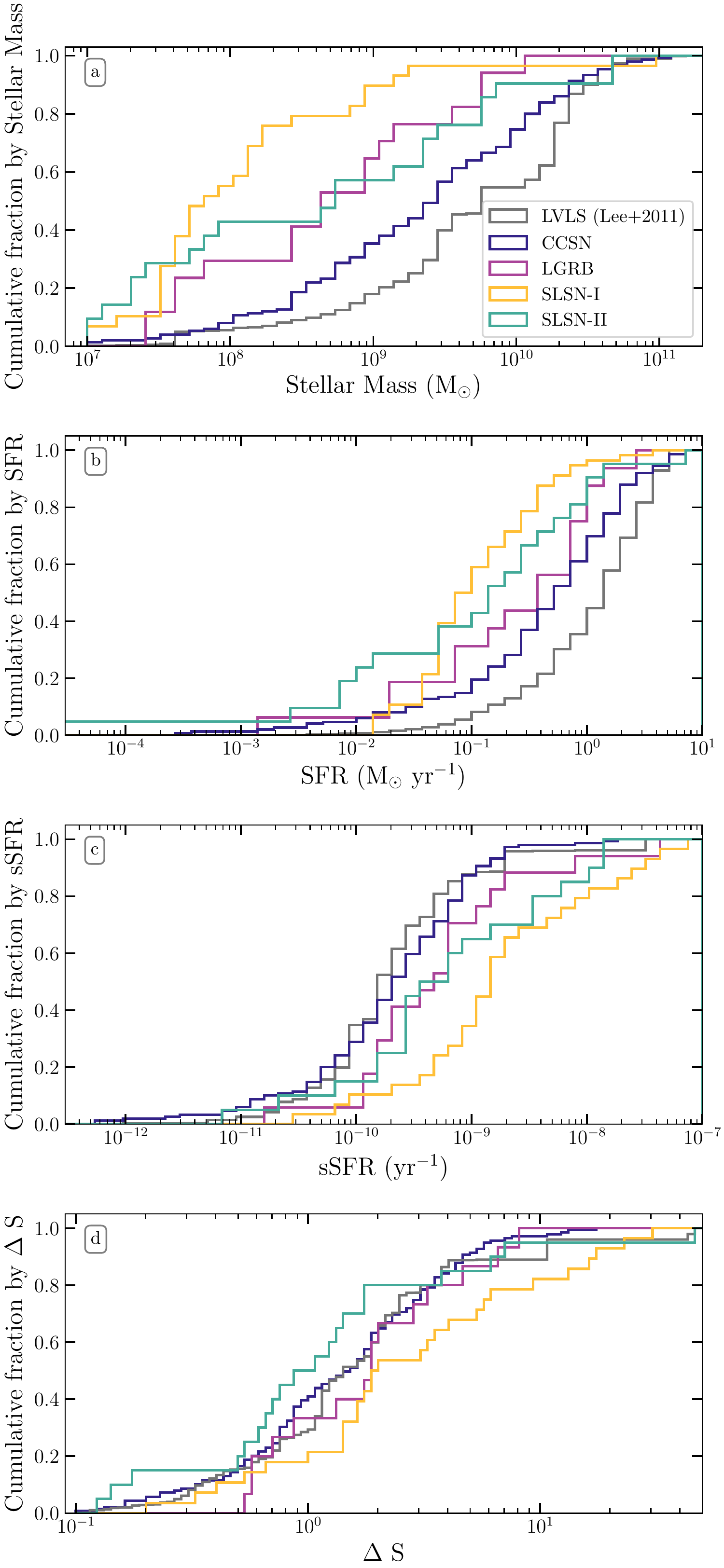}
\label{fig:cumulative_stacked}
\end{figure}

In this section we present the integrated galaxy properties derived from the SED fitting for nearby SLSN, LGRBs and the ASAS-SN CCSN. Basic statistical properties of each sample are summarized in Table~\ref{tab:statprop}. Uncertainties (1-$\sigma$) are calculated using a simple bootstrap.

\subsection{Basic properties of CCSN hosts and comparisons to nearby star-forming galaxies}

A key goal of our study is to produce a uniform and unbiased (by galaxy mass) sample of CCSN hosts, providing a galaxy-luminosity-independent tracer of the sites of star-formation in the local Universe. However, since all of our SN samples are selected via an optical search, highly dusty starbursts and SN environments are likely to be missed in this analysis (see also Appendix \ref{sec:biases} for a description of the possible biases in the SN samples). While our primary motivation for this exercise will be to compare this sample to `exotic' supernova types (SLSNe and LGRBs) in order to constrain their progenitors, our CCSN sample is also useful for studying the nature of star-formation at low-redshift: few galaxy surveys are complete beyond the dwarf galaxy $\lesssim$10$^{9}$ M$_\odot$ limit, with those that are typically confined to small volumes. 

In Fig.~\ref{fig:panel_sfr}a, we present the distribution of SFR vs. stellar mass for core-collapse SNe as compared to galaxies from the Local Volume Legacy (LVL) survey of a volume-complete sample of galaxies within $\sim$11 Mpc. The stellar masses of the LVL galaxies are derived from SED fits (Johnson et al. in prep) and star formation rates derived from H$\alpha$ flux \citep{Lee2011}\footnote{Note that this SFR indicator is different from the one employed in our SED analysis; we provide it as a visual comparison indicator and because it has been employed as the comparison sample in earlier transient host studies (in particular, \citealt{2016Perley}). We also statistically compare the LVL vs. CCSN sample using both H$\alpha$ and UV SFRs.}. Most LVL galaxies are observed to populate the main sequence of star-forming galaxies, where mass and star-formation rate are strongly correlated in a fairly narrow band of specific star-formation rate between 10$^{-9}$--10$^{-10}$ yr$^{-1}$. 

If the SN rate strictly tracks the star-formation rate, then the distribution of SN host masses should follow the distribution of galaxy masses, re-weighted by star-formation rate. As expected, CCSNe populate star-forming galaxies across their entire mass distribution--probing large spiral galaxies with stellar masses $\sim$10$^{11}$ M$_{\odot}$ down to the low-mass dwarf galaxy regime with stellar masses of $\sim$10$^{7}$ M$_{\odot}$. However, the SN host mass distribution is similar to the SFR-weighted galaxy mass distribution but they are not strictly consistent: the median SFR-weighted log stellar mass of LVL galaxies is 9.8(0.1), 0.3 dex higher than the median mass of CCSN hosts 9.5(0.1) (the associated Anderson--Darling $p$-value is $p_{\rm AD}<0.001$). This may be a result of small-scale inhomogeneities associated with the LVL covering a small volume (e.g., an overabundance of large galaxies due to large-scale structure) and demonstrates the importance of obtaining a sample selected via SNe. Similarly small but statistically significant differences are also seen in other parameters (SFR, sSFR, and sequence offset). We find the median stellar mass 9.5(0.1) is slightly higher in comparison to the Dark Energy Survey CCSN sample of 47 objects (9.4) \citep{Wiseman2020}, but still within the uncertainties of the measurements. 

A few CCSN galaxies in Fig.~\ref{fig:panel_sfr}a show very low star-formation rates despite high masses -- specifically 14de and 16am. Morphologically, these galaxies are not classical spiral galaxies, neither are they elliptical galaxies. These galaxies do have red colours and the uncertainties on the SFRs derived for these galaxies are likely underestimated by our SED fitting procedure at minimum. Genuine elliptical galaxies are expected to contribute very little to the cosmic supernova rate, although previous examples have been reported \citep[e.g.,][]{Irani2019}.

The fraction of star-formation in very faint or very rare galaxies that are poorly probed by traditional flux- or volume-limited galaxy surveys is of particular interest. Using our sample, we measure the fraction of CCSNe in dwarf galaxies and the fraction in `starburst' galaxies. We use the Bayesian beta distribution quantile technique to derive the 1-$\sigma$ uncertainties following methods outlined in \cite{Cameron2011}. We find 33$^{\Plus 4}_{\Minus 4}$ per cent of CCSNe (50/150 from our sample) occur in dwarf galaxies with stellar masses less than 10$^9$ M$_{\odot}$ and 7$^{\Plus 3}_{\Minus 2}$ per cent of CCSNe (11/150 from our sample) occur in dwarf galaxies with stellar masses less than 10$^8$ M$_{\odot}$. These fractions are substantial, emphasizing the importance of dwarf galaxies to the ongoing star formation rate density in the local Universe, together with potential future chemical enrichment of their environments. However, only 2$^{\Plus 2}_{\Minus 1}$ per cent (3/150) of CCSN hosts are undergoing very rapid star-formation in a starburst galaxy (sSFR>10$^{-8}$ yr$^{-1}$), all of which are dwarf galaxies. Thus, we find the vast majority of star-formation in the local Universe does not occur in starbursting galaxies. This is in agreement with the LVL survey \citep{Lee2009} which found that only a few per cent of the galaxies are now in a bursting mode (defined in their analysis as having a H$\,\alpha$ equivalent width $>$100 \AA). \citet{Brinchmann2004} estimated that $\sim$20 per cent of local star-formation occurs in starburst galaxies using a volume-corrected sample of galaxies from SDSS DR2, although their definition of a starburst differs from ours and is much more generous (they require that the ratio of between the present SFR and the mean past SFR ($b$) is 2--3, which corresponds to a specific star-formation rate threshold of approximately 10$^{-9.75}$ yr$^{-1}$).  The fraction of strongly starbursting galaxies in SDSS is clearly much lower (see e.g., their figure 22), but cannot easily be quantified because most such star-formation is in galaxies with stellar masses below the SDSS completeness limit.

\subsection{Basic properties of exotic SN hosts}

In Fig.~\ref{fig:panel_sfr}b--d we also plot the mass and SFRs of the `exotic' SN samples in comparison to local galaxies. These populations are clearly quite different from ordinary CCSNe. The peak of the SLSN-I host mass distribution is much lower than that of the CCSN population, with a median log stellar mass of 7.9(0.2), though notably, there are a few outliers in galaxies with relatively high masses (PTF10uhf, SN2017egm and PTF09q). SLSN-II and LGRBs with observed associated SNe lie intermediate between the SLSN-I and CCSN samples with median logarithmic mass of 8.8(0.5) and 8.7(0.2) respectively (SN-less LGRBs have masses more consistent with CCSN with a median logarithmic stellar mass of 9.6(0.9), although this is poorly constrained). 

Unlike CCSNe, SLSNe and LGRBs frequently populate galaxies above the galaxy main sequence with a median logarithmic sSFR of -8.6(0.1) for SLSNe-I, -9.2(0.3) for SLSNe-II, and -9.1(0.1) for LGRB SNe. SN-less LGRBs have sSFR of -9.6(0.4) which is more consistent with the CCSN population. This effect can be seen more clearly in Fig.~\ref{fig:panel_specsfr}. which shows specific star formation vs. stellar mass. The impartially selected CCSNe are consistent with star-forming local galaxies, whereas $\sim$70 per cent of SLSNe-I lie above the star-forming galaxy main sequence with specific star-formation rates exceeding 10$^{-9}$ yr$^{-1}$. This places many SLSN-I hosts in the top left of this diagram, with 8 hosts with specific star formation rates exceeding 10$^{-8}$ yr$^{-1}$, which is much more than expected if the SLSN rate purely traces SFR \citep[this has also been noted by others; e.g.,][]{Lunnan2014, 2016Perley, Schulze2018}. These galaxies (with specific star formation significantly above this main sequence) are sometimes referred to as starbursts. There are 8 ($\sim$30 percent) SLSN-I galaxies with specific star formation rates exceeding 10$^{-8}$ yr$^{-1}$ (which we will define as a `starburst' for the purpose of this paper). This is in qualitative agreement with other studies, such as in \cite{Leloudas2015} where $\sim$50 per cent of SLSNe-I were found in EELGs indicative of an intense starburst episode within the galaxy. \citet{2016Perley} and \citet{Schulze2018} also noted that many SLSN-I host galaxies in PTF and SUSHIES samples are undergoing intense star-formation. 

\begin{table}
\caption{Two sample Anderson--Darling probabilities between CCSNe, the LVL weighted by SFR, SN host galaxy samples (SLSNe-I, SLSNe-II, LGRB-SNe and SN-less LGRBs) and between LGRBs with and without supernova. We empirically re-scale all star formation rates to $z$ = 0 for all host galaxy samples (CCSNe, SLSNe-I, SLSNe-II and LGRBs) using the procedure in \ref{subsec:redshiftcorrection}. Samples that differ at $p_{\rm AD}<0.05$ for that parameter are in boldface. The combined sample size of the two comparisons are given in the effective size column.}
\begin{threeparttable}
\label{tab:adtest}
\begin{tabular}{p{0.12\linewidth} p{0.37\linewidth} p{0.15\linewidth}p{0.20\linewidth}}
\hline
Parameter  & Comparison & $p_{\rm AD}$--value & Effective size\\
\hline
Mass
& CCSN--LVL     & \bf $<$1e-03 & 350 \\
\hline
& CCSN--SLSN-I     & \bf $<$1e-03 &179 \\
& CCSN--SLSN-II    & \bf 1.90e-03    &171\\
& CCSN--LGRB SN    & \bf 5.67e-03    &162\\
& CCSN--SN-less LGRB & $>$0.25       &155\\
\hline
&  LGRB SN--SN-less LGRB &$>$0.25 & 17 \\
&  LGRB SN--SLSN-I & 0.058        & 41 \\
\hline
SFR
& CCSN--LVL (UV)       & \bf $<$1e-03& 350\\
\hline
& CCSN--SLSN-I      & \bf $<$1e-03 &179 \\
& CCSN--SLSN-II     & \bf 4.96e-03 &171\\
& CCSN--LGRB SN     &   $>$0.25  &162\\
& CCSN--SN-less LGRB  &$>$0.25 &155\\
\hline
& LGRB SN--SN-less LGRB &$>$0.25  & 17 \\
& LGRB SN--SLSN-I & 0.051 & 41 \\
\hline
sSFR
& CCSN--LVL (UV)       & \bf $<$1e-03 & 350 \\
\hline
& CCSN--SLSN-I     &   \bf 1.05e-03  &179 \\
& CCSN--SLSN-II    &   \bf 3.36e-03  &171\\
& CCSN--LGRB SN    &   \bf 2.49e-02  &162\\
& CCSN--SN-less LGRB & 0.059  &155\\
\hline
 & LGRB SN--SN-less LGRB & \bf 3.5e-02  & 17 \\
 & LGRB SN--SLSN-I LGRB & 0.11  & 41 \\
\hline
$\Delta$SFR
& CCSN--LVL (UV)	 & \bf $<$1e-03 &350\\
\hline
& CCSN--SLSN-I     & \bf $<$1e-03 &179 \\
& CCSN--SLSN-II    &    $>$0.25  &171\\
& CCSN--LGRB SN    &    0.19  &162\\
& CCSN--SN-less LGRB & $>$0.25  &155\\
\hline
& LGRB SN--SN-less LGRB & $>$0.25  & 17 \\
& LGRB SN--SN-SLSN-I & $>$0.25  & 41 \\
\hline
\end{tabular}     
\end{threeparttable}
\end{table}

\subsection{Relative Rates of SN Subtypes}\label{sec:explrelrate}

While we can qualitatively observe that the distributions of certain samples in Fig.~\ref{fig:panel_redshift}--~\ref{fig:panel_specsfr} seem similar or dissimilar, this is not a statistical statement. We employ several different methods to quantify the significance and model the nature of these apparent differences below.

\subsubsection{Cumulative Distribution Tests}

In Fig.~\ref{fig:cumulative_stacked}, we show the cumulative distributions of mass, star-formation rate, specific star formation rate and sequence offset for each of our galaxy samples. The step size of local galaxies in LVL are weighted by star-formation to create a population consistent with one that traces star-formation. The CCSNe and LVL samples have remarkably similar sSFR and  $\Delta$S distributions, while the rarer SN subtypes seem to show different distributions in most properties. These differences can be tested formally using Anderson--Darling tests.

We compute the Anderson--Darling (AD) statistic, and associated $p$-value, for each pair of samples and for each parameter of interest: stellar mass, SFR, sSFR and the sequence offset parameter ($\Delta$S). The results are summarized in Table \ref{tab:adtest}. SLSN-I are statistically distinct from the CCSN in every parameter ($p_{\rm AD}>$0.05): mass ($p_{\rm AD}<$1e-03), SFR ($p_{\rm AD}<$1e-03) , sSFR ($p_{\rm AD}$=1e-03) and $\Delta$S ($p_{\rm AD}<$1e-03). This population shows the most divergent properties out of all galaxy samples. SLSN-II fall intermediately between these two populations and are statistically distinct from CCSN in terms of mass ($p_{\rm AD}$=2e-03), SFR ($p_{\rm AD}$=5e-03), and sSFR ($p_{\rm AD}$=3e-03). 

\subsubsection{Relative rate formalism for univariate comparisons}

While the Anderson--Darling tests above confirm that differences exist between some distributions, they do not tell us anything about the degree or quantitative nature of the differences between any two distributions.

To gain further insight into the differences between the distributions of different samples, we define a new quantity which we refer to as the \emph{relative} rate (designated $\Re$). This quantity measures how more frequent a specific type of SN (type `A', typically an exotic class of SN) is compared to another type of SN (type `B', typically a normal class of SN) in a specific type of galaxy, compared to the Universe as a whole. Expressed in terms of a single parameter $y$ (which can be mass, SFR, etc.), it is the ratio of the inferred probability density functions of the two SN types: 

\begin{equation}
\Re_{\rm A\big/B}\big(y\big) \Equals \frac{\rm PDF_{\rm A}\big(y\big)}{\rm PDF_{\rm B}\big(y\big)}
\end{equation}

A relative rate $\Re$ = 1 for all values of $y$ would indicate that the distributions over $y$ for A and B are identical (although the \emph{absolute} rates may not be the same). Otherwise, regions over $y$ with $\Re>1$ indicate environments where production of SNe of type A is enhanced relative to B; regions with $\Re<1$ indicate environments where production of type A is suppressed relative to B.

In practice, we use a sliding-window method to estimate $\Re$ for each parameter of interest (stellar mass, star-formation rate, specific star-formation rate, or sequence offset). The PDF function for each parameter for each sample (A or B) is estimated by calculating the proportion $P_i$ of host galaxies in that sample with parameter values within $\pm$0.5 dex of a grid of bin centres, $y_i$. If the number of galaxies within $\pm$0.5 dex of $y_i$ is $n_i$ and the sample size is $N$, this is then (for sample A):

\begin{equation}
P_{\rm A}\big(y_{i}\big) \Equals \frac{\rm n_{{\rm A},i}}{\rm N_{\rm A}}
\end{equation}

The (estimated) \emph{relative} rate of one transient compared to another, $\Re_{\rm A \big/ B}$, is then the ratio of the two $P$ arrays:  

\begin{equation}
\Re_{\rm A \big/ B}\big(y_{i}\big) \Equals \frac{P_{\rm A}\big(y_{i}\big)}{P_{\rm B}\big(y_{i}\big)} \Equals \frac{\rm n_{{\rm A},i}\rm N_{\rm B}}{\rm n_{{\rm B},i} N \rm _{\rm A}}
\end{equation}

The bin centres are defined in logarithmic intervals of 0.1 dex, such that every 10th window has no overlap with the first. For example, the window is evaluated between a mass of 1$\times$10$^{6}$ M$_{\odot}$ to 1$\times$10$^{7}$ M$_{\odot}$ (centred at 3.16$\times$10$^{6}$ M$_{\odot}$), then at 1.26$\times$10$^{6}$ M$_{\odot}$ to 1.26$\times$10$^{7}$ M$_{\odot}$ (centred at 3.98$\times$10$^{6}$ M$_{\odot}$), etc. Note that because windows within 1 dex overlap, values of $\Re$ within 1 dex of each other are not fully independent.

To calculate the confidence intervals on the relative rate we draw a new CCSN sample and a new SLSN sample from the original samples (with replacement) for 1000 bootstrap iterations. We derive a relative rate for each bootstrap iteration and determine the 2-$\sigma$ uncertainties based on the bootstrapped relative rate function.

\subsubsection{Relative rate formalism for bivariate comparisons}

Testing on a single parameter at a time will not be able to distinguish between fundamental differences vs. those that originate purely due to correlations with other parameters: many galaxy parameters (e.g., SFR and stellar mass) are strongly correlated, making it is difficult to tell which parameter is more directly related to the special conditions that appear necessary for SLSN production or LGRB production.

However, our relative-rate formalism above can be extended to ascertain whether a difference in distributions associated with a control parameter (e.g., stellar mass) can completely explain an observed difference in distributions for another parameter (e.g., SFR). To test this, we \emph{reweight} the comparison sample (sample `B'). The weights for each galaxy in the comparison sample are interpolated from the relative-rate for the control parameter. For example, the host masses are weighted based on the relative rate weights for the sSFR. We use the same confidence intervals derived from the univariate bootstrap procedure and rescale them using the same factor to the weighted relative rate. We then test whether the relative rate for sample `A' vs. the reweighted sample `B' is consistent with a constant $\Re$=1 over the entire range of the test parameter.

\subsection{SLSNe-I vs. CCSNe}

\begin{figure*}
\includegraphics[width=\textwidth]{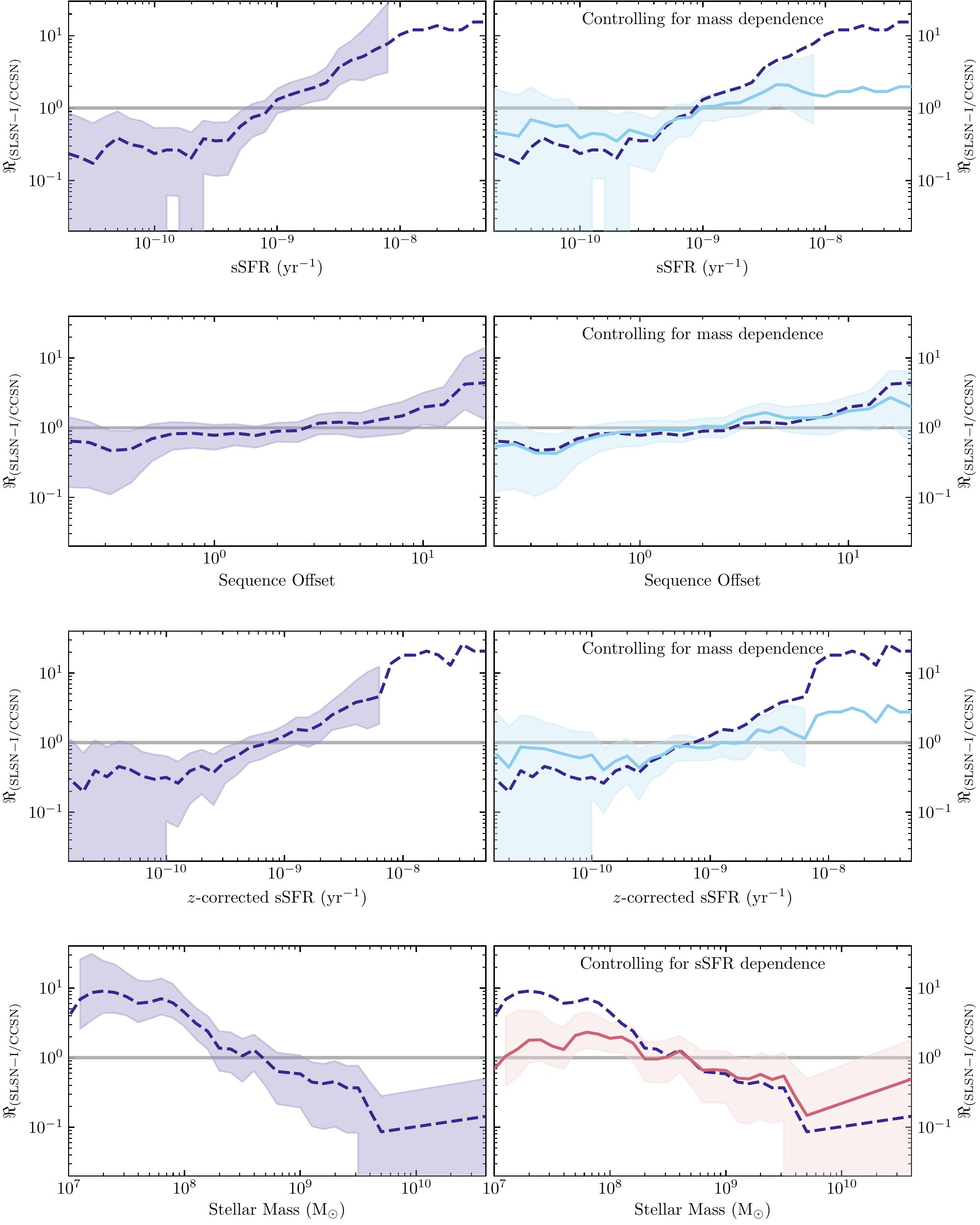}
 \caption{Relative rates of SLSNe-I to CCSNe for various host galaxy parameters. Left panels show the relative rates in purple given by the dashed lines for specific star formation rate, sequence offset (redshift corrected to z$\sim$0) and stellar mass in a moving window function with a width of 1 dex. The window function moves such that after it has moved 10 times it has no overlap with the first window. 2-$\sigma$ confidence intervals are shown in a lighter shade and when there are too few of either samples, the confidence intervals are not shown for these regions. Right panels show the same quantity, but after controlling for the modelled dependence on the alternative variable (stellar mass for SFR-related quantities, or SFR for mass-related quantities). Light blue lines are mass-controlled rates and red lines are the sSFR controlled rates.}
\label{fig:specsfrvsmass_ssfr_mass}
\end{figure*}

The relative rate, $\Re$, of SLSNe-I vs. CCSNe is plotted in the left panels of Fig.~\ref{fig:specsfrvsmass_ssfr_mass} as purple dashed lines with the 2-$\sigma$ confidence intervals in a lighter colour against sSFR, sequence offset, redshift corrected sSFR scaled to z$\sim$0 and stellar mass. The grey line indicates the same relative rate. 

SLSNe-I are enhanced in galaxies with sSFR exceeding 10$^{-9}$ yr$^{-1}$ (after correcting for redshift evolution) and strongly enhanced (by a factor of $\sim$10) for sSFR exceeding 10$^{-8}$ yr$^{-1}$. The rate is also enhanced for galaxies with a sequence offset parameter $\Delta \rm S>5$, which corresponds to galaxies with $\rm SFR>5$ times that predicted of galaxies on the main sequence with the same stellar mass and redshift. The bottom left panel shows that the rate is increased for galaxies with stellar mass less than 2$\times$10$^{8}$ M$_\odot$.

To investigate whether SLSN host galaxy mass (a proxy for metallicity) or specific star formation rate (a proxy for star-formation intensity) is more closely related to the factor driving the production of these events, we must correct for the co-variation between these two parameters. As described above, we remove the effects of a possible dependence in the relative rate of SLSNe to CCSNe as a function of specific star formation rate by controlling for the mass dependence in order to see whether specific star formation rate alone can explain the over-abundance of SLSN-I relative to CCSN. We also do the reverse, in order to see whether a specific star formation rate dependence alone would explain the observed apparent mass dependence in the relative rate.)

The right panels of Fig.~\ref{fig:specsfrvsmass_ssfr_mass} show the original relative rates as a purple dashed line. The light blue and red solid lines show the rates when one controls for mass dependence or sSFR dependence respectively. The covariance-corrected rates do appear to broadly level off (at a 2-$\sigma$ confidence level) to an equal rate (grey line), suggesting that either mass dependence or sSFR alone can explain the difference in relative rates between the CCSNe and SLSNe in our sample. However, in rows three and four, the covariance-corrected rates possibly significant deviance from an equal rate at sSFR $>$8$\times$10$^{-9}$ yr$^{-1}$ and from the mass at $<$2$\times$10$^8$ M$_\odot$.  This may hint that the rate of SLSNe-I production is increased as a result of high sSFR \textit{and} low stellar mass. A larger sample size should help to solidify this claim.  

\subsection{LGRBs vs. CCSNe}
\begin{figure*}
 \includegraphics[width=\textwidth]{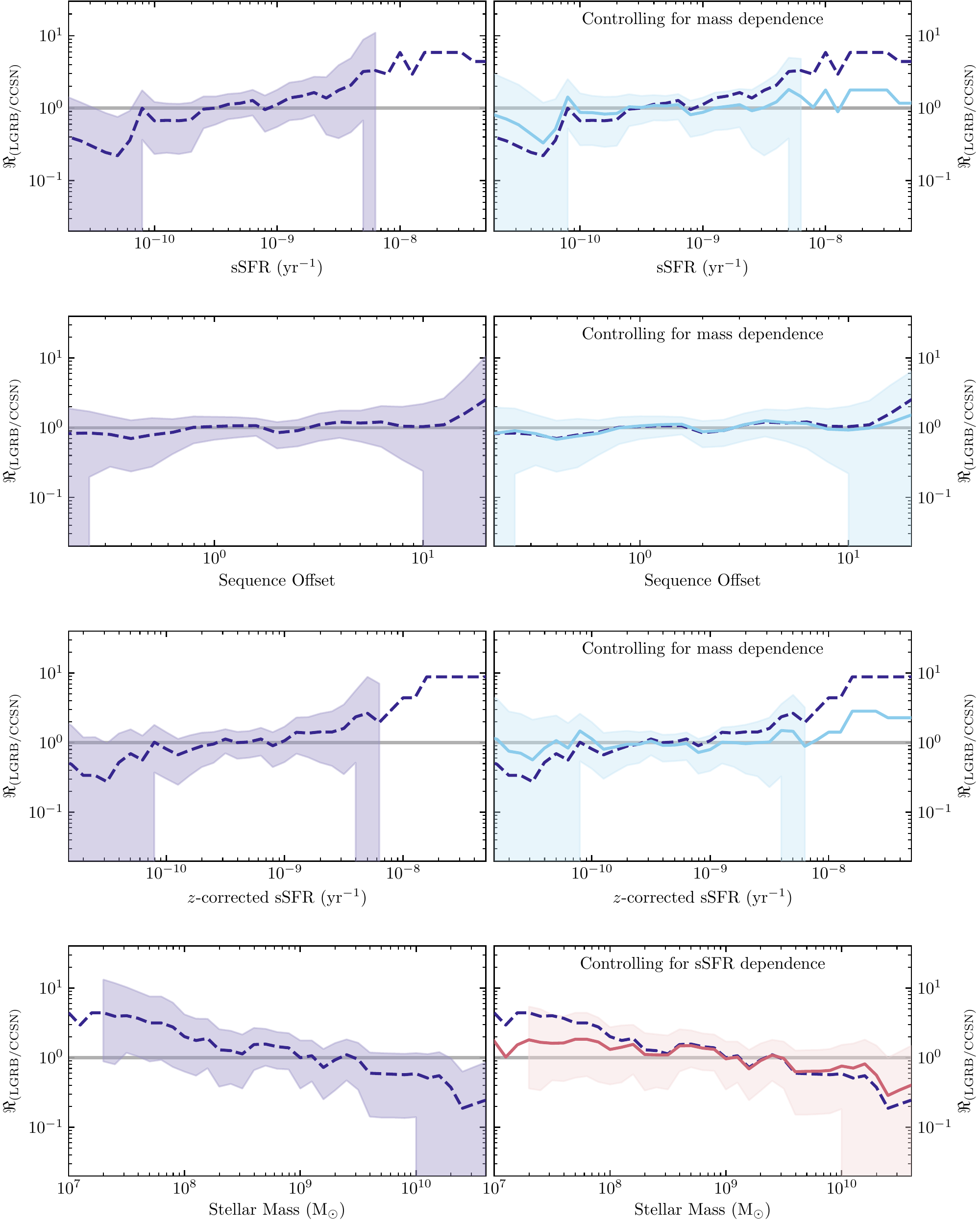}
 \caption{Relative rates of LGRBs to CCSNe for various host galaxy parameters. Left panels show the relative rates in purple given by the dashed lines for specific star formation rate, sequence offset (redshift corrected to z$\sim$0) and stellar mass in a moving window function with a width of 1 dex. The window function moves such that after it has moved 10 times it has no overlap with the first window. 2-$\sigma$ confidence intervals are shown in a lighter shade and when there are too few of either samples, the confidence intervals are not shown for these regions. Right panels show the same quantities, but after controlling for the alternative variable as in Fig.~\ref{fig:specsfrvsmass_ssfr_mass}. Light blue lines are mass-controlled rates and red lines are the sSFR controlled rates.}
\label{fig:specsfrvsmass_ssfr_mass_grb}
\end{figure*}

Using the same method as described above, we also calculate the relative rate $\Re$ of LGRBs vs. CCSNe in Fig.~\ref{fig:specsfrvsmass_ssfr_mass_grb}. Given the rather limited low-$z$ LGRB sample the results are generally less constraining than for SLSNe, and we cannot conclusively (for any 1-dex bin) state that $\Re \ne 1$ for LGRBs versus SNe given this analysis.

Formally, the relative rate of LGRBs is enhanced in galaxies with sSFRs exceeding 10$^{-9}$ yr$^{-1}$ (after correcting for redshift evolution) by a factor of $\sim$3; it is enhanced in galaxies with sequence offsets $>$2 by a factor of approximately 2, and it is enhanced in low-mass dwarfs $<10^8$ M$_\odot$ by a factor of approximately 2.5. As with SLSNe, these effects are degenerate and given the small sample sizes, we cannot yet determine which parameter (if any) is the primary cause of the differences.
  
\subsection{SLSNe-I vs. LGRBs}

We can also compare the LGRB and SLSN-I host populations directly against each other. In our work, we find that SLSNe-I and LGRBs are statistically consistent with being drawn from the same galaxy populations in terms of all measured parameters (see Table~\ref{tab:adtest}), similar to the findings of \citet{Japelj2018}. However, the AD values for mass ($p_{\rm AD}$=0.058) and SFR ($p_{\rm AD}$=0.051) are right on the threshold ($p_{\rm AD}$=0.05) for a statistically distinct population. We do find that SLSNe-I seem to be in less massive galaxies in comparison to LGRBs. SLSNe-I have a median logarithmic stellar mass of 7.9(0.2), while LGRB SNe have a median stellar mass of 8.7(0.2). This is a similar conclusion to that found in \citet{Lunnan2014},\citet{Leloudas2015} and \citet{Schulze2018}. However, we note that due to our selection of nearby events, our sample size for LGRBs is smaller than in these studies. 

In terms of sSFR, we do not find any statistical differences ($p_{\rm AD}$=0.11). Our results are fully consistent with those of \citet{Leloudas2015} who found the median sSFR (SFR determined via spectroscopic line measurements) was more strongly star-forming in SLSN-I compared to LGRBs with logarithmic sSFR of --8.53 yr$^{\Minus1}$ for SLSNe-I and --9.15 yr$^{\Minus1}$ for LGRBs. We find sSFRs --8.6(0.1) yr$^{\Minus1}$ for SLSNe-I and --9.1(0.1) yr$^{\Minus1}$ for LGRB-SNe. LGRBs and SLSNe-I both have a higher median logarithmic sSFR than CCSNe --9.6(0.1). There is a 0.5 dex difference between the median sSFR of SLSNe-I and LGRBs (although this difference statistically significant from the AD test), which is in agreement with \citet{Leloudas2015,Schulze2018}, but the comparison is somewhat limited by smaller sample of low-redshift LGRBs. 

\subsection{SN-less LGRBs vs. LGRB-SNe}

To address whether the sub-population of `SN-less' LGRBs may represent a distinct class from the remainder of LGRBs, we compare the host properties of the five events above to the remainder of the sample (Table.~\ref{tab:statprop}). While some SN-less LGRB hosts are individually unusual, their cumulative properties are not significantly different from the hosts of LGRBs with confirmed SNe (see Table \ref{tab:adtest}), although the redshift corrected sSFR may show some difference ($p_{\rm AD}$=0.04). However, this comparison is not strongly constraining given the small size of the SN-less sample (5 objects) and the possibility that some of these events hosted ordinary LGRB-SNe which were dust-obscured\footnote{A more detailed discussion of this issue can be found in Appendix \ref{sec:biases}.}.

\newpage 
 
\section{Summary and conclusions}\label{sec:conclusions}
In this paper we presented photometric observations of 150 galaxies hosting CCSN discovered or observed by the ASAS-SN in order to provide the most comprehensive and unbiased comparison sample of CCSN host galaxies, and compared the properties of this sample to 70 nearby ($z<0.3$) SLSNe and LGRB SN hosts analyzed using a consistent methodology.Our key conclusions are summarized as follows:
\begin{enumerate}[I.]
\item CCSNe generally exhibit similar star-formation properties to star-formation-weighted local galaxies (LVL), consistent with the expectation that CCSNe should trace star-formation. However, we find the CCSN-selected galaxy stellar mass distribution to be weighted towards slightly lower-mass galaxies 9.5(0.1) than the SFR-weighted LVL galaxy stellar mass distribution with median log stellar mass of 9.8, possibly indicating that the local-volume sample is not truly representative of the average distribution of low-redshift galaxies. 

\item 33$^{\Plus 4}_{\Minus 4}$ per cent of CCSNe (50/150) from our sample) occur in dwarf galaxies with stellar masses less than 10$^9$ M$_{\odot}$ and 7$^{\Plus 3}_{\Minus 2}$ per cent of CCSNe (11/150 from our sample) occur in dwarf galaxies with stellar masses less than 10$^8$ M$_{\odot}$, representing a substantial fraction of the population. 

\item Only a few per cent (2$^{\Plus 2}_{\Minus 1}$) of CCSN hosts are undergoing starbursts with rapid star-formation sSFR>10$^8$ yr$^{-1}$, all of which are dwarf galaxies with stellar masses $<$10$^9$ M$_{\odot}$.

\item LGRB SN and SLSN-I host populations exhibit similar host galaxy properties. The peak of their host mass distributions is clearly much lower and spans a much smaller mass range than the CCSN population which trace star-formation (with median logarithmic mass of 8.7(0.2) for LGRB SNe, 7.9(0.2) for SLSNe-I, and 9.5(0.1) for CCSNe. LGRB SNe explode, on average, in higher mass galaxies than SLSNe-I. This lends further support to models in which LGRBs and SLSNe-I form only in certain environmental conditions related to low-mass and metallicity.

\item We do not find statistically significant differences between LGRB-SN and SN-less LGRBs. However, this comparison is limited by the small sample size of SN-less events (only 5).

\item Many (8/29) SLSNe-I are found in starbursts. This greater fraction is consistent with an intrinsic preference for starbursting galaxies, but is also consistent with a strong SLSN-I mass dependence in covariance with a larger starburst fraction in dwarf galaxies. We cannot yet conclusively identify or rule out a role for intense star-formation in increasing the SLSN-I rate in starbursting dwarf galaxies.
\end{enumerate}

Here we provided an unbiased sample of photometrically-derived properties of CCSN host galaxies, and directly compared them to a consistent analysis of all known SLSN and LGRB host galaxies. These catalogues are all included in a machine readable format and could be used for host preferences of events broad applicability, including unusual classes of object--such as the emerging category of fast blue transients \citep[e.g.,][]{Drout2014}. 

In future work, we will increase our sample size, to try to better disentangle the role of sSFR and stellar mass in SLSN-I production. We will also gather spectroscopy of the dwarf galaxies hosting CCSN and narrow-band H$\,\alpha$ to study their chemical abundances and star-formation histories in more detail and thus obtain a deeper understanding of star formation in dwarf galaxies hosting CCSN. New, deeper all sky-surveys such as ATLAS \citep{Tonry2018} and ZTF \citep{Fremling2018rcf}, and eventually LSST, will increase the sample size of host unbiased samples of CCSNe and SLSNe significantly -- as will comprehensive analysis of completed surveys \citep[e.g., iPTF; ][]{Schulze2020}.

\section*{Data availability}\label{sec:dataavailability}
The photometry used in this work is compiled into one machine-readable table and is available in the online supplementary material.

\section*{Acknowledgements}\label{sec:acknowledgements}

First we thank the referee for the careful and thorough read of our paper.  In particular for their carefully structured comments, and helpful grammar suggestions which simplified the process of revising our study and have substantially improved the clarity and readability of the paper. We would also like to thank N. Blagorodnova and S. Kulkarni for their Keck data and S. Kim and S. Schulze for kindly allowing us to use their Magellan data for LGRB\,150518A. We acknowledge important feedback from K. Stanek who pointed out that we should include the `recovered' ASAS-SN events in addition to those `discovered' by ASAS-SN. This comment led to and increased CCSN sample size, better sample selection and has substantially improved the paper. We also acknowledge useful feedback from R. Lunnan, M. Modjaz, S. Schulze, S. Vergani, J. Japelj, A. Gal-Yam, and useful conversations with A. Wetzel and D. Bersier. We also thank E. Stanway and D. Bersier for are careful read of this paper and for their feedback. DAP thanks the Aspen Centre for Physics for providing a stimulating host environment during the completion of this work. The ACP is supported by National Science Foundation grant PHY-1607611. This work was partially supported by a grant from the Simons Foundation.

Some of the data presented herein were obtained at the W. M. Keck Observatory, which is operated as a scientific partnership among the California Institute of Technology, the University of California and the National Aeronautics and Space Administration. The Observatory was made possible by the generous financial support of the W. M. Keck Foundation. We recognise and acknowledge the very significant cultural role and reverence that the summit of Maunakea has always had within the indigenous Hawaiian community. We are most fortunate to have the opportunity to conduct observations from this mountain. This work is based in part on observations made with the \textit{Spitzer Space Telescope}, which is operated by the Jet Propulsion Laboratory, California Institute of Technology under a contract with NASA. Support for this work was provided by NASA through an award issued by JPL/Caltech.  This paper includes data gathered with the 6.5 meter Magellan Telescopes located at Las Campanas Observatory, Chile.




\bibliographystyle{mnras}

\bibliography{host_galaxies} 


\FloatBarrier
\appendix

\section{Host galaxy photometry}
In Table \ref{tab:phot} we present the photometry used in the spectral energy distribution fits for all the host galaxies in our sample. In Table \ref{tab:lgrbphot} and \ref{tab:slsn} we present the references to all the photometry gathered from the literature.

\begin{table*}
\caption{Photometry of all galaxy samples including ASAS-SN CCSN, LGRBs and SLSN used in our analysis. These data include new photometry, photometry from public data and data gathered from the literature. Only the first few lines are shown; the full table will be made available online.}
\begin{threeparttable}
\label{tab:phot}
\renewcommand\TPTminimum{\linewidth}
\makebox[\linewidth]{
\begin{tabular}{ccccccccc}
\hline
Type & Name & Filter & Mag & Unc & System & Extinction & Instrument & Ref  \\
\hline
CCSN & ASAS-SN13co &  $u'$  & 16.54 & 0.05  & std & no & SDSS  & (1) \\
CCSN & ASAS-SN13co &  $g'$  & 15.39 & 0.01  & std & no & SDSS  & (1) \\
CCSN & ASAS-SN13co &  $r'$  & 14.88 & 0.01  & std & no & SDSS  & (1) \\
CCSN & ASAS-SN13co &  $i'$  & 14.57 & 0.01  & std & no & SDSS  & (1) \\
CCSN & ASAS-SN13co &  $z'$  & 14.39 & 0.01  & std & no & SDSS  & (1) \\
CCSN & ASAS-SN13co &  $y'$  & 14.34 & 0.05  & std & no & PS1   & (4) \\
CCSN & ASAS-SN13co &  $J $  & 13.31 & 0.03  & std & no & 2MASS & (2) \\
CCSN & ASAS-SN13co &  $H $  & 12.79 & 0.07  & std & no & 2MASS & (2) \\
CCSN & ASAS-SN13co &  $K_s$ & 12.34 & 0.11  & std & no & 2MASS & (2) \\ 
\hline
\end{tabular}}
\begin{tablenotes}[flushleft]
\item \textbf{References:} [1] Nasa Sloan Atlas; \citet{Blanton2011}, [2] 2MASS; this work, [3] \textit{GALEX}; \citet{Martin2005}, [4] PS1; this work, [5] 2MASS Extended Source Catalogue; \citet{2MASS2012}, [6] SDSS; this work [7], 2MASS Large Galaxy Atlas; \citet{Jarrett2003}.
\item \textbf{Notes.} All photometry is available online in a machine-readable form. Magnitudes are expressed in the conventional frame, this is indicated as  `std' under the System column, unless given in AB form in the literature where is indicated as `AB'. For SDSS $gri$ and PS1 filters, `std' is identical to `AB'. Magnitudes are not corrected for foreground extinction and under the Extinction column as  `no', unless unless corrected for Galactic foreground extinction in the literature, indicated by  `yes'.
\end{tablenotes}
\end{threeparttable}
\end{table*}

\begin{table}
\caption{LGRB photometry sources. For cases where LGRBs do have SNe, but there is no known SN name designation on TNS, we use SN in the name column.}
\begin{threeparttable}
\label{tab:lgrbphot}
\renewcommand\TPTminimum{\linewidth}
\makebox[\linewidth]{
\begin{tabular}{lll}
\hline
LGRB & SN name &  Reference   \\
\hline
980425           & 1998bw  & [2,3,4] 	\\
020903           & SN      & [1,5,6] 	\\
030329A          & 2003dh  & [1,7]      \\
031203           & 2003lw  & [1,8,9,10] \\
050826           & --      & [1,11,12] 	\\
060218           & 2006aj  & [1,13,14]   \\
060505           & --      & [4,14,15,16]  \\
060614           & --      & [1,14,17,18] \\
080517           & --      & [1,19] 	\\
100316D          & 2010bh  & [20,21] 	\\
111225A          & --	   & [1,22]   	\\
120422A          & 2012bz  & [1,23]  	\\
130702A          & 2013dx  & [1,24]  	\\
150518A          & SN      & [1] 		\\
150818A          & SN      & [1] 		\\
161219B          & 2016jca & [1,25] 	\\
171205A          & 2017iuk & [1,26,27] 	\\
\hline
\end{tabular}}
\begin{tablenotes}[flushleft]
\item  \textbf{References:}

[1] This work, [2] \cite{Michalowski2009}, [3] \cite{Michalowski2014}, [4] \cite{Castroceron2010}, [5] \cite{Bersier2006}, [6] \cite{Wainwright2007}, [7] \cite{Gorosabel2005}, [8] \cite{Margutti2007}, [9] \cite{Mazzali2006}, [10] \cite{Prochaska2004}, [11] \cite{Ovaldsen2007}, [12] \cite{Mirabel2007}, [13] \cite{Sollerman2006}, [14] \cite{Hjorth2012}, [15] \cite{Thone2008}, [16] \cite{Wright2010}, [17] \cite{Mangano2007}, [18] \citet{GalYam2006}, [19] \cite{Stanway2015}, [20] \cite{Starling2011}, [21] \citet{Michalowski2015}, [22] \cite{Niino2017}, [23] \cite{Schulze2014}, [24] \cite{Toy2016}, [25] \cite{Cano2017review}, [26] \citet{Bianchi2011}, [27] \citet{2MASS2012}.
\item \textbf{Notes.} New photometry measurements for the LGRBs are detailed in Section \ref{sec:newphotometry} and is included in online machine-readable table. 
\end{tablenotes}
\end{threeparttable}
\end{table}

\begin{table}
\caption{SLSN archival photometry sources}
\begin{threeparttable}
\label{tab:slsn}
\renewcommand\TPTminimum{\linewidth}
\makebox[\linewidth]{
\begin{tabular}{llcl}
\hline
SLSN & TNS Name & Type & Reference   \\
\hline
LSQ12dlf  &     ---  & I  & [1,2]   \\ 
LSQ14an   &     ---  & I  & [1,3]   \\ 
LSQ14mo   &     ---  & I  & [1]    	\\ 
MLS121104 &     ---  & I  & [1,4]   \\
PTF09as   & SN2009cb & I  & [5]    	\\
PTF09cnd  &	   ---   & I  & [5,6]   \\
PTF10aagc & 	---	 & I  & [5]   	\\
PTF10bfz  & 	---  & I  & [5]  	\\
PTF10cwr  &	SN2010gx & I  & [1,4,5,6] \\
PTF10hgi  &	SN2010md & I  & [1,4,5,6] \\
PTF10nmn  & 	---	 & I  & [5]       \\
PTF10uhf  & 	---	 & I  & [5,7] 	  \\
PTF10vwg  &	SN2010hy & I  & [5]     \\
PTF11dij  &	SN2011ke & I  & [1,5,6] \\
PTF11hrq  & 	---	 & I  & [5,7] 	\\
PTF11rks  &	SN2011kg & I  & [1,5]   \\
PTF12dam  &	   ---   & I  & [1,5,7] \\
SN1999as  &	   ---   & I  & [1,5,6,7] \\  
SN2005ap  &	   ---   & I  & [1,4,6,8] \\  
SN2007bi  &	   ---   & I  & [1,6] \\ 
SN2010kd &      ---  & I  & [1,3]    	\\
SN2011ep  &     ---  & I  & [1,3]    	\\
SN2011kf  &	   ---   & I  & [1,4,6] \\
SN2012il  &	   ---   & I  & [1,4,6] \\
SN2013dg  &     ---  & I  & [1]    	\\
SN2015bn  &     ---  & I  & [1,3]    	\\
SSS120810 &     ---  & I  & [1]    	\\
\hline
CSS100217 &     ---  & II & [1]    	\\ 
CSS121015 &     ---  & II & [1]    	\\ 
PTF10fel  & 	---	 & II & [5,7]   	\\
PTF10qaf  &	    ---  & II & [5]         \\
PTF10qwu  & 	---	 & II & [5]     	\\
PTF10scc  & 	---	 & II & [5]     	\\
PTF10tpz  & 	---	 & II & [5,9]   	\\
PTF10yyc  & 	---	 & II & [5]     	\\
PTF12gwu  & 	---  & II & [5]     	\\
PTF12mkp  & 	---  & II & [5]     	\\
PTF12mue  & 	---  & II & [5]     	\\
SN1999bd  &	    ---  & II & [1,6,7,10] \\
SN2003ma  &     ---  & II & [11,12]   	\\
SN2006gy  &	    ---  & II & [6,7] \\
SN2006tf  &     ---  & II & [1,3]    	\\
SN2007bw  &     ---  & II & [1]    	\\
SN2008am  &	    ---  & II & [1,3,6,9,13] \\ 
SN2008es  &	    ---  & II & [1,6] \\ 
SN2008fz  &     ---  & II & [6] \\ 
SN2009nm  &     ---  & II & [1]    	\\
SN2013hx  &     ---  & II & [1]    	\\
\hline
\end{tabular}}
\begin{tablenotes}[flushleft]
\item \textbf{References:} [1] \cite{Schulze2018}, [2] \cite{Nicholl2014}, [3]  \cite{Bianchi2011}, [4] \cite{Lunnan2014}, [5] \cite{2016Perley}, [6] \cite{Angus2016}, [7]  \cite{Cutri2013}, [8] \cite{Adami2006}, [9] \cite{Cutri2014}, [10] \cite{Neill2011}, [11] \citet{Kato2007}, [12] \citet{Rest20112003ma}, [13] \cite{Lawrence2007}, [14] \cite{Blanton2011} , [15] \citet{2MASS2012}.
\end{tablenotes}
\end{threeparttable}
\end{table}

\section{Derived physical parameters and spectral energy distribution fits}\label{sec:seds}

In this section, we show tables of physical properties, including stellar masses and star-formation rates derived from the host galaxy photometry using methods described in Section \ref{subsec:sedfitting}. Redshifts were derived from the host galaxy, or the supernova if there was no host redshift available. The `discovered' ASAS-SN CCSN host galaxies can be found in the first two pages of Table~\ref{tab:asassntable} and the `recovered' ASAS-SN host galaxies in the third page of Table~\ref{tab:asassntable}. The physical parameters for LGRBs can be found in Table~\ref{tab:lgrbtable} and the SLSNe in Table~\ref{tab:slsnetable}. In addition, we present the SED fits to the photometry for the CCSN sample in Fig.~\ref{fig:ccsn_sed_1_2}, the LBRB sample in Fig.~\ref{fig:grbsed} and the SLSN sample in Fig.~\ref{fig:combined_slsnI_II}. 

\begin{table*}
\centering
\caption{Properties of ASAS-SN CCSN host galaxies, including physical parameters derived from the SED fitting procedure.}
 \begin{threeparttable}
\label{tab:asassntable}
\begin{tabular}{lllllllrrrrrrr} 
\hline

\multicolumn{1}{c}{ASAS-SN}&\multicolumn{1}{c}{Class}&\multicolumn{1}{c}{$\alpha$(2000)}&\multicolumn{1}{c}{$\delta$(2000)}&\multicolumn{1}{c}{$z_{\rm sn}$}&\multicolumn{1}{c}{$z_{\rm host}$} &\multicolumn{1}{c}{Distance$^{\dagger}$}&\multicolumn{1}{c}{E($B$-$V$)}&\multicolumn{1}{c}{log$_{10}\big($M$_{*}\big)$}&\multicolumn{1}{c}{SFR} &\multicolumn{1}{c}{sSFR$^{\ddagger}$} \\

&&&&&& \multicolumn{1}{c}{Mpc} && \multicolumn{1}{c}{M$_{\odot}$}& \multicolumn{1}{c}{M$_{\odot}$ yr$^{-1}$ } &\multicolumn{1}{c}{yr $^{-1}$} \\
\hline
13co & IIP     & 21:40:38.74 & +06:30:36.87  & 0.023 & 0.0234 & 100.2$\pm$7.0 & 0.052             & 9.85 $^{\Plus 0.05}_{\Minus 0.05}$ &  0.800 $^{\Plus 0.224}_{\Minus 0.193}$ & -9.96 $^{\Plus 0.14}_{\Minus 0.10}$  \\
13dn & II      & 12:52:58.20 & +32:25:09.30  & 0.023 & 0.0228 & 105.6$\pm$7.4 & 0.014             & 9.01 $^{\Plus 0.04}_{\Minus 0.04}$ &  1.791 $^{\Plus 0.021}_{\Minus 1.086}$ & -8.69 $^{\Plus 0.01}_{\Minus 0.46}$  \\
14at & II      & 17:55:05.43 & +18:15:26.45  & 0.010 & 0.0104 & 52.0$\pm$3.6  & 0.073             & 8.05 $^{\Plus 0.90}_{\Minus 0.35}$ &  0.002 $^{\Plus 0.033}_{\Minus 0.001}$ & -10.65 $^{\Plus 0.19}_{\Minus 1.29}$  \\
14az & IIb     & 23:44:48.00 & --02:07:03.17 & 0.007 & 0.0067 & 29.3$\pm$2.1  & 0.028             & 8.00 $^{\Plus 0.01}_{\Minus 0.01}$ &  0.012 $^{\Plus 0.002}_{\Minus 0.001}$ & -9.90 $^{\Plus 0.07}_{\Minus 0.01}$  \\
14bf & IIP     & 13:58:12.75 & +17:31:53.66  & 0.022 & 0.0225 & 105.4$\pm$7.4 & 0.026             & 9.97 $^{\Plus 0.02}_{\Minus 0.01}$ &  0.198 $^{\Plus 0.001}_{\Minus 0.096}$ & -10.70 $^{\Plus 0.01}_{\Minus 0.24}$  \\
14bu & II      & 11:18:41.03 & +25:09:59.88  & 0.025 & 0.0255 & 115.9$\pm$8.1 & 0.014             & 8.59 $^{\Plus 0.05}_{\Minus 0.12}$ &  0.141 $^{\Plus 0.126}_{\Minus 0.025}$ & -9.46 $^{\Plus 0.38}_{\Minus 0.12}$  \\
14de & Ic      & 10:40:39.33 & +39:03:52.70  & 0.029 & 0.0293 & 130.7$\pm$9.2 & 0.015             & 10.44 $^{\Plus 0.01}_{\Minus 0.11}$ &  0.019 $^{\Plus 0.001}_{\Minus 0.005}$ & -12.20 $^{\Plus 0.02}_{\Minus 0.07}$  \\
14di & II      & 02:01:46.39 & +26:32:41.96  & 0.017 & 0.0167 & 69.6$\pm$4.9  & 0.064             & 10.09 $^{\Plus 0.03}_{\Minus 0.08}$ &  0.780 $^{\Plus 0.857}_{\Minus 0.080}$ & -10.17 $^{\Plus 0.27}_{\Minus 0.04}$  \\
14dl & II      & 12:21:51.38 & --24:09:54.00 & 0.014 & 0.0139 & 65.2$\pm$4.6  & 0.068             & 10.91 $^{\Plus 0.01}_{\Minus 0.04}$ &  3.589 $^{\Plus 0.058}_{\Minus 0.323}$ & -10.30 $^{\Plus 0.01}_{\Minus 0.04}$  \\
14dq & II      & 21:57:59.97 & +24:16:08.10  & 0.010 & 0.0104 & 46.9$\pm$3.3  & 0.062             & 8.01 $^{\Plus 0.07}_{\Minus 0.23}$ &  0.344 $^{\Plus 0.291}_{\Minus 0.046}$ & -8.46 $^{\Plus 0.46}_{\Minus 0.13}$  \\
14fj & II      & 14:40:39.50 & +38:37:58.55  & 0.013 & 0.0125 & 61.8$\pm$4.3  & 0.014             & 9.72 $^{\Plus 0.12}_{\Minus 0.02}$ &  0.897 $^{\Plus 0.008}_{\Minus 0.352}$ & -9.81 $^{\Plus 0.01}_{\Minus 0.32}$  \\
14gm & II      & 00:59:47.83 & --07:34:19.30 & 0.005 & 0.0055 & 23.2$\pm$1.6  & 0.099             & 9.47 $^{\Plus 0.11}_{\Minus 0.02}$ &  0.443 $^{\Plus 0.171}_{\Minus 0.039}$ & -9.80 $^{\Plus 0.01}_{\Minus 0.01}$  \\
14il & IIn     & 00:45:32.55 & --14:15:34.60 & 0.022 & 0.0220 & 92.3$\pm$6.5  & 0.020             & 9.62 $^{\Plus 0.06}_{\Minus 0.04}$ &  0.622 $^{\Plus 0.231}_{\Minus 0.187}$ & -9.90 $^{\Plus 0.25}_{\Minus 0.11}$  \\
14jb & IIP     & 22:23:16.12 & --28:58:30.78 & 0.006 & 0.0060 & 27.3$\pm$1.9  & 0.017             & 8.46 $^{\Plus 0.11}_{\Minus 0.01}$ &  0.251 $^{\Plus 0.006}_{\Minus 0.138}$ & -9.11 $^{\Plus 0.01}_{\Minus 0.45}$  \\
14jh & II      & 08:40:44.27 & +57:15:04.91  & 0.018 & 0.0175 & 78.4$\pm$5.5  & 0.055             & 8.92 $^{\Plus 0.05}_{\Minus 0.04}$ &  0.807 $^{\Plus 0.161}_{\Minus 0.214}$ & -9.06 $^{\Plus 0.08}_{\Minus 0.05}$  \\
14kg & II      & 01:44:38.38 & +35:48:20.45  & 0.014 & 0.0145 & 60.8$\pm$4.3  & 0.041             & 9.79 $^{\Plus 0.04}_{\Minus 0.02}$ &  0.089 $^{\Plus 0.052}_{\Minus 0.001}$ & -10.90 $^{\Plus 0.24}_{\Minus 0.01}$  \\
14ma & IIP     & 23:55:09.13 & +10:12:54.21  & 0.014 & 0.0137$^{[1]}$ & 59.3$\pm$4.7  & 0.091        & 8.57 $^{\Plus 0.04}_{\Minus 0.03}$ &  0.001 $^{\Plus 0.001}_{\Minus 0.001}$ & -12.16 $^{\Plus 0.38}_{\Minus 0.11}$  \\
14ms & Ibn     & 13:04:08.69 & +52:18:46.50  & 0.054 & 0.0540$^{[1]}$ & 241.0$\pm$19.3 & 0.010        & 7.58 $^{\Plus 0.22}_{\Minus 0.32}$ &  0.030 $^{\Plus 0.015}_{\Minus 0.012}$ & -9.12 $^{\Plus 0.55}_{\Minus 0.47}$  \\
15bd & IIb     & 15:54:38.33 & +16:36:38.06  & 0.008 & 0.0078 & 42.1$\pm$2.9  & 0.030             & 7.73 $^{\Plus 0.03}_{\Minus 0.06}$ &  0.019 $^{\Plus 0.004}_{\Minus 0.005}$ & -9.56 $^{\Plus 0.07}_{\Minus 0.08}$  \\
15ed & Ibn     & 16:48:25.16 & +50:59:30.72  & 0.049 & 0.04866$^{[1]}$ & 216.4$\pm$17.3 & 0.022                  & 11.02 $^{\Plus 0.01}_{\Minus 0.01}$ &  1.259 $^{\Plus 0.003}_{\Minus 0.001}$ & -10.90 $^{\Plus 0.01}_{\Minus 0.01}$  \\
15fi & II      & 16:31:48.80 & +20:24:38.50  & 0.017 & 0.0172 & 82.1$\pm$5.8 & 0.050              & 9.12 $^{\Plus 0.01}_{\Minus 0.17}$ &  0.341 $^{\Plus 0.121}_{\Minus 0.091}$ & -9.60 $^{\Plus 0.20}_{\Minus 0.01}$  \\
15fz & II      & 13:35:25.14 & +01:24:33.00  & 0.017 & 0.0175 & 84.4$\pm$5.9 & 0.022              & 10.48 $^{\Plus 0.01}_{\Minus 0.05}$ &  9.311 $^{\Plus 0.261}_{\Minus 1.368}$ & -9.53 $^{\Plus 0.02}_{\Minus 0.05}$  \\
15ik & IIn     & 11:02:04.75 & +03:30:02.66  & 0.035 & 0.0346$^{[2]}$ & 152.2$\pm$12.2 & 0.048     & 8.48 $^{\Plus 0.01}_{\Minus 0.04}$ &  0.081 $^{\Plus 0.003}_{\Minus 0.007}$ & -9.60 $^{\Plus 0.01}_{\Minus 0.01}$  \\
15ir & II      & 10:48:30.30 & --21:38:07.95 & 0.013 & 0.0127 & 59.2$\pm$4.2 & 0.056              & 9.14 $^{\Plus 0.26}_{\Minus 0.02}$ &  1.469 $^{\Plus 0.545}_{\Minus 0.250}$ & -9.02 $^{\Plus 0.02}_{\Minus 0.04}$  \\
15kz & IIP     & 13:37:18.67 & --28:39:23.55 & 0.008 & 0.0080 & 30.5$\pm$2.2 & 0.053              & 7.95 $^{\Plus 0.11}_{\Minus 0.06}$ &  0.789 $^{\Plus 0.072}_{\Minus 0.420}$ & -7.98 $^{\Plus 0.04}_{\Minus 0.55}$  \\
15lf & IIn     & 12:06:45.56 & +67:09:24.00  & 0.008 & 0.0084 & 41.9$\pm$2.9 & 0.016              & 9.65 $^{\Plus 0.01}_{\Minus 0.01}$ &  2.570 $^{\Plus 0.024}_{\Minus 0.495}$ & -9.30 $^{\Plus 0.01}_{\Minus 0.01}$  \\
15ln & II      & 00:53:41.40 & +18:05:29.00  & 0.015 & 0.0150 & 62.8$\pm$4.4 & 0.042              & 9.40 $^{\Plus 0.03}_{\Minus 0.11}$ &  0.230 $^{\Plus 0.077}_{\Minus 0.007}$ & -10.04 $^{\Plus 0.25}_{\Minus 0.03}$  \\
15mj & Ib      & 14:02:15.64 & +33:39:40.29  & 0.034 & 0.0344 &155.0$\pm$10.9& 0.013              & 9.38 $^{\Plus 0.04}_{\Minus 0.02}$ &  0.566 $^{\Plus 0.618}_{\Minus 0.102}$ & -9.63 $^{\Plus 0.34}_{\Minus 0.16}$  \\
15mm & II      & 15:25:23.50 & +29:10:24.50  & 0.021 & 0.0215 &100.5$\pm$7.1 & 0.021              & 9.53 $^{\Plus 0.04}_{\Minus 0.16}$ &  0.386 $^{\Plus 0.281}_{\Minus 0.149}$ & -9.96 $^{\Plus 0.41}_{\Minus 0.09}$  \\
15no & Ic      & 15:38:25.30 & +46:54:06.60  & 0.043 & 0.03638$^{[1]}$ &160.3$\pm$12.8 & 0.015    & 8.46 $^{\Plus 0.09}_{\Minus 0.03}$ &  0.351 $^{\Plus 0.022}_{\Minus 0.215}$ & -8.88 $^{\Plus 0.03}_{\Minus 0.55}$  \\
15nx & II-pec  & 04:43:53.19 & --09:42:11.22 & 0.026 & 0.02823$^{[1]}$ & 123.6$\pm$9.9 & 0.074    & 8.99 $^{\Plus 0.09}_{\Minus 0.16}$ &  0.264 $^{\Plus 0.068}_{\Minus 0.131}$ & -9.61 $^{\Plus 0.22}_{\Minus 0.31}$  \\
15ov & II      & 03:30:59.15 & --18:33:23.19 & 0.025 & 0.0255 &106.7$\pm$7.5 & 0.034              & 9.46 $^{\Plus 0.01}_{\Minus 0.03}$ &  1.807 $^{\Plus 0.038}_{\Minus 0.061}$ & -9.20 $^{\Plus 0.01}_{\Minus 0.01}$  \\
15qh & II      & 22:45:13.22 & --22:43:39.82 & 0.010 & 0.0102 & 45.0$\pm$3.2 & 0.026              & 9.93 $^{\Plus 0.01}_{\Minus 0.18}$ &  0.519 $^{\Plus 1.715}_{\Minus 0.012}$ & -10.29 $^{\Plus 0.88}_{\Minus 0.01}$  \\
15rb & IIn     & 10:08:08.24 & +19:17:59.38  & 0.034 & 0.0336 &149.0$\pm$10.4& 0.023              & 8.01 $^{\Plus 0.03}_{\Minus 0.18}$ &  0.015 $^{\Plus 0.001}_{\Minus 0.006}$ & -9.88 $^{\Plus 0.09}_{\Minus 0.08}$  \\
15tm & IIP     & 23:27:35.60 & +29:24:31.17  & 0.016 &   ---  & 69.4$\pm$21.9 & 0.116             & 9.18 $^{\Plus 0.07}_{\Minus 0.09}$ &  0.123 $^{\Plus 0.013}_{\Minus 0.016}$ & -10.13 $^{\Plus 0.12}_{\Minus 0.05}$  \\
15tw & IIP     & 12:50:28.05 & --10:50:29.15 & 0.008 & 0.0080 & 36.7$\pm$2.6 & 0.040              & 9.52 $^{\Plus 0.05}_{\Minus 0.03}$ &  0.510 $^{\Plus 0.001}_{\Minus 0.012}$ & -9.80 $^{\Plus 0.02}_{\Minus 0.04}$  \\
15ua & IIn     & 13:34:54.47 & +10:59:04.69  & 0.061 &    --- &273.7$\pm$23.4 & 0.026             & 8.55 $^{\Plus 0.16}_{\Minus 0.14}$ &  0.143 $^{\Plus 0.065}_{\Minus 0.049}$ & -9.42 $^{\Plus 0.27}_{\Minus 0.29}$  \\
15ug & II      & 06:45:01.68 & +63:14:59.89  & 0.022 &  0.0221$^{[1]}$ & 96.3$\pm$7.7& 0.077     & 7.44 $^{\Plus 0.08}_{\Minus 0.18}$ &  0.032 $^{\Plus 0.009}_{\Minus 0.030}$ & -8.96 $^{\Plus 0.27}_{\Minus 1.28}$  \\
15un & II      & 02:40:41.38 & +16:49:51.82  & 0.029 & 0.0292 &121.7$\pm$8.5 & 0.074              & 9.61 $^{\Plus 0.01}_{\Minus 0.04}$ &  3.304 $^{\Plus 0.277}_{\Minus 0.319}$ & -9.01 $^{\Plus 0.01}_{\Minus 0.06}$  \\
15uo & IIn-pec & 01:17:00.00 & --04:56:34.10 & 0.038 &    --- &167.6$\pm$22.6 & 0.040             & 8.76 $^{\Plus 0.07}_{\Minus 0.27}$ &  0.399 $^{\Plus 0.143}_{\Minus 0.047}$ & -9.18 $^{\Plus 0.40}_{\Minus 0.11}$  \\
15uy & IIb     & 14:32:15.31 & +26:19:32.02  & 0.016 & 0.0160 & 77.5$\pm$5.5 & 0.017              & 9.35 $^{\Plus 0.10}_{\Minus 0.04}$ &  0.762 $^{\Plus 0.113}_{\Minus 0.106}$ & -9.46 $^{\Plus 0.05}_{\Minus 0.14}$  \\
16ab & II      & 11:55:04.25 & +01:43:06.77  & 0.004 & 0.0043 & 26.0$\pm$1.9 & 0.019              & 7.85 $^{\Plus 0.01}_{\Minus 0.01}$ &  0.159 $^{\Plus 0.007}_{\Minus 0.001}$ & -8.70 $^{\Plus 0.07}_{\Minus 0.01}$  \\
16ai & IIP     & 14:39:44.73 & +23:23:43.27  & 0.015 & 0.0149 & 72.8$\pm$5.1 & 0.028              & 8.12 $^{\Plus 0.01}_{\Minus 0.01}$ &  0.007 $^{\Plus 0.006}_{\Minus 0.001}$ & -10.29 $^{\Plus 0.28}_{\Minus 0.01}$  \\
16al & IIP     & 15:00:27.47 & --13:33:09.00 & 0.009 & 0.0093 & 44.2$\pm$3.1 & 0.088              & 7.52 $^{\Plus 0.05}_{\Minus 0.14}$ &  0.552 $^{\Plus 0.176}_{\Minus 0.107}$ & -7.75 $^{\Plus 0.13}_{\Minus 0.09}$  \\
16am & II      & 04:45:21.28 & +73:23:41.09  & 0.015 & 0.0150 & 66.3$\pm$4.7 & 0.151              & 10.41 $^{\Plus 0.10}_{\Minus 0.05}$ &  0.001 $^{\Plus 0.001}_{\Minus 0.001}$ & -13.90 $^{\Plus 0.66}_{\Minus 0.15}$  \\
16at & II      & 12:55:15.50 & +00:05:59.70  & 0.004 & 0.0044 & 26.7$\pm$1.9 & 0.020              & 9.30 $^{\Plus 0.01}_{\Minus 0.01}$ &  0.794 $^{\Plus 0.001}_{\Minus 0.001}$ & -9.40 $^{\Plus 0.01}_{\Minus 0.01}$  \\
16ba & II      & 09:42:29.22 & --16:58:26.88 & 0.014 & 0.0139 & 64.6$\pm$4.5 & 0.058              & 9.26 $^{\Plus 0.03}_{\Minus 0.05}$ &  0.316 $^{\Plus 0.031}_{\Minus 0.029}$ & -9.72 $^{\Plus 0.01}_{\Minus 0.03}$  \\
16bm & II      & 11:51:56.24 & --13:25:03.07 & 0.007 & 0.0068$^{[1]}$  & 29.3$\pm$2.3 & 0.038   & 8.07 $^{\Plus 0.19}_{\Minus 0.24}$ &  0.320 $^{\Plus 0.113}_{\Minus 0.114}$ & -8.58 $^{\Plus 0.37}_{\Minus 0.40}$  \\
16cr & II      & 11:42:34.65 & --25:54:45.22 & 0.014 &   ---  & 60.6$\pm$21.8& 0.042              & 8.12 $^{\Plus 0.07}_{\Minus 0.11}$ &  0.049 $^{\Plus 0.054}_{\Minus 0.036}$ & -9.52 $^{\Plus 0.45}_{\Minus 0.40}$  \\
16dm & IIP     & 11:37:20.64 & --04:54:36.84 & 0.018 & 0.0183 & 86.7$\pm$6.1 & 0.046              & 9.42 $^{\Plus 0.08}_{\Minus 0.02}$ &  0.462 $^{\Plus 0.039}_{\Minus 0.092}$ & -9.70 $^{\Plus 0.01}_{\Minus 0.26}$  \\
16eh & II      & 15:40:29.23 & +00:54:36.38  & 0.012 & 0.0117 & 58.6$\pm$4.1 & 0.263              & 8.45 $^{\Plus 0.01}_{\Minus 0.26}$ &  0.234 $^{\Plus 0.010}_{\Minus 0.063}$ & -9.04 $^{\Plus 0.04}_{\Minus 0.03}$  \\
16ek & IIb     & 07:20:24.16 & +32:51:02.58  & 0.014 &   ---  & 60.6$\pm$21.8& 0.052              & 8.95 $^{\Plus 0.03}_{\Minus 0.01}$ &  0.959 $^{\Plus 0.013}_{\Minus 0.139}$ & -9.01 $^{\Plus 0.01}_{\Minus 0.05}$  \\
16el & II      & 08:56:39.08 & +52:06:10.01  & 0.014 & 0.0135 & 61.9$\pm$4.4 & 0.017              & 10.13 $^{\Plus 0.03}_{\Minus 0.02}$ &  2.748 $^{\Plus 0.032}_{\Minus 0.038}$ & -9.70 $^{\Plus 0.01}_{\Minus 0.01}$  \\
16eu & IIP     & 08:44:11.05 & +34:42:55.80  & 0.014 & 0.0141 & 64.4$\pm$4.5 & 0.028              & 10.20 $^{\Plus 0.01}_{\Minus 0.01}$ &  5.012 $^{\Plus 0.001}_{\Minus 0.080}$ & -9.50 $^{\Plus 0.01}_{\Minus 0.01}$  \\
\hline
\end{tabular}
\begin{tablenotes}[flushleft]
\item \textbf{Notes.} SN Type, Host Galaxy, and Discovery Reference columns come from 	             \url{http://www.astronomy.ohio-state.edu/~assassin/sn_list.txt}, except where noted.  
\item $^{\dagger}$ Hubble flow distances are derived from the host galaxy if available in NED and the uncertainty is derived from the velocity calculator which accounts for the Virgo Cluster, Great Attractor and Shapley Supercluster infall velocities. If a redshift is not available in NED, we search the literature for redshifts derived from a host galaxy spectrum or narrow emission lines from the SN spectrum, in these cases we adopt an 8 per cent uncertainty in the distance (since this is the maximum uncertainty derived for the NED velocity uncertainties). Finally, if the host galaxy redshift is unknown, we use the SN redshift and give the luminosity distance, with an uncertainty on the redshift of $z$=0.005.
 \item $^{\ddagger}$ sSFR is based on the PDF marginalised over all the other parameters in the SED fit. Thus it is slightly different from the derived SFR/Mass.
 \item [1] Host redshift was not obtained from NED, but from another source. For 14m,14ms and 15nx the redshift was derived from narrow emission lines of the host galaxy in the SN spectrum \citep{Zhang2014,Vallely2018,Bose2018}. For 15ed and 15no the redshift was derived from unresolved emission lines in the host galaxy spectrum \citep{Pastorello2015,Benetti2018}

 \item [2] Host redshift was derived from spectroscopy of the host galaxies in Taggart et al. (in prep).
\end{tablenotes}
\end{threeparttable}
\end{table*}
\begin{table*}
\centering
\contcaption{}
\begin{threeparttable}
\label{tab:asassntable2}
\begin{tabular}{lllllllrrrrrrr} 
\hline
\multicolumn{1}{c}{ASAS-SN}&\multicolumn{1}{c}{Class}&\multicolumn{1}{c}{$\alpha$(2000)}&\multicolumn{1}{c}{$\delta$(2000)}&\multicolumn{1}{c}{$z_{\rm sn}$}&\multicolumn{1}{c}{$z_{\rm host}$} &\multicolumn{1}{c}{Distance$^{\dagger}$}&\multicolumn{1}{c}{E($B$-$V$)}&\multicolumn{1}{c}{log$_{10}\big($M$_{*}\big)$}&\multicolumn{1}{c}{SFR} &\multicolumn{1}{c}{sSFR$^{\ddagger}$} \\

&&&&&& \multicolumn{1}{c}{Mpc} && \multicolumn{1}{c}{M$_{\odot}$}& \multicolumn{1}{c}{M$_{\odot}$ yr$^{-1}$ } &\multicolumn{1}{c}{yr $^{-1}$} \\
\hline
16fp & Ib/c-BL & 21:59:04.14 & +18:11:10.50  & 0.004 & 0.0037 & 18.8$\pm$1.3 & 0.075       & 8.73 $^{\Plus 0.01}_{\Minus 0.03}$ &  0.216 $^{\Plus 0.004}_{\Minus 0.002}$ & -9.37 $^{\Plus 0.05}_{\Minus 0.04}$  \\
16fq & IIP     & 11:20:19.09 & +12:58:57.20  & 0.002 & 0.00243 & 11.5$\pm$0.8 & 0.030      & 10.56 $^{\Plus 0.05}_{\Minus 0.04}$ &  4.887 $^{\Plus 0.325}_{\Minus 0.969}$ & -9.87 $^{\Plus 0.08}_{\Minus 0.07}$  \\
16ft & II      & 23:56:13.74 & --00:32:28.44 & 0.022 & 0.0222 & 93.6$\pm$6.6 & 0.028       & 9.79 $^{\Plus 0.01}_{\Minus 0.02}$ &  0.294 $^{\Plus 0.016}_{\Minus 0.029}$ & -10.30 $^{\Plus 0.01}_{\Minus 0.01}$  \\
16gn & IIn  & 12:06:57.59 & +27:18:04.93  & 0.056 & 0.0560 & 246.4$\pm$17.3 & 0.018        & 10.13 $^{\Plus 0.03}_{\Minus 0.05}$ &  1.236 $^{\Plus 0.845}_{\Minus 0.421}$ & -9.99 $^{\Plus 0.16}_{\Minus 0.22}$  \\
16go & II   & 13:02:44.26 & --26:56:26.81 & 0.016 & 0.0161 & 80.1$\pm$5.6   & 0.070        & 9.78 $^{\Plus 0.01}_{\Minus 0.01}$ &  0.561 $^{\Plus 0.004}_{\Minus 0.046}$ & -10.00 $^{\Plus 0.01}_{\Minus 0.03}$  \\
16gy & II   & 02:21:22.77 & +16:33:54.56  & 0.014 & 0.0137 & 56.7$\pm$4.0   & 0.129        & 10.11 $^{\Plus 0.12}_{\Minus 0.01}$ &  1.140 $^{\Plus 0.614}_{\Minus 0.005}$ & -10.00 $^{\Plus 0.05}_{\Minus 0.01}$ \\
16hy & II   & 15:26:29.52 & +41:44:03.32  & 0.008 & 0.0078 & 41.5$\pm$2.9   & 0.022        & 8.65 $^{\Plus 0.05}_{\Minus 0.01}$ &  0.127 $^{\Plus 0.066}_{\Minus 0.017}$ & -9.60 $^{\Plus 0.18}_{\Minus 0.01}$  \\
16in & IIn  & 04:59:30.07 & --28:51:39.43 & 0.016 & 0.0161 & 68.5$\pm$4.8   & 0.012        & 9.27 $^{\Plus 0.04}_{\Minus 0.12}$ &  0.508 $^{\Plus 0.717}_{\Minus 0.097}$ & -9.60 $^{\Plus 0.51}_{\Minus 0.05}$  \\
16ll & II   & 19:00:32.43 & +54:34:09.70  & 0.028 & 0.026$^{[1]}$ & 113.6$\pm$9.1 & 0.054  & 9.27 $^{\Plus 0.13}_{\Minus 0.04}$ &  1.102 $^{\Plus 0.175}_{\Minus 0.654}$ & -9.17 $^{\Plus 0.04}_{\Minus 0.62}$  \\
16mz & II   & 12:04:16.91 & +21:48:03.30  & 0.021 & 0.0215 & 99.8$\pm$7.0   & 0.020        & 9.02 $^{\Plus 0.04}_{\Minus 0.01}$ &  0.011 $^{\Plus 0.025}_{\Minus 0.010}$ & -10.95 $^{\Plus 0.63}_{\Minus 1.23}$  \\
16ns & II   & 10:04:18.59 & +43:25:29.13  & 0.038$^{[2]}$ & ---  & 167.6$\pm$22.8 & 0.011  & 8.48 $^{\Plus 0.12}_{\Minus 0.15}$ &  0.200 $^{\Plus 0.067}_{\Minus 0.166}$ & -9.19 $^{\Plus 0.23}_{\Minus 0.79}$  \\
17ai & Ib   & 12:07:18.83 & +16:50:26.02  & 0.023 & 0.0231 & 107.0$\pm$7.5  & 0.041        & 9.46 $^{\Plus 0.02}_{\Minus 0.13}$ &  2.624 $^{\Plus 0.136}_{\Minus 0.857}$ & -9.01 $^{\Plus 0.01}_{\Minus 0.09}$  \\
17br & IIP  & 15:52:00.31 & +66:18:55.27  & 0.026 & ---    & 113.6$\pm$22.2 & 0.024        & 8.76 $^{\Plus 0.05}_{\Minus 0.03}$ &  0.002 $^{\Plus 0.008}_{\Minus 0.001}$ & -11.96 $^{\Plus 0.96}_{\Minus 0.73}$  \\
17bw & II  & 16:58:37.69  & +50:29:26.50 & 0.01 & 0.01020 &  44.0$\pm$3.1   &  0.019       & 8.62 $^{\Plus 0.18}_{\Minus 0.02}$ &  0.445 $^{\Plus 0.057}_{\Minus 0.029}$ & -8.99 $^{\Plus 0.06}_{\Minus 0.01}$  \\
17cl & II   & 05:02:19.58 & --10:21:22.78 & 0.013 & 0.0133 & 56.2$\pm$3.9   & 0.078        & 10.49 $^{\Plus 0.01}_{\Minus 0.29}$ &  0.593 $^{\Plus 0.017}_{\Minus 0.084}$ & -10.70 $^{\Plus 0.01}_{\Minus 0.01}$ \\
17ds & II   & 08:03:55.21 & +26:31:12.73  & 0.022 & 0.0217 & 95.2$\pm$6.7   & 0.038        & 10.03 $^{\Plus 0.04}_{\Minus 0.02}$ &  0.565 $^{\Plus 0.063}_{\Minus 0.049}$ & -10.28 $^{\Plus 0.08}_{\Minus 0.03}$ \\
17dv & Ib/c & 09:52:31.22 & --21:57:54.59 & 0.029 &    --- & 127.0$\pm$22.3 & 0.037        & 8.60 $^{\Plus 0.04}_{\Minus 0.02}$ &  0.045 $^{\Plus 0.053}_{\Minus 0.043}$ & -9.92 $^{\Plus 0.21}_{\Minus 1.42}$  \\
17fy & IIn  & 09:03:32.47 & --21:20:02.73 & 0.018 & 0.0182 & 82.7$\pm$5.8   & 0.140        & 9.59 $^{\Plus 0.22}_{\Minus 0.00}$ &  0.766 $^{\Plus 0.258}_{\Minus 0.313}$ & -9.80 $^{\Plus 0.05}_{\Minus 0.11}$  \\
17gi & Ibn  & 14:14:48.94 & --29:33:37.01 & 0.020 &    --- & 87.0$\pm$22.0  & 0.057        & 8.73 $^{\Plus 0.06}_{\Minus 0.13}$ &  0.107 $^{\Plus 0.044}_{\Minus 0.028}$ & -9.76 $^{\Plus 0.24}_{\Minus 0.15}$  \\
17he & II   & 09:45:48.36 & --14:22:05.60 & 0.008 & 0.0081 & 37.4$\pm$2.6   & 0.053        & 9.50 $^{\Plus 0.05}_{\Minus 0.05}$ &  7.096 $^{\Plus 0.811}_{\Minus 0.727}$ & -8.65 $^{\Plus 0.05}_{\Minus 0.04}$  \\
17ia & IIP  & 13:10:59.29 & +78:24:37.16  & 0.023 & 0.0234 & 100.3$\pm$7.1  & 0.033        & 10.41 $^{\Plus 0.01}_{\Minus 0.01}$ &  2.075 $^{\Plus 0.058}_{\Minus 0.052}$ & -10.10 $^{\Plus 0.01}_{\Minus 0.01}$ \\
17is & II   & 02:11:06.94 & +03:50:36.63  & 0.011 & 0.0105 & 44.9$\pm$3.1   & 0.036        & 9.46 $^{\Plus 0.02}_{\Minus 0.13}$ &  2.965 $^{\Plus 0.283}_{\Minus 1.166}$ & -9.03 $^{\Plus 0.03}_{\Minus 0.07}$  \\
17jp & II   & 02:54:02.09 & +02:58:07.71  & 0.010 & 0.0102 & 41.9$\pm$2.9   & 0.095        & 10.40 $^{\Plus 0.01}_{\Minus 0.16}$ &  1.135 $^{\Plus 0.221}_{\Minus 0.258}$ & -10.30 $^{\Plus 0.11}_{\Minus 0.01}$ \\
17nb & II   & 07:27:37.32 & +35:36:30.64  & 0.016 &  ---   & 69.4$\pm$21.9  & 0.048        & 8.81 $^{\Plus 0.05}_{\Minus 0.06}$ &  0.659 $^{\Plus 0.033}_{\Minus 0.587}$ & -9.00 $^{\Plus 0.06}_{\Minus 0.92}$  \\
17oj & II   & 21:44:22.95 & --29:54:59.30 & 0.016 & 0.01874   & 82.6$\pm$22.0  & 0.040     & 9.19 $^{\Plus 0.10}_{\Minus 0.12}$ &  2.228 $^{\Plus 0.834}_{\Minus 0.953}$ & -8.83 $^{\Plus 0.23}_{\Minus 0.36}$  \\
17om & II   & 03:34:11.10 & --13:56:09.37 & 0.08  &  ---   & 363.9$\pm$24.0 & 0.042        & 9.97 $^{\Plus 0.04}_{\Minus 0.12}$ &  6.546 $^{\Plus 0.420}_{\Minus 1.906}$ & -9.27 $^{\Plus 0.10}_{\Minus 0.04}$  \\
17os & II   & 04:33:05.88 & --26:07:41.34 & 0.032 & 0.0323 & 138.4$\pm$9.8  & 0.035        & 8.74 $^{\Plus 0.03}_{\Minus 0.10}$ &  0.778 $^{\Plus 0.063}_{\Minus 0.668}$ & -8.86 $^{\Plus 0.07}_{\Minus 0.79}$  \\
17qp & II   & 20:28:49.80 & --04:22:57.29 & 0.01$^{[2]}$ & ---& 43.3$\pm$21.7  & 0.050     & 7.02 $^{\Plus 0.01}_{\Minus 0.01}$ &  0.005 $^{\Plus 0.001}_{\Minus 0.001}$ & -8.60 $^{\Plus 0.30}_{\Minus 0.18}$  \\
17qt & II   & 02:27:36.59 & --20:42:56.18 & 0.036 &    --- & 158.6$\pm$22.5 & 0.023        & 9.68 $^{\Plus 0.04}_{\Minus 0.03}$ &  1.854 $^{\Plus 0.193}_{\Minus 0.354}$ & -9.40 $^{\Plus 0.02}_{\Minus 0.05}$  \\
17rl & Ib/c & 07:15:00.04 & +46:22:43.79  & 0.045 &    --- & 199.5$\pm$22.8 & 0.073        & 9.57 $^{\Plus 0.02}_{\Minus 0.02}$ &  1.343 $^{\Plus 0.140}_{\Minus 0.587}$ & -9.42 $^{\Plus 0.03}_{\Minus 0.28}$  \\
\hline
\end{tabular}
\begin{tablenotes}[flushleft]
\item \textbf{Notes.} SN Type, Host Galaxy, and Discovery Reference columns come from \url{http://www.astronomy.ohio-state.edu/~assassin/sn_list.txt}, except where noted. 
\item $^{\dagger}$ Hubble flow distances are derived from the host galaxy if available in NED and the uncertainty is derived from the velocity calculator which accounts for the Virgo Cluster, Great Attractor and Shapley Supercluster infall velocities. If a redshift is not available in NED, we search the literature for redshifts derived from a host galaxy spectrum or narrow emission lines from the SN spectrum, in these cases we adopt an 8 per cent uncertainty in the distance (since this is the maximum uncertainty derived for the NED velocity uncertainties). Finally, if the host galaxy redshift is unknown, we use the SN redshift and give the luminosity distance, with an uncertainty on the redshift of $z$=0.005. 
 \item $^{\ddagger}$ sSFR is based on the PDF marginalised over all the other parameters in the SED fit. Thus it is slightly different from the derived SFR/Mass.
 \item [1] Host redshift was not obtained from NED, but from another source. For 16ll, the redshift was derived from narrow emission lines of the host galaxy in the SN spectrum \citep{Tomasella2016}. For 16ns, there is no available host galaxy redshift, therefore we use the best estimate SN redshift of $z$=0.038 \citet{Turatto2016}. Finally for 17qp, we use the best available redshift estimate from \citet{Benetti2017} with a 50 per cent uncertainty.  
  \item [2] For these cases, SN redshift was not obtained from the ASAS-SN website, but from another source. For 16ns, there is no available host galaxy redshift, therefore we use the best estimate SN redshift of $z$=0.038 \citet{Turatto2016}. Finally for 17qp, we use the best available redshift estimate from \citet{Benetti2017} with a 50 per cent uncertainty.  
\end{tablenotes}
\end{threeparttable}
\end{table*}

\begin{table*}
\centering
\contcaption{}
\begin{threeparttable}
\label{tab:asassntable3}
\begin{tabular}{lllllllrrrrrrr} 
\hline
\multicolumn{1}{c}{Name}&\multicolumn{1}{c}{Class}&\multicolumn{1}{c}{$\alpha$(2000)}&\multicolumn{1}{c}{$\delta$(2000)}&\multicolumn{1}{c}{$z_{\rm sn}$}&\multicolumn{1}{c}{$z_{\rm host}$} &\multicolumn{1}{c}{Distance$^{\dagger}$}&\multicolumn{1}{c}{E($B$-$V$)}&\multicolumn{1}{c}{log$_{10}\big($M$_{*}\big)$}&\multicolumn{1}{c}{SFR} &\multicolumn{1}{c}{sSFR$^{\ddagger}$} \\
&&&&&& \multicolumn{1}{c}{Mpc} && \multicolumn{1}{c}{M$_{\odot}$}& \multicolumn{1}{c}{M$_{\odot}$ yr$^{-1}$ } &\multicolumn{1}{c}{yr $^{-1}$} \\
\hline
SN2014ce             & II    & 23:27:40.86 & +23:35:21.4  & 0.011000 & 0.011000& 49.00$\pm$3.43   & 0.039 & 9.43 $^{\Plus 0.04}_{\Minus 0.01}$ &  5.598 $^{\Plus 0.832}_{\Minus 0.051}$ & -8.70 $^{\Plus 0.01}_{\Minus 0.01}$ & \\
SN2014cw             & II    & 22:15:26.55 & -10:28:34.6  & 0.006000 & --- &     25.82$\pm$21.70  & 0.051 & 7.72 $^{\Plus 0.09}_{\Minus 0.07}$ &  0.038 $^{\Plus 0.015}_{\Minus 0.010}$ & -9.17 $^{\Plus 0.16}_{\Minus 0.18}$ & \\
SN2014cy             & IIP   & 23:44:16.03 & +10:46:12.5  & 0.005547 & 0.005547& 24.65$\pm$1.73   & 0.049 & 10.18 $^{\Plus 0.06}_{\Minus 0.17}$ &  1.227 $^{\Plus 0.097}_{\Minus 0.161}$ & -10.08 $^{\Plus 0.14}_{\Minus 0.07}$ & \\
SN2014eb             & II    & 09:52:55.58 & +42:50:51.1  & 0.016000 & 0.016000& 73.37$\pm$5.14   & 0.011 & 10.46 $^{\Plus 0.05}_{\Minus 0.09}$ &  1.589 $^{\Plus 0.393}_{\Minus 0.195}$ & -10.27 $^{\Plus 0.17}_{\Minus 0.08}$ & \\
SN2014eh             & Ic    & 20:25:03.86 & -24:49:13.3  & 0.010614 & 0.010614& 49.45$\pm$3.46   & 0.056 & 10.61 $^{\Plus 0.08}_{\Minus 0.04}$ &  13.270 $^{\Plus 1.048}_{\Minus 4.019}$ & -9.45 $^{\Plus 0.06}_{\Minus 0.36}$ & \\
SN2015da             & IIn   & 13:52:24.11 & +39:41:28.6  & 0.00722  & 0.00722 & 39.11$\pm$2.75   & 0.013 & 9.97 $^{\Plus 0.01}_{\Minus 0.01}$ &  0.116 $^{\Plus 0.002}_{\Minus 0.002}$ & -10.90 $^{\Plus 0.01}_{\Minus 0.01}$ & \\
SN2015U              & Ibn   & 07:28:53.87 & +33:49:10.6  & 0.01379  & 0.01379 & 61.36$\pm$4.30   & 0.051 & 10.78 $^{\Plus 0.02}_{\Minus 0.04}$ &  0.169 $^{\Plus 0.075}_{\Minus 0.074}$ & -11.56 $^{\Plus 0.19}_{\Minus 0.33}$ & \\
LSQ15xp              & IIP   & 11:32:42.79 & -16:44:01.2  & 0.01226  & 0.01226 & 57.34$\pm$4.07   & 0.034 & 8.79 $^{\Plus 0.01}_{\Minus 0.05}$ &  1.560 $^{\Plus 0.073}_{\Minus 0.134}$ & -8.60 $^{\Plus 0.04}_{\Minus 0.03}$ & \\
PS15si               & IIn   & 11:10:22.93 & -04:21:31.5  & 0.05390  & 0.05390 & 237.74$\pm$16.67 & 0.046 & 8.92 $^{\Plus 0.05}_{\Minus 0.06}$ &  0.522 $^{\Plus 0.113}_{\Minus 0.257}$ & -9.20 $^{\Plus 0.05}_{\Minus 0.26}$ & \\
SN2015V              & IIP   & 17:49:27.05 & +36:08:36.0  & 0.00457  & 0.00457 & 26.64$\pm$1.87   & 0.034 & 8.53 $^{\Plus 0.07}_{\Minus 0.03}$ &  0.314 $^{\Plus 0.005}_{\Minus 0.005}$ & -9.09 $^{\Plus 0.05}_{\Minus 0.01}$ & \\
SN2015Y              & IIb   & 09:02:37.87 & +25:56:04.2  & 0.00817  & 0.00817 & 39.50$\pm$2.77   & 0.034 & 10.37 $^{\Plus 0.01}_{\Minus 0.02}$ &  0.288 $^{\Plus 0.003}_{\Minus 0.009}$ & -10.90 $^{\Plus 0.01}_{\Minus 0.01}$ & \\
PSNJ14372160+3634018 & II    & 14:37:21.60 & +36:34:01.8  & 0.01409  & 0.01409 & 68.72$\pm$4.81   & 0.015 & 10.57 $^{\Plus 0.02}_{\Minus 0.01}$ &  1.406 $^{\Plus 0.003}_{\Minus 0.171}$ & -10.50 $^{\Plus 0.01}_{\Minus 0.03}$ & \\
SN2015Q              & Ib    & 11:47:35.08 & +55:58:14.7  & 0.00803  & 0.00803 & 40.99$\pm$2.87   & 0.010 & 10.03 $^{\Plus 0.04}_{\Minus 0.03}$ &  2.254 $^{\Plus 0.526}_{\Minus 0.061}$ & -9.69 $^{\Plus 0.09}_{\Minus 0.02}$ & \\
PSNJ17292918+7542390 & II    & 17:29:29.18 & +75:42:39.0  & 0.00438  & 0.00438 & 24.68$\pm$1.73   & 0.036 & 9.40 $^{\Plus 0.03}_{\Minus 0.01}$ &  1.567 $^{\Plus 0.135}_{\Minus 0.128}$ & -9.18 $^{\Plus 0.05}_{\Minus 0.03}$ & \\
PSNJ22460504-1059484 & Ib    & 22:46:05.04 & -10:59:48.4  & 0.00895  & 0.00895 & 39.55$\pm$2.77   & 0.054 & 10.33 $^{\Plus 0.03}_{\Minus 0.25}$ &  1.758 $^{\Plus 0.318}_{\Minus 0.323}$ & -10.07 $^{\Plus 0.17}_{\Minus 0.02}$ & \\
SN2015ah             & Ib    & 23:00:24.63 & +01:37:36.8  & 0.01613  & 0.01613 & 69.18$\pm$4.91   & 0.071 & 9.95 $^{\Plus 0.02}_{\Minus 0.10}$ &  1.507 $^{\Plus 0.597}_{\Minus 0.081}$ & -9.81 $^{\Plus 0.29}_{\Minus 0.04}$ & \\
PSNJ22411479-2147421 & Ib    & 22:41:14.79 & -21:47:42.1  & 0.01495  & 0.01495 & 65.10$\pm$4.56   & 0.026 & 9.38 $^{\Plus 0.01}_{\Minus 0.01}$ &  0.875 $^{\Plus 0.016}_{\Minus 0.550}$ & -9.41 $^{\Plus 0.01}_{\Minus 0.48}$ & \\
SN2015ap             & Ib    & 02:05:13.32 & +06:06:08.4  & 0.01138  & 0.01138 & 47.02$\pm$3.29   & 0.038 & 9.46 $^{\Plus 0.01}_{\Minus 0.04}$ &  3.837 $^{\Plus 0.190}_{\Minus 0.173}$ & -8.87 $^{\Plus 0.05}_{\Minus 0.02}$ & \\
SN2015aq             & II    & 09:25:44.53 & +34:16:36.1  & 0.00550  & 0.00550 & 28.74$\pm$2.01   & 0.015 & 9.05 $^{\Plus 0.01}_{\Minus 0.01}$ &  0.050 $^{\Plus 0.001}_{\Minus 0.001}$ & -10.30 $^{\Plus 0.01}_{\Minus 0.01}$ & \\
SN2015ay             & II    & 01:09:46.77 & +13:18:28.9  & 0.01407  & 0.01407 & 58.61$\pm$4.10   & 0.024 & 9.43 $^{\Plus 0.11}_{\Minus 0.04}$ &  0.708 $^{\Plus 0.021}_{\Minus 0.189}$ & -9.58 $^{\Plus 0.05}_{\Minus 0.26}$ & \\
SN2015as             & II    & 10:08:11.37 & +51:50:40.9  & 0.00365  & 0.00365 & 21.30$\pm$1.49   & 0.009 & 8.53 $^{\Plus 0.11}_{\Minus 0.09}$ &  0.479 $^{\Plus 0.023}_{\Minus 0.100}$ & -8.81 $^{\Plus 0.10}_{\Minus 0.17}$ & \\
SN2015ba             & II    & 14:32:29.19 & +49:53:34.5  & 0.00795  & 0.00795 & 41.79$\pm$2.93   & 0.016 & 10.40 $^{\Plus 0.01}_{\Minus 0.01}$ &  0.178 $^{\Plus 0.002}_{\Minus 0.002}$ & -11.10 $^{\Plus 0.01}_{\Minus 0.01}$ & \\
SN2015bf             & IIn   & 23:24:49.03 & +15:16:52.0  & 0.01423  & 0.01423 & 60.76$\pm$4.26   & 0.059 & 10.25 $^{\Plus 0.27}_{\Minus 0.01}$ &  0.899 $^{\Plus 0.320}_{\Minus 0.018}$ & -10.31 $^{\Plus 0.03}_{\Minus 0.16}$ & \\ 
SN2016C              & IIP   & 13:38:05.30 & -17:51:15.3  & 0.00452  & 0.00452 & 20.32$\pm$1.42   & 0.079 & 10.23 $^{\Plus 0.07}_{\Minus 0.09}$ &  2.655 $^{\Plus 1.272}_{\Minus 0.836}$ & -9.85 $^{\Plus 0.32}_{\Minus 0.19}$ & \\
SN2016P              & Ic-BL & 13:57:31.10 & +06:05:51.0  & 0.01462  & 0.01462 & 71.73$\pm$5.02   & 0.024 & 10.23 $^{\Plus 0.01}_{\Minus 0.02}$ &  4.519 $^{\Plus 0.116}_{\Minus 0.113}$ & -9.60 $^{\Plus 0.01}_{\Minus 0.01}$ & \\
SN2016afa            & II    & 15:36:32.47 & +16:36:36.7  & 0.00653  & 0.00653 & 35.70$\pm$2.50   & 0.048 & 10.10 $^{\Plus 0.05}_{\Minus 0.03}$ &  2.239 $^{\Plus 0.440}_{\Minus 0.031}$ & -9.80 $^{\Plus 0.10}_{\Minus 0.01}$ & \\
SN2016bam            & II    & 07:46:52.72 & +39:01:21.8  & 0.01350  & 0.01335 & 59.93$\pm$4.20   & 0.045 & 10.19 $^{\Plus 0.07}_{\Minus 0.02}$ &  7.998 $^{\Plus 0.552}_{\Minus 1.544}$ & -9.28 $^{\Plus 0.04}_{\Minus 0.16}$ & \\
SN2016bau            & Ib    & 11:20:59.02 & +53:10:25.6  & 0.00386  & 0.00386 & 23.16$\pm$1.62   & 0.015 & 9.52 $^{\Plus 0.05}_{\Minus 0.02}$ &  2.559 $^{\Plus 0.228}_{\Minus 0.933}$ & -9.08 $^{\Plus 0.04}_{\Minus 0.27}$ & \\
SN2016bdu            & IIn   & 13:10:13.95 & +32:31:14.1  & 0.01700  & 0.01700 & 73.79$\pm$22.07  & 0.013 & 7.04 $^{\Plus 0.09}_{\Minus 0.01}$ &  0.004 $^{\Plus 0.001}_{\Minus 0.002}$ & -8.89 $^{\Plus 0.25}_{\Minus 0.66}$ & \\
SN2016bir            & IIb   & 13:14:05.90 & +33:55:09.7  & 0.03535  & 0.03535 & 155.61$\pm$22.68 & 0.010 & 9.03 $^{\Plus 0.12}_{\Minus 0.07}$ &  1.766 $^{\Plus 0.062}_{\Minus 0.425}$ & -8.77 $^{\Plus 0.07}_{\Minus 0.25}$ & \\
SN2016bkv            & IIb   & 10:18:19.31 & +41:25:39.3  & 0.00198  & 0.00198 & 10.55$\pm$0.74   & 0.015 & 8.97 $^{\Plus 0.01}_{\Minus 0.01}$ &  0.140 $^{\Plus 0.003}_{\Minus 0.010}$ & -9.90 $^{\Plus 0.01}_{\Minus 0.06}$ & \\
SN2016ccm            & IIP   & 14:09:58.91 & +17:45:49.4  & 0.01816  & 0.01816 & 87.04$\pm$6.09   & 0.022 & 11.17 $^{\Plus 0.01}_{\Minus 0.02}$ &  0.641 $^{\Plus 0.161}_{\Minus 0.016}$ & -11.39 $^{\Plus 0.16}_{\Minus 0.01}$ & \\
SN2016gfy            & II    & 07:26:45.93 & +85:45:51.2  & 0.00806  & 0.00806 & 38.73$\pm$2.71   & 0.088 & 10.06 $^{\Plus 0.02}_{\Minus 0.04}$ &  7.096 $^{\Plus 0.921}_{\Minus 1.552}$ & -9.17 $^{\Plus 0.05}_{\Minus 0.16}$ & \\
SN2016gkg            & IIb   & 01:34:14.46 & -29:26:25.0  & 0.00490  & 0.00490 & 20.53$\pm$1.44   & 0.017 & 10.75 $^{\Plus 0.06}_{\Minus 0.04}$ &  3.508 $^{\Plus 1.224}_{\Minus 2.501}$ & -10.10 $^{\Plus 0.08}_{\Minus 0.76}$ & \\
SN2016hbd            & IIP   & 02:56:06.21 & +27:42:06.8  & 0.02159  & 0.02159 & 90.00$\pm$6.30   & 0.122 & 9.10 $^{\Plus 0.06}_{\Minus 0.01}$ &  0.482 $^{\Plus 0.121}_{\Minus 0.414}$ & -9.43 $^{\Plus 0.08}_{\Minus 0.87}$ & \\
SN2016hgm            & II    & 01:22:11.73 & +00:57:07.8  & 0.00780  & 0.00780 & 32.53$\pm$2.28   & 0.029 & 9.53 $^{\Plus 0.01}_{\Minus 0.01}$ &  1.371 $^{\Plus 0.013}_{\Minus 0.019}$ & -9.40 $^{\Plus 0.01}_{\Minus 0.01}$ & \\
SN2016hvu            & IIP   & 22:35:55.56 & +20:19:12.6  & 0.01852  & 0.01852 & 79.75$\pm$5.58   & 0.042 & 10.10 $^{\Plus 0.12}_{\Minus 0.01}$ &  6.310 $^{\Plus 0.675}_{\Minus 2.737}$ & -9.30 $^{\Plus 0.01}_{\Minus 0.40}$ & \\
SN2016idl            & IIn   & 10:06:29.13 & +22:26:43.8  & 0.05800  & 0.05800 & 259.67$\pm$23.42 & 0.029 & 7.16 $^{\Plus 0.13}_{\Minus 0.02}$ &  0.038 $^{\Plus 0.001}_{\Minus 0.022}$ & -7.67 $^{\Plus 0.12}_{\Minus 0.38}$ & \\
SN2016iyy            & II    & 06:56:34.62 & +46:53:38.6  & 0.02860  & ---     & 125.25$\pm$22.46 & 0.060 & 9.15 $^{\Plus 0.08}_{\Minus 0.16}$ &  0.449 $^{\Plus 0.131}_{\Minus 0.094}$ & -9.55 $^{\Plus 0.28}_{\Minus 0.15}$ & \\
SN2016jft            & IIP   & 09:43:55.87 & +41:41:17.8  & 0.01750  & 0.01750 & 79.68$\pm$5.59   & 0.014 & 10.37 $^{\Plus 0.01}_{\Minus 0.09}$ &  1.542 $^{\Plus 0.043}_{\Minus 0.435}$ & -10.20 $^{\Plus 0.01}_{\Minus 0.01}$ & \\
SN2016jfu            & IIP   & 12:54:42.60 & +28:56:26.0  & 0.00829  & 0.00829 & 43.50$\pm$3.05   & 0.011 & 10.14 $^{\Plus 0.05}_{\Minus 0.01}$ &  2.799 $^{\Plus 0.006}_{\Minus 0.076}$ & -9.70 $^{\Plus 0.01}_{\Minus 0.03}$ & \\
SN2017ati            & IIb   & 09:49:56.7  & +67:10:59.56 & 0.013050 & 0.013050& 56.47$\pm$21.94  & 0.106 & 8.34 $^{\Plus 0.03}_{\Minus 0.04}$ &  0.224 $^{\Plus 0.036}_{\Minus 0.097}$ & -8.96 $^{\Plus 0.02}_{\Minus 0.33}$ & \\
SN2017ays            & II    & 12:12:45.99 & +00:24:28.11 & 0.020515 & 0.020515& 89.29$\pm$22.19  & 0.025 & 8.78 $^{\Plus 0.03}_{\Minus 0.04}$ &  0.604 $^{\Plus 0.020}_{\Minus 0.043}$ & -9.02 $^{\Plus 0.05}_{\Minus 0.03}$ & \\
SN2017bgu            & Ib    & 16:55:59.47 & +42:33:36.01 & 0.008503 & 0.008503& 45.30$\pm$3.17   & 0.019 & 8.33 $^{\Plus 0.03}_{\Minus 0.04}$ &  0.171 $^{\Plus 0.004}_{\Minus 0.124}$ & -9.07 $^{\Plus 0.04}_{\Minus 0.66}$ & \\
SN2017byz            & II    & 11:23:30.78 & -08:39:11.84 & 0.012285 & 0.012285& 58.52$\pm$4.10   & 0.037 & 10.27 $^{\Plus 0.01}_{\Minus 0.06}$ &  6.310 $^{\Plus 0.001}_{\Minus 1.832}$ & -9.50 $^{\Plus 0.01}_{\Minus 0.10}$ & \\
SN2017cat            & II    & 17:58:52.09 & +34:00:09.32 & 0.024894 & 0.024894& 112.77$\pm$7.9   & 0.042 & 10.63 $^{\Plus 0.02}_{\Minus 0.01}$ &  2.512 $^{\Plus 0.001}_{\Minus 0.216}$ & -10.20 $^{\Plus 0.01}_{\Minus 0.08}$ & \\
SN2017cfa            & IIP   & 09:57:3.89  & -07:52:51.14 & 0.014063 & 0.014063& 65.71$\pm$4.62   & 0.088 & 9.58 $^{\Plus 0.04}_{\Minus 0.01}$ &  1.429 $^{\Plus 0.156}_{\Minus 0.033}$ & -9.40 $^{\Plus 0.02}_{\Minus 0.01}$ & \\
SN2017cik            & IIn   & 07:54:13.07 & +21:47:36.49 & 0.015474 & 0.015474& 67.08$\pm$22.02  & 0.054 & 8.48 $^{\Plus 0.09}_{\Minus 0.01}$ &  0.129 $^{\Plus 0.093}_{\Minus 0.003}$ & -9.39 $^{\Plus 0.14}_{\Minus 0.01}$ & \\
SN2017cjb            & II    & 12:53:50.45 & +09:42:17.70 & 0.009443 & 0.009443& 47.67$\pm$3.34   & 0.020 & 9.95 $^{\Plus 0.04}_{\Minus 0.02}$ &  2.553 $^{\Plus 0.024}_{\Minus 0.475}$ & -9.60 $^{\Plus 0.04}_{\Minus 0.08}$ & \\
SN2017cjd            & Ic    & 11:50:30.17 & -18:35:44.96 & 0.023000 & 0.023000& 100.30$\pm$22.27 & 0.032 & 8.89 $^{\Plus 0.08}_{\Minus 0.01}$ &  0.687 $^{\Plus 0.046}_{\Minus 0.426}$ & -9.10 $^{\Plus 0.01}_{\Minus 0.49}$ & \\
SN2017czd            & II    & 14:51:47.05 & +43:38:40.96 & 0.008410 & 0.008410& 43.81$\pm$3.07   & 0.022 & 9.13 $^{\Plus 0.01}_{\Minus 0.04}$ &  0.280 $^{\Plus 0.001}_{\Minus 0.016}$ & -9.70 $^{\Plus 0.03}_{\Minus 0.01}$ & \\
SN2017dcc            & Ic    & 12:49:4.89  & -12:12:22.42 & 0.024500 & 0.024500& 106.96$\pm$22.32 & 0.041 & 9.18 $^{\Plus 0.04}_{\Minus 0.07}$ &  1.094 $^{\Plus 0.122}_{\Minus 0.374}$ & -9.16 $^{\Plus 0.04}_{\Minus 0.05}$ & \\
SN2017ein            & Ic    & 11:52:53.25 & +44:07:26.20 & 0.002699 & 0.002699& 16.20$\pm$1.14   & 0.019 & 9.60 $^{\Plus 0.01}_{\Minus 0.26}$ &  1.016 $^{\Plus 0.028}_{\Minus 0.409}$ & -9.60 $^{\Plus 0.01}_{\Minus 0.01}$ & \\
SN2017ewx            & Ib    & 14:02:16.52 & +07:40:44.21 & 0.015217 & 0.015217& 74.44$\pm$5.21   & 0.024 & 10.39 $^{\Plus 0.03}_{\Minus 0.06}$ &  0.356 $^{\Plus 0.090}_{\Minus 0.039}$ & -10.84 $^{\Plus 0.13}_{\Minus 0.06}$ & \\
SN2017faa            & II    & 13:19:03.90 & -02:30:45.81 & 0.018480 & 0.018480& 88.81$\pm$6.22   & 0.030 & 9.69 $^{\Plus 0.18}_{\Minus 0.02}$ &  1.875 $^{\Plus 0.120}_{\Minus 1.463}$ & -9.43 $^{\Plus 0.03}_{\Minus 0.76}$ & \\
SN2017fek            & IIb   & 20:21:47.44 & -10:43:53.27 & 0.033000 & 0.033000& 145.01$\pm$22.61 & 0.066 & 10.04 $^{\Plus 0.10}_{\Minus 0.15}$ &  2.410 $^{\Plus 1.146}_{\Minus 0.764}$ & -9.71 $^{\Plus 0.34}_{\Minus 0.26}$ & \\
SN2017fem            & IIP   & 14:32:27.32 & +27:25:36.75 & 0.014337 & 0.014337& 70.23$\pm$4.94   & 0.019 & 9.51 $^{\Plus 0.05}_{\Minus 0.07}$ &  1.047 $^{\Plus 0.660}_{\Minus 0.202}$ & -9.51 $^{\Plus 0.33}_{\Minus 0.13}$ & \\
SN2017gmr            & II    & 02:35:30.15 & -09:21:14.95 & 0.005037 & 0.005037& 20.77$\pm$1.46   & 0.024 & 9.59 $^{\Plus 0.08}_{\Minus 0.34}$ &  3.162 $^{\Plus 1.792}_{\Minus 0.599}$ & -9.12 $^{\Plus 0.52}_{\Minus 0.13}$ & \\
SN2017grn            & II    & 23:31:53.6  & -05:00:43.40 & 0.017312 & 0.017312& 73.59$\pm$5.15   & 0.040 & 10.87 $^{\Plus 0.01}_{\Minus 0.01}$ &  0.575 $^{\Plus 0.761}_{\Minus 0.074}$ & -11.14 $^{\Plus 0.44}_{\Minus 0.06}$ & \\
SN2017hca            & II    & 08:49:41.07 & -08:05:31.25 & 0.013403 & 0.013403& 61.35$\pm$4.30   & 0.037 & 8.97 $^{\Plus 0.01}_{\Minus 0.03}$ &  0.298 $^{\Plus 0.033}_{\Minus 0.178}$ & -9.49 $^{\Plus 0.08}_{\Minus 0.38}$ & \\
SN2017hcc            & IIn   & 00:03:50.58 & -11:28:28.78 & 0.017300 & 0.017300& 75.11$\pm$22.08  & 0.029 & 8.45 $^{\Plus 0.02}_{\Minus 0.03}$ &  0.500 $^{\Plus 0.012}_{\Minus 0.090}$ & -8.74 $^{\Plus 0.02}_{\Minus 0.10}$ & \\
SN2017hcd            & IIn   & 01:42:51.83 & +31:28:56.57 & 0.034847 & 0.034847& 146.17$\pm$10.23 & 0.054 & 9.97 $^{\Plus 0.01}_{\Minus 0.01}$ &  2.512 $^{\Plus 0.106}_{\Minus 0.097}$ & -9.60 $^{\Plus 0.02}_{\Minus 0.02}$ & \\
SN2017hky            & II    & 11:23:30.51 & +63:21:59.43 & 0.009725 & 0.009725& 47.44$\pm$3.32   & 0.011 & 8.75 $^{\Plus 0.18}_{\Minus 0.01}$ &  0.070 $^{\Plus 0.002}_{\Minus 0.030}$ & -9.90 $^{\Plus 0.01}_{\Minus 0.33}$ & \\
SN2017iro            & Ib/c  & 14:06:23.11 & +50:43:20.20 & 0.006191 & 0.006191& 34.32$\pm$2.40   & 0.016 & 10.01 $^{\Plus 0.04}_{\Minus 0.14}$ &  0.859 $^{\Plus 0.305}_{\Minus 0.170}$ & -10.09 $^{\Plus 0.21}_{\Minus 0.07}$ & \\
SN2017ivu            & IIP   & 15:36:32.7  & +16:36:19.40 & 0.006528 & 0.006528& 35.70$\pm$2.50   & 0.048 & 10.10 $^{\Plus 0.05}_{\Minus 0.03}$ &  2.239 $^{\Plus 0.440}_{\Minus 0.031}$ & -9.80 $^{\Plus 0.10}_{\Minus 0.01}$ & \\
SN2017jbj            & II    & 00:48:5.42  & -02:47:22.40 & 0.013492 & 0.013492& 56.42$\pm$3.95   & 0.040 & 10.76 $^{\Plus 0.01}_{\Minus 0.12}$ &  2.280 $^{\Plus 0.158}_{\Minus 0.873}$ & -10.38 $^{\Plus 0.06}_{\Minus 0.24}$ & \\
\hline
\end{tabular}
\begin{tablenotes}[flushleft]
\item \textbf{Notes.}
SN2016afa and SN2017ivu have the same host galaxy NGC 5962. 
\end{tablenotes}
\end{threeparttable}
\end{table*}

\begin{table*}
\centering
\caption{Photometric properties of LGRB host galaxies.}
\label{tab:lgrbtable}
\begin{threeparttable}
\label{tab:lgrbtable}
\begin{tabular}{llllllrrr} 
\hline
\multicolumn{1}{c}{LGRB}&\multicolumn{1}{c}{Class}&\multicolumn{1}{c}{$\alpha$(2000)}&\multicolumn{1}{c}{$\delta$(2000)}&\multicolumn{1}{c}{$z$}&\multicolumn{1}{c}{E($B$-$V$)}&\multicolumn{1}{c}{log$_{10}\big($M$_{*}\big)$}&\multicolumn{1}{c}{SFR} &\multicolumn{1}{c}{sSFR$^{\ddagger}$} \\
&&&&&& \multicolumn{1}{c}{M$_{\odot}$}& \multicolumn{1}{c}{M$_{\odot}$ yr $^{-1}$} &\multicolumn{1}{c}{yr $^{-1}$} \\
\hline
980425   &	 SN  	 & 19:35:03.12 & --52:50:44.88  & 0.009  & 0.060  & 8.48 $^{\Plus 0.40}_{\Minus 0.11}$ &  0.113 $^{\Plus 0.319}_{\Minus 0.010}$   & -9.43 $^{\Plus 0.23}_{\Minus 0.02}$ \\
020903   &	 SN      & 22:48:42.24 & --20:46:09.12  & 0.251  & 0.030  & 8.65 $^{\Plus 0.10}_{\Minus 0.23}$ &  1.014 $^{\Plus 0.465}_{\Minus 0.188}$   & -8.66 $^{\Plus 0.41}_{\Minus 0.19}$ \\
030329A  &	 SN  	 & 10:44:49.99 & +21:31:17.76   & 0.169  & 0.030  & 7.71 $^{\Plus 0.03}_{\Minus 0.08}$ &  0.086 $^{\Plus 0.016}_{\Minus 0.006}$   & -8.79 $^{\Plus 0.15}_{\Minus 0.06}$ \\
031203   &	 SN  	 & 08:02:29.04 & --39:51:11.88  & 0.105  & 0.937  & 8.50 $^{\Plus 0.01}_{\Minus 0.01}$ &  15.520 $^{\Plus 0.844}_{\Minus 0.318}$  & -7.30 $^{\Plus 0.02}_{\Minus 0.01}$ \\
050826   &	 SN-less & 05:51:01.58 & --02:38:35.88  & 0.296  & 0.600  & 9.80 $^{\Plus 0.06}_{\Minus 0.11}$ &  1.923 $^{\Plus 0.764}_{\Minus 1.722}$   & -9.61 $^{\Plus 0.31}_{\Minus 0.80}$ \\
060218   &	 SN   	 & 03:21:39.67 & +16:52:01.92   & 0.033  & 0.150  & 7.48 $^{\Plus 0.04}_{\Minus 0.08}$ &  0.039 $^{\Plus 0.015}_{\Minus 0.011}$   & -8.93 $^{\Plus 0.27}_{\Minus 0.19}$ \\
060505   &	 SN-less & 22:07:03.43 & --27:48:51.84  & 0.089  & 0.020  & 9.57 $^{\Plus 0.02}_{\Minus 0.06}$ &  0.766 $^{\Plus 0.097}_{\Minus 0.012}$   & -9.69 $^{\Plus 0.08}_{\Minus 0.03}$ \\
060614   &	 SN-less & 21:23:32.11 & --53:01:36.12  & 0.126  & 0.020  & 7.91 $^{\Plus 0.07}_{\Minus 0.04}$ &  0.002 $^{\Plus 0.015}_{\Minus 0.002}$   & -10.52 $^{\Plus 0.69}_{\Minus 1.09}$ \\
080517   &	 SN-less & 06:48:58.06 & +50:44:05.64   & 0.089  & 0.110  & 9.80 $^{\Plus 0.03}_{\Minus 0.02}$ &  1.500 $^{\Plus 0.096}_{\Minus 0.249}$   & -9.62 $^{\Plus 0.03}_{\Minus 0.07}$ \\
100316D  &	 SN   	 & 07:10:30.53 & --56:15:19.80  & 0.059  & 0.120  & 8.94 $^{\Plus 0.03}_{\Minus 0.08}$ &  0.762 $^{\Plus 0.254}_{\Minus 0.086}$   & -9.10 $^{\Plus 0.24}_{\Minus 0.06}$ \\
111225A  &	 SN-less & 00:52:37.22	& +51:34:19.5   & 0.297  & 0.229  & 7.42 $^{\Plus 0.23}_{\Minus 0.17}$ &  0.259 $^{\Plus 0.086}_{\Minus 0.094}$   & -8.00 $^{\Plus 0.31}_{\Minus 0.42}$ \\
120422A  &	 SN   	 & 09:07:38.42 & +14:01:07.68   & 0.283  & 0.030  & 9.04 $^{\Plus 0.03}_{\Minus 0.03}$ &  1.219 $^{\Plus 0.233}_{\Minus 0.280}$   & -9.01 $^{\Plus 0.09}_{\Minus 0.06}$ \\
130702A  &	 SN   	 & 14:29:14.78 & +15:46:26.40   & 0.145  & 0.040  & 7.68 $^{\Plus 0.03}_{\Minus 0.17}$ &  0.032 $^{\Plus 0.012}_{\Minus 0.017}$   & -9.19 $^{\Plus 0.26}_{\Minus 0.33}$ \\
150518A  &	 SN      & 15:36:48.27 & +16:19:47.1    & 0.256  & 0.046  & 9.14 $^{\Plus 0.08}_{\Minus 0.05}$ &  1.086 $^{\Plus 0.434}_{\Minus 0.319}$   & -9.12 $^{\Plus 0.25}_{\Minus 0.21}$ \\
150818A  &	 SN      & 15:21:25.43 & +68:20:33.0    & 0.282  & 0.021  & 8.67 $^{\Plus 0.13}_{\Minus 0.30}$ &  1.489 $^{\Plus 0.906}_{\Minus 0.611}$   & -8.49 $^{\Plus 0.51}_{\Minus 0.35}$ \\
161219B  &	 SN	     & 06:06:51.43 & --26:47:29.52  & 0.148  & 0.028  & 9.03 $^{\Plus 0.07}_{\Minus 0.14}$ &  0.228 $^{\Plus 0.176}_{\Minus 0.059}$   & -9.79 $^{\Plus 0.49}_{\Minus 0.16}$ \\
171205A   &	 SN	     & 11:09:39.52 & --12:35:18.34  & 0.037  & 0.045  & 10.11 $^{\Plus 0.02}_{\Minus 0.08}$ & 2.897 $^{\Plus 0.158}_{\Minus 0.261}$   & -9.63 $^{\Plus 0.03}_{\Minus 0.02}$ \\
\hline
\end{tabular}
\begin{tablenotes}[flushleft]
\item \textbf{Notes.}$^{\ddagger}$sSFR is based on the PDF marginalised over all the other parameters in the SED fit. Thus it is slightly different from the derived SFR/Mass.
\end{tablenotes}
\end{threeparttable}
\end{table*}

\begin{table*}
\centering
\caption{Photometric properties of the SLSN host galaxies.}
\label{tab:slsnphottable}
\begin{threeparttable}
\label{tab:slsnetable}
\begin{tabular}{llllllrrr} 
\hline
\multicolumn{1}{c}{SLSN}&\multicolumn{1}{c}{Class}&\multicolumn{1}{c}{$\alpha$(2000)}&\multicolumn{1}{c}{$\delta$(2000)}&\multicolumn{1}{c}{$z$}&\multicolumn{1}{c}{E($B$-$V$)}&\multicolumn{1}{c}{log$_{10}\big($M$_{*}\big)$}&\multicolumn{1}{c}{SFR} &\multicolumn{1}{c}{sSFR$^{\ddagger}$} \\
&&&&&& \multicolumn{1}{c}{M$_{\odot}$}& \multicolumn{1}{c}{M$_{\odot}$ yr $^{-1}$} &\multicolumn{1}{c}{yr $^{-1}$} \\
\hline
LSQ12dlf	& I & 01:50:29.80 & --21:48:45.4 &   0.255 	& 0.011 &   7.64 $^{\Plus 0.24}_{\Minus 0.39}$ &  0.030 $^{\Plus 0.022}_{\Minus 0.010}$  & -9.17 $^{\Plus 0.73}_{\Minus 0.46}$ \\
LSQ14an 	& I & 12:53:47.83 & --29:31:27.2 &   0.163	& 0.074 &   8.20 $^{\Plus 0.08}_{\Minus 0.19}$ &  1.791 $^{\Plus 2.941}_{\Minus 0.358}$  & -7.97 $^{\Plus 0.62}_{\Minus 0.13}$ \\
LSQ14mo 	& I & 10:22:41.53 & --16:55:14.4 &   0.256	& 0.065 &   7.64 $^{\Plus 0.16}_{\Minus 0.23}$ &  0.331 $^{\Plus 0.101}_{\Minus 0.133}$  & -8.11 $^{\Plus 0.33}_{\Minus 0.39}$ \\
MLS121104 	& I & 02:16:42.51 & +20:40:08.5 &    0.30    & 0.150 &  9.28 $^{\Plus 0.14}_{\Minus 0.38}$ &  3.828 $^{\Plus 2.411}_{\Minus 1.485}$  & -8.70 $^{\Plus 0.61}_{\Minus 0.37}$ \\
PTF09as 	& I & 12:59:15.864 & +27:16:40.58 &  0.187	& 0.008 &   8.52 $^{\Plus 0.11}_{\Minus 0.23}$ &  0.265 $^{\Plus 0.028}_{\Minus 0.029}$  & -9.12 $^{\Plus 0.26}_{\Minus 0.14}$ \\
PTF09cnd 	& I & 16:12:08.839 & +51:29:16.01 &  0.258 	& 0.021 &   8.29 $^{\Plus 0.05}_{\Minus 0.04}$ &  0.284 $^{\Plus 0.083}_{\Minus 0.101}$  & -8.82 $^{\Plus 0.16}_{\Minus 0.26}$ \\
PTF10aagc 	& I & 09:39:56.923 & +21:43:17.09 &  0.206	& 0.022 &   8.89 $^{\Plus 0.04}_{\Minus 0.08}$ &  1.076 $^{\Plus 0.215}_{\Minus 0.297}$  & -8.82 $^{\Plus 0.10}_{\Minus 0.21}$ \\
PTF10bfz 	& I & 12:54:41.288 & +15:24:17.08 &  0.169	& 0.018 &   7.59 $^{\Plus 0.16}_{\Minus 0.06}$ &  1.052 $^{\Plus 0.073}_{\Minus 0.526}$  & -7.57 $^{\Plus 0.09}_{\Minus 0.45}$ \\
PTF10cwr 	& I & 11:25:46.73  & --08:49:41.9  & 0.231	& 0.035 &   7.71 $^{\Plus 0.17}_{\Minus 0.09}$ &  0.553 $^{\Plus 0.073}_{\Minus 0.395}$  & -7.95 $^{\Plus 0.14}_{\Minus 0.72}$ \\
PTF10hgi 	& I & 16:37:47.074 & +06:12:31.83 &  0.099	& 0.074 &   7.88 $^{\Plus 0.03}_{\Minus 0.04}$ &  0.101 $^{\Plus 0.003}_{\Minus 0.006}$  & -8.86 $^{\Plus 0.05}_{\Minus 0.08}$ \\
PTF10nmn 	& I & 15:50:02.809 & --07:24:42.38 & 0.123	& 0.138 &   7.94 $^{\Plus 0.06}_{\Minus 0.11}$ &  0.248 $^{\Plus 0.280}_{\Minus 0.066}$  & -8.55 $^{\Plus 0.45}_{\Minus 0.18}$ \\
PTF10uhf 	& I & 16:52:46.696 & +47:36:21.76 &  0.289	& 0.018 &   11.08 $^{\Plus 0.01}_{\Minus 0.06}$ & 7.278 $^{\Plus 0.596}_{\Minus 0.941}$  & -10.21 $^{\Plus 0.03}_{\Minus 0.04}$ \\ 
PTF10vwg 	& I & 18:59:32.881 & +19:24:25.74 &  0.1901 & 0.467 &   7.59 $^{\Plus 0.07}_{\Minus 0.08}$ &  0.078 $^{\Plus 0.030}_{\Minus 0.027}$  & -8.58 $^{\Plus 0.02}_{\Minus 0.01}$ \\
PTF11dij 	& I & 13:50:57.798 & +26:16:42.44 &  0.143	& 0.011 &   7.01 $^{\Plus 0.01}_{\Minus 0.01}$ &  0.327 $^{\Plus 0.108}_{\Minus 0.009}$  & -7.34 $^{\Plus 0.27}_{\Minus 0.03}$ \\
PTF11hrq 	& I & 00:51:47.22  & --26:25:10.0 &  0.057	& 0.012 &   8.18 $^{\Plus 0.14}_{\Minus 0.06}$ &  0.366 $^{\Plus 0.050}_{\Minus 0.210}$  & -8.59 $^{\Plus 0.10}_{\Minus 0.55}$ \\
PTF11rks 	& I & 01:39:45.528 & +29:55:27.43 &  0.19 	& 0.038 &   9.02 $^{\Plus 0.03}_{\Minus 0.05}$ &  0.741 $^{\Plus 0.128}_{\Minus 0.056}$  & -9.15 $^{\Plus 0.11}_{\Minus 0.05}$ \\
PTF12dam 	& I & 14:24:46.228 & +46:13:48.64 &  0.108 	& 0.100 &   8.14 $^{\Plus 0.01}_{\Minus 0.01}$ &  14.490 $^{\Plus 0.578}_{\Minus 0.199}$ & -7.00 $^{\Plus 0.01}_{\Minus 0.01}$  \\ 
SN1999as 	& I & 09:16:30.86 & +13:39:02.2 &    0.127	& 0.096 &   9.04 $^{\Plus 0.03}_{\Minus 0.03}$ &  0.379 $^{\Plus 0.170}_{\Minus 0.202}$  & -9.49 $^{\Plus 0.18}_{\Minus 0.29}$ \\
SN2005ap 	& I & 13:01:14.83 & +27:43:32.3 &    0.283	& 0.026 &   7.73 $^{\Plus 0.05}_{\Minus 0.09}$ &  0.129 $^{\Plus 0.020}_{\Minus 0.018}$  & -8.62 $^{\Plus 0.13}_{\Minus 0.11}$ \\
SN2007bi 	& I & 13:19:20.00 & +08:55:44.0 &    0.128	& 0.084 &   7.55 $^{\Plus 0.14}_{\Minus 0.23}$ &  0.070 $^{\Plus 0.026}_{\Minus 0.030}$  & -8.72 $^{\Plus 0.36}_{\Minus 0.38}$ \\
SN2010kd 	& I & 12:08:01.11 & +49:13:31.1 &    0.101	& 0.021 &   7.21 $^{\Plus 0.12}_{\Minus 0.03}$ &  0.135 $^{\Plus 0.014}_{\Minus 0.081}$  & -7.50 $^{\Plus 0.09}_{\Minus 1.04}$ \\
SN2011ep 	& I & 17:03:41.78 & +32:45:52.6 &    0.28	& 0.020 &   7.75 $^{\Plus 0.32}_{\Minus 0.26}$ &  0.916 $^{\Plus 0.059}_{\Minus 0.349}$  & -7.75 $^{\Plus 0.15}_{\Minus 0.46}$ \\
SN2011kf 	& I & 14:36:57.53 & +16:30:56.6 &    0.245	& 0.069 &   7.52 $^{\Plus 0.11}_{\Minus 0.50}$ &  0.144 $^{\Plus 0.649}_{\Minus 0.021}$  & -8.38 $^{\Plus 1.36}_{\Minus 0.15}$ \\
SN2012il 	& I & 09:46:12.91 & +19:50:28.7 &    0.175	& 0.069 &   8.11 $^{\Plus 0.02}_{\Minus 0.03}$ &  0.142 $^{\Plus 0.016}_{\Minus 0.012}$  & -8.95 $^{\Plus 0.06}_{\Minus 0.11}$ \\
SN2013dg	& I & 13:18:41.38 & --07:04:43.1 &   0.265  & 0.042 &   7.59 $^{\Plus 0.09}_{\Minus 0.02}$ &  0.021 $^{\Plus 0.005}_{\Minus 0.001}$ & -8.59 $^{\Plus 0.03}_{\Minus 0.18}$  \\
SN2015bn	& I & 11:33:41.57 & +00:43:32.2 &    0.11	& 0.022 &   7.75 $^{\Plus 0.21}_{\Minus 0.23}$ &  0.076 $^{\Plus 0.018}_{\Minus 0.072}$  & -8.86 $^{\Plus 0.28}_{\Minus 1.37}$ \\
SSS120810 	& I & 23:18:01.82  & --56:09:25.7 &  0.156 	& 0.017 &   7.02 $^{\Plus 0.18}_{\Minus 0.01}$ &  0.562 $^{\Plus 0.362}_{\Minus 0.235}$  & -7.22 $^{\Plus 0.21}_{\Minus 0.34}$ \\
\hline
PTF09q   & I$^{\dagger}$ &   12:24:50.11 & +08:25:58.8  & 0.09   & 0.021  & 10.45 $^{\Plus 0.01}_{\Minus 0.01}$ &  3.155 $^{\Plus 0.007}_{\Minus 0.017}$ & -9.90 $^{\Plus 0.01}_{\Minus 0.01}$ \\
PTF10gvb & I$^{\dagger}$ &   12:15:32.28 & +40:18:09.5  & 0.098  & 0.022  & 8.76 $^{\Plus 0.08}_{\Minus 0.17}$ &  0.330 $^{\Plus 0.097}_{\Minus 0.187}$ & -9.25 $^{\Plus 0.22}_{\Minus 0.37}$ \\
PTF11mnb & I$^{\dagger}$ &   00:34:13.25 & +02:48:31.4  & 0.0603 & 0.016  & 8.67 $^{\Plus 0.06}_{\Minus 0.07}$ &  0.297 $^{\Plus 0.184}_{\Minus 0.099}$ & -9.20 $^{\Plus 0.28}_{\Minus 0.23}$\\ 
PTF12gty & I             &   16:01:15.23 & +21:23:17.4  & 0.1768 & 0.061  & 8.24 $^{\Plus 0.07}_{\Minus 0.18}$ &  0.025 $^{\Plus 0.007}_{\Minus 0.005}$ & -9.81 $^{\Plus 0.13}_{\Minus 0.01}$  \\
PTF12hni & I             &   22:31:55.86 & --06:47:49.0 & 0.1056 & 0.054  & 9.15 $^{\Plus 0.05}_{\Minus 0.03}$ &  0.142 $^{\Plus 0.700}_{\Minus 0.045}$ & -9.97 $^{\Plus 0.66}_{\Minus 0.24}$ \\  
\hline 
CSS100217 &  II & 10:29:12.56  & +40:42:20.0   & 0.147	& 0.013 &  9.84 $^{\Plus 0.02}_{\Minus 0.03}$ &  11.510 $^{\Plus 1.827}_{\Minus 2.175}$ &  -9.63 $^{\Plus 0.43}_{\Minus 0.35}$\\  
CSS121015 &  II & 00:42:44.34  & +13:28:26.5   & 0.286	& 0.076 &  7.91 $^{\Plus 0.22}_{\Minus 0.29}$ &  0.516 $^{\Plus 0.195}_{\Minus 0.238}$  &  -8.21 $^{\Plus 0.44}_{\Minus 0.46}$\\  
PTF10fel &   II & 16:27:31.103 & +51:21:43.45  & 0.234	& 0.017 &  9.87 $^{\Plus 0.04}_{\Minus 0.06}$ &  0.863 $^{\Plus 0.223}_{\Minus 0.259}$  &  -9.93 $^{\Plus 0.13}_{\Minus 0.16}$\\  
PTF10qaf &   II & 23:35:42.887 & +10:46:32.57  & 0.284	& 0.070 &  9.24 $^{\Plus 0.03}_{\Minus 0.12}$ &  0.498 $^{\Plus 0.187}_{\Minus 0.022}$  &  -9.54 $^{\Plus 0.20}_{\Minus 0.05}$\\  
PTF10qwu & 	 II & 16:51:10.572 & +28:18:07.62  & 0.226	& 0.040 &  7.34 $^{\Plus 0.10}_{\Minus 0.15}$ &  0.230 $^{\Plus 0.045}_{\Minus 0.072}$  &  -7.99 $^{\Plus 0.22}_{\Minus 0.30}$\\  
PTF10scc &   II & 23:28:10.495 & +28:38:31.10  & 0.242	& 0.093 &  7.16 $^{\Plus 0.01}_{\Minus 0.04}$ &  0.018 $^{\Plus 0.002}_{\Minus 0.007}$  &  -7.64 $^{\Plus 0.10}_{\Minus 0.19}$\\  
PTF10tpz &   II & 21:58:31.74  & --15:33:02.6  & 0.040	& 0.041 &  10.68 $^{\Plus 0.10}_{\Minus 0.05}$ &  0.458 $^{\Plus 1.316}_{\Minus 0.256}$ & -11.09 $^{\Plus 0.69}_{\Minus 0.40}$\\ 
PTF10yyc &   II & 04:39:17.297 & --00:20:54.5  & 0.214	& 0.041 &  9.77 $^{\Plus 0.04}_{\Minus 0.09}$ &  0.230 $^{\Plus 0.178}_{\Minus 0.066}$  & -10.45 $^{\Plus 0.38}_{\Minus 0.12}$\\  
PTF12gwu &   II & 15:02:32.876 & +08:03:49.47  & 0.275	& 0.033 &  7.81 $^{\Plus 0.19}_{\Minus 0.15}$ &  0.106 $^{\Plus 0.018}_{\Minus 0.034}$  &  -8.78 $^{\Plus 0.21}_{\Minus 0.44}$\\  
PTF12mkp &   II & 08:28:35.092 & +65:10:55.60  & 0.153	& 0.046 &  7.36 $^{\Plus 0.16}_{\Minus 0.21}$ &  0.005 $^{\Plus 0.003}_{\Minus 0.002}$  &  -9.45 $^{\Plus 1.11}_{\Minus 0.43}$\\  
PTF12mue &   II & 03:18:51.072 & --11:49:13.55 & 0.279	& 0.062 &  8.76 $^{\Plus 0.08}_{\Minus 0.12}$ &  0.382 $^{\Plus 0.288}_{\Minus 0.096}$  &  -9.17 $^{\Plus 0.39}_{\Minus 0.20}$\\  
SN1999bd & 	 II & 09:30:29.17  & +16:26:07.8   & 0.151	& 0.096 &  9.50 $^{\Plus 0.20}_{\Minus 0.02}$ &  1.380 $^{\Plus 0.092}_{\Minus 0.650}$  &  -9.33 $^{\Plus 0.02}_{\Minus 0.53}$\\ 
SN2003ma & 	 II & 05:31:01.88  & --70:04:15.9  & 0.289	& 0.348 &  8.76 $^{\Plus 0.03}_{\Minus 0.06}$ &  14.290 $^{\Plus 0.743}_{\Minus 0.293}$ &  -7.60 $^{\Plus 0.08}_{\Minus 0.04}$\\  
SN2006gy & 	 II & 03:17:27.06  & +41:24:19.5   & 0.019	& 0.493 &  10.76 $^{\Plus 0.14}_{\Minus 0.04}$ &  0.001 $^{\Plus 0.001}_{\Minus 0.001}$ & -16.48 $^{\Plus 0.79}_{\Minus 0.73}$\\  
SN2006tf & 	 II & 12:46:15.82  & +11:25:56.3   & 0.074	& 0.023 &  7.97 $^{\Plus 0.04}_{\Minus 0.06}$ &  0.075 $^{\Plus 0.014}_{\Minus 0.011}$  &  -9.11 $^{\Plus 0.12}_{\Minus 0.08}$\\  
SN2007bw & 	 II & 17:11:01.99  & +24:30:36.4   & 0.14	& 0.046 &  9.42 $^{\Plus 0.03}_{\Minus 0.07}$ &  1.199 $^{\Plus 2.038}_{\Minus 1.043}$  &  -9.33 $^{\Plus 0.46}_{\Minus 0.93}$\\  
SN2008am & 	 II & 12:28:36.25  & +15:35:49.1   & 0.234	& 0.078 &  9.38 $^{\Plus 0.04}_{\Minus 0.03}$ &  2.761 $^{\Plus 0.667}_{\Minus 0.706}$  &  -8.96 $^{\Plus 0.13}_{\Minus 0.15}$\\  
SN2008es & 	 II & 11:56:49.13  & +54:27:25.7   & 0.205	& 0.037 &  7.02 $^{\Plus 0.01}_{\Minus 0.01}$ &  0.013 $^{\Plus 0.005}_{\Minus 0.005}$  &  -7.78 $^{\Plus 0.41}_{\Minus 0.59}$\\  
SN2008fz &	 II & 23:16:16.60  & +11:42:47.5   & 0.133	& 0.132 &  7.02 $^{\Plus 0.01}_{\Minus 0.01}$ &  0.011 $^{\Plus 0.003}_{\Minus 0.008}$  &  -8.26 $^{\Plus 0.26}_{\Minus 0.84}$\\  
SN2009nm & 	 II & 10:05:24.54  & +51:16:38.7   & 0.21	& 0.011 &  8.63 $^{\Plus 0.34}_{\Minus 0.42}$ &  0.200 $^{\Plus 0.107}_{\Minus 0.091}$  &  -9.40 $^{\Plus 0.67}_{\Minus 0.60}$\\  
SN2013hx &	 II & 01:35:32.83  & --57:57:50.6  & 0.13	& 0.022 &  7.49 $^{\Plus 0.20}_{\Minus 0.14}$ &  0.019 $^{\Plus 0.008}_{\Minus 0.012}$  &  -8.66 $^{\Plus 0.53}_{\Minus 0.99}$\\   
\hline
\end{tabular}
\begin{tablenotes}[flushleft]
\item \textbf{Notes.}$^{\dagger}$ Possible SLSN-I are indicated 
\item$^{\ddagger}$sSFR is based on the PDF marginalised over all the other parameters in the SED fit. Thus it is slightly different from the derived SFR/Mass.
\end{tablenotes}
\end{threeparttable}
\end{table*}


\begin{figure*}
\includegraphics[width=\textwidth]{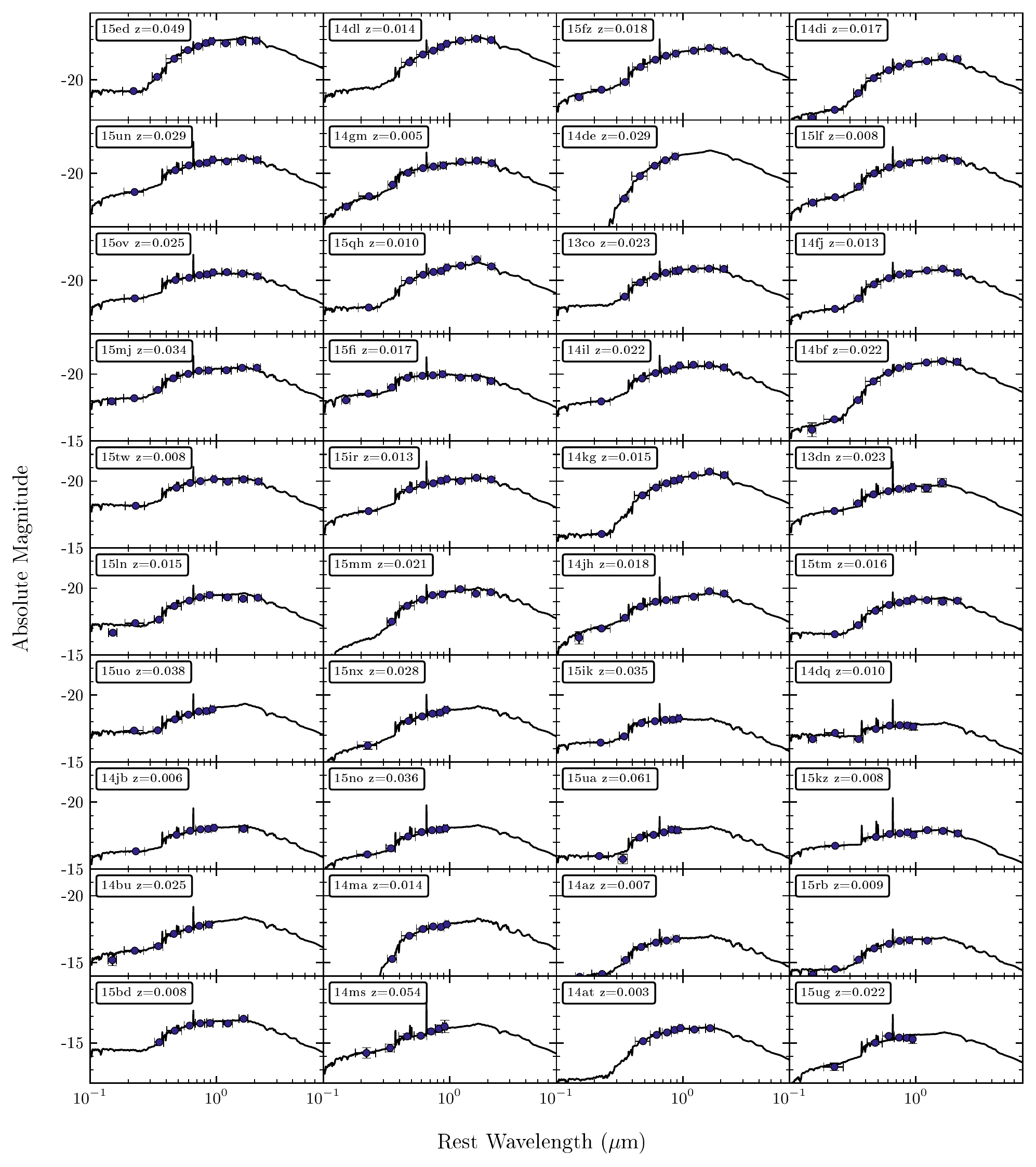}
\caption{Spectral energy distributions of the CCSN host galaxy sample. The multi-band photometry is shown as blue  markers and error bars show photometric uncertainties. The best-fitting SED model is displayed by the black curve, fitted to our data using the procedure outlined in Section ~\ref{subsec:sedfitting}. Galaxies are ordered in terms of their luminosity as measured in the \textit{r}-band via the SED. The absolute magnitude axis uses  appropriate limits for each row.}
\label{fig:ccsn_sed_1_2}
\end{figure*}

\begin{figure*}
\includegraphics[width=\textwidth]{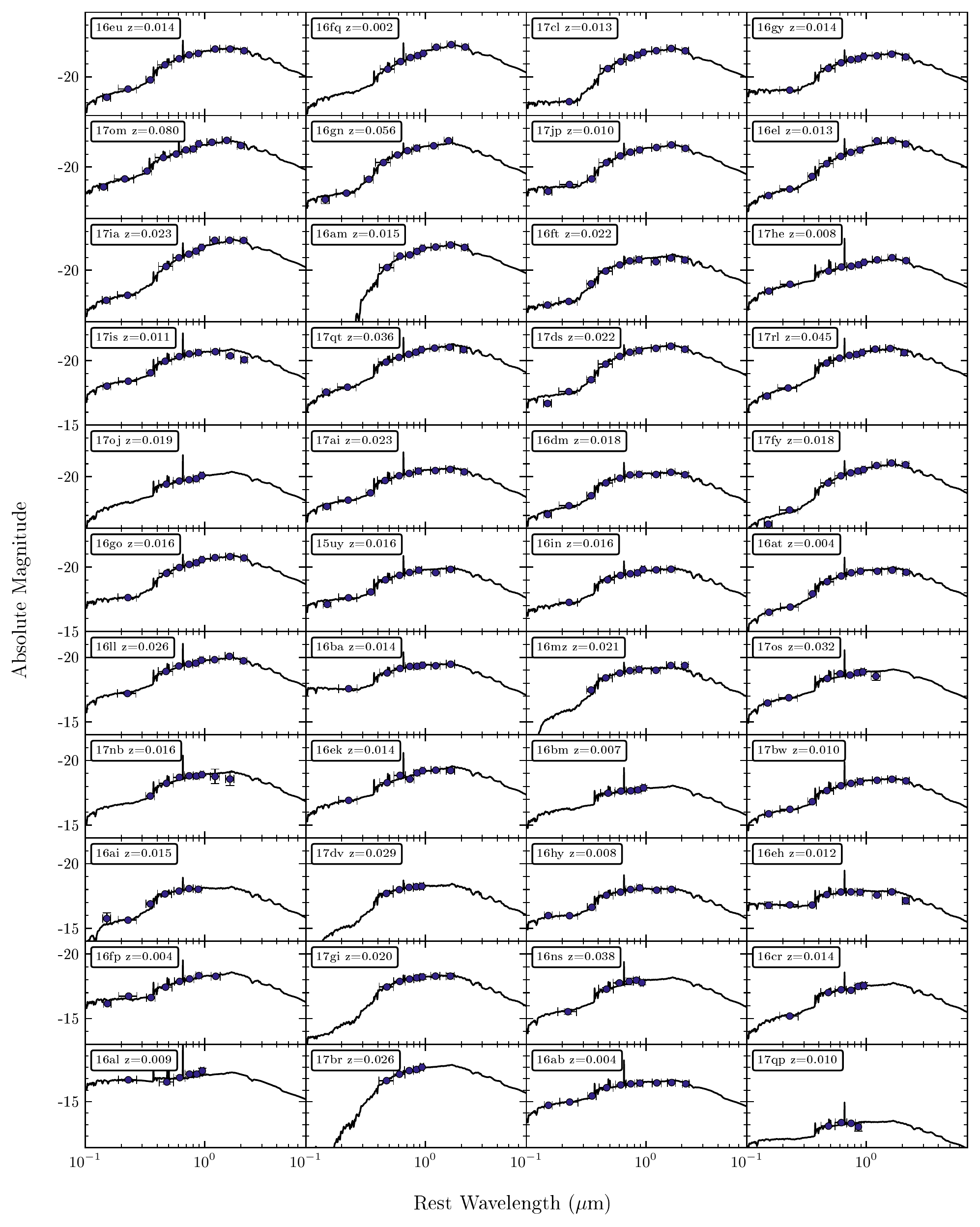}
\contcaption{}
\label{fig:ccsn_sed_3_4}
\end{figure*}

\begin{figure*}
\includegraphics[width=\textwidth]{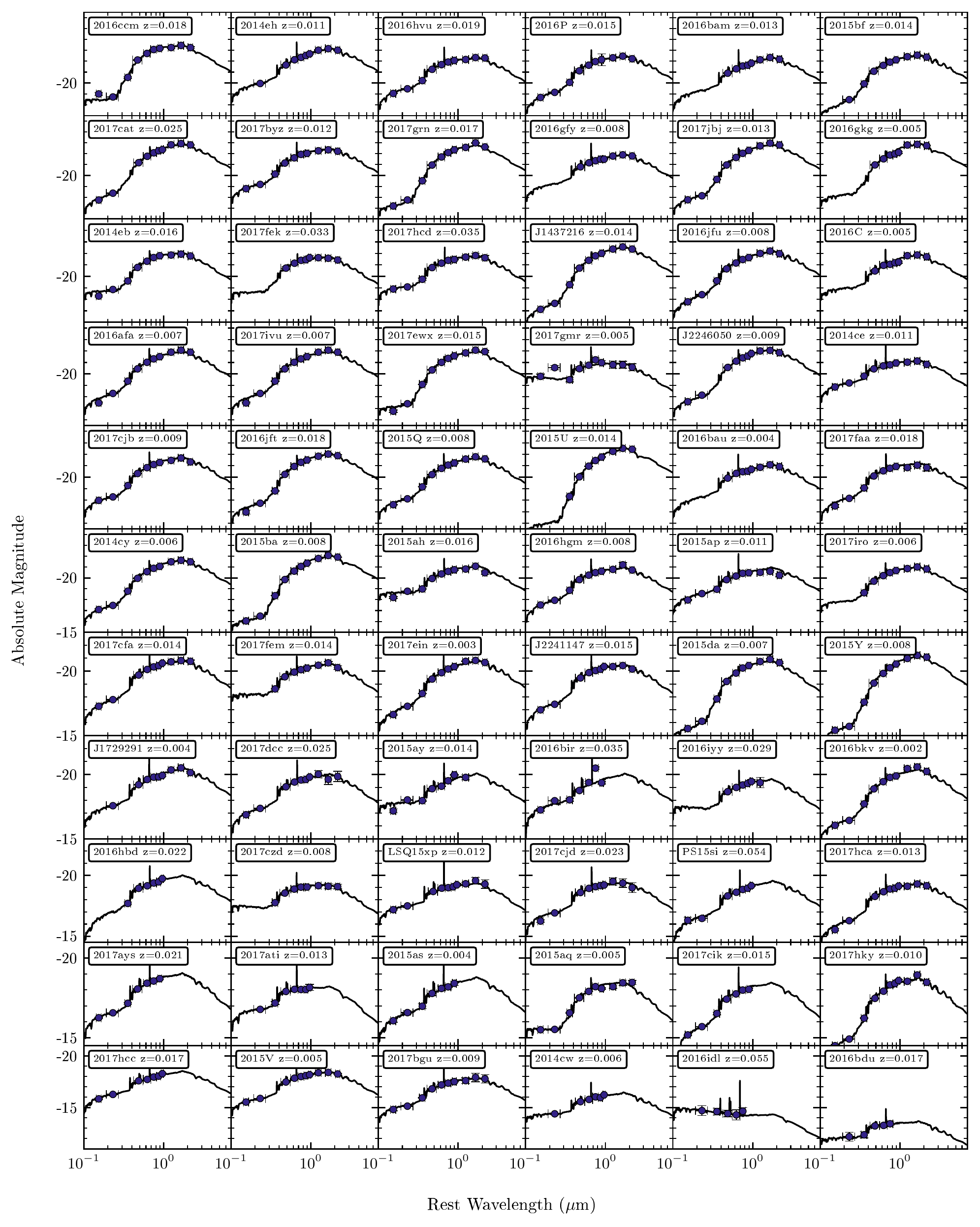}
\contcaption{}
\label{fig:ccsn_sed_5_6}
\end{figure*}

\begin{figure*}
\includegraphics[width=\textwidth]{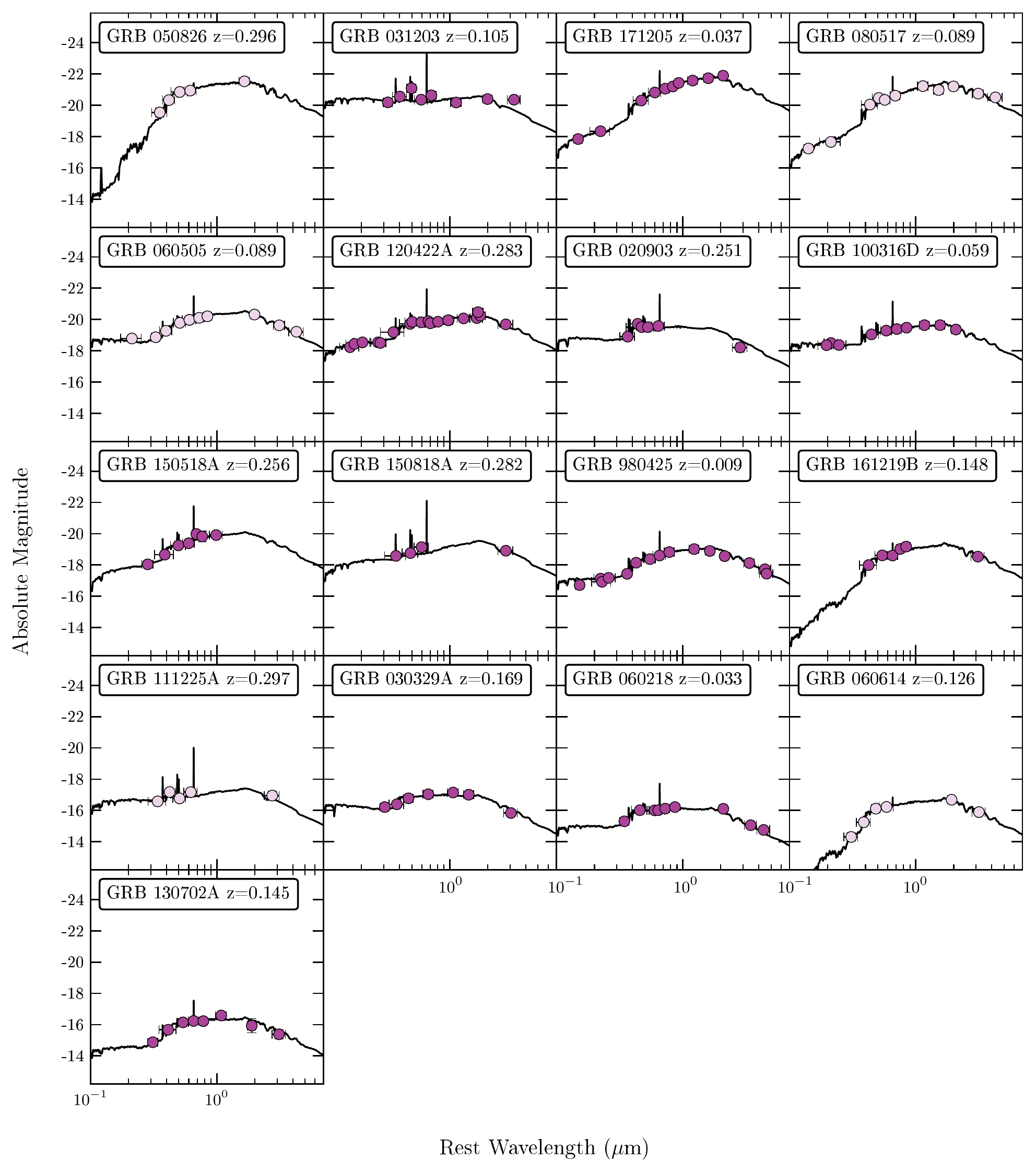}
\caption{Spectral energy distributions of the LGRB sample. The multi-band photometry for LGRBs with SN are dark purple markers whereas LGRBs without SN are light purple and error bars show photometric uncertainties. The best-fitting SED model is displayed by the black curve, fitted to our data using the procedure outlined in Section ~\ref{subsec:sedfitting}. Galaxies are ordered in terms of their luminosity as measured in the $r$-band via the SED. The absolute magnitude axis uses  appropriate limits for each row.}
\label{fig:grbsed}
\end{figure*}

\begin{figure*}
	\includegraphics[width=\textwidth]{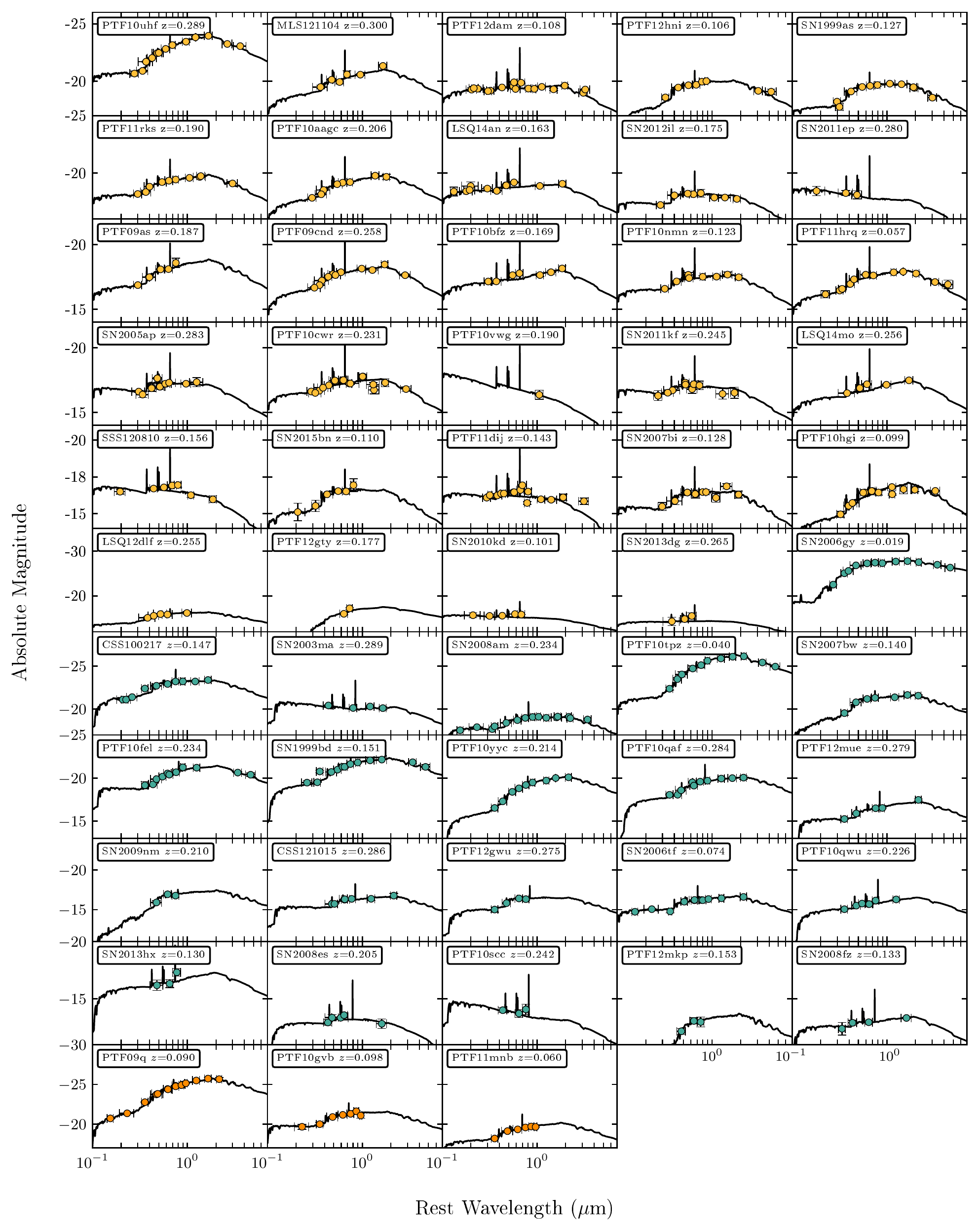}
    \caption{Spectral energy distributions of the SLSN sample. The multi-band photometry for SLSN-I are yellow markers, green for SLSN-II, and orange for possible SLSN-I and error bars show photometric uncertainties. The best-fitting SED model is displayed by the black curve, fitted to our data using the procedure outlined in Section ~\ref{subsec:sedfitting}. Galaxies are ordered in terms of their luminosity as measured in the \textit{r}-band via the SED. The absolute magnitude axis uses  appropriate limits for each row.}
    \label{fig:combined_slsnI_II} 
\end{figure*}


\section{Comparisons with previous work}\label{sec:comparisons}

We can gain further insights by examining the fraction of CCSNe exploding in dwarf galaxies. There have been a handful of studies of CCSNe hosted by dwarf galaxies discovered in untargeted surveys \citep{Young2010,Prieto2012}. Recently, \cite{Anderson2018} have began a large-scale study of supernovae in dwarf galaxies to get follow-up of dwarf galaxies (fainter than --18.5 in the \textit{B}-band) hosting SNe of type II, but have not specifically measured the fraction of all SNe in galaxies of this type. 
\citet{Arcavi2010} presented the first sample of (72) spectroscopically classified CCSN host galaxies from the Palomar Transient Factory (PTF), an untargeted (but spectroscopically incomplete) survey for supernovae. They studied the integrated properties of the hosts and found 22 per cent of CCSNe are in `dwarf' galaxies, which they define in terms of \textit{r-}band luminosity (M$_r$ $>=-18$ mag). We find a substantial fraction (33 per cent) of CCSN hosts are in galaxies with stellar masses $<1\times$10$^{9}$ M$_\odot$, which is broadly consistent with their conclusion that a large (but non-majority) fraction of star-formation occurs in subluminous galaxies.

Spectroscopic studies of CCSN host galaxies show diversity within subclasses of the CCSN population. Specifically, recent studies suggest that highly stripped progenitor stars that explode as Ic supernovae may be found more often in higher-metallicity galaxies \citep{Prieto2008,Leloudas2011,Modjaz2011} with higher sSFRs \citep{Kelly2012} than non-stripped envelope CCSNe. In addition, a sequence of decreasing metallicity has been established from Ic--(Ib/IIb) \citep{Modjaz2019}. On the other hand, the Ic-BL subclass may prefer metal-poor galaxies \citep{Modjaz2019}. The ASAS-SN sample is dominated by Type II SNe (98) with only a small number of stripped-enveloped events (19 Ib/Ic and 10 IIb and 2 Ib/c-BL), so we are unable to address these differences. 

\section{Potential Biases}\label{sec:biases}

Several biases may affect the conclusions drawn from our SN host galaxy samples. We briefly summarize these below. 

\subsection{CCSNe}  

Firstly, there may be biases in the CCSN sample associated with their discovery and follow-up. However, ASAS-SN is an untargeted survey, so the biases associated with targeted surveys are minimal and since the survey is shallow and their discovery numbers are relatively small, almost all ASAS-SN discoveries are followed-up spectroscopically. During 2013--2017 ASAS-SN discovered 595 SNe, of which 585 have known classifications (98 per cent were spectroscopically confirmed) on the ASAS-SN list on their website.\footnote{\url{http://www.astronomy.ohio-state.edu/~assassin/sn_list.txt}} In addition, ASAS-SN discovers one quarter of its SN in catalogued host galaxies without known redshifts \citep{Holoien2017a}, indicating that ASAS-SN is less biased against finding supernovae in uncatalogued hosts than previous low-redshift SN surveys; it also finds supernova closer to the galaxy nuclei than preceding projects. 

The ASAS-SN survey is effectively flux-limited, so luminous types of CCSNe will be overpopulated in our sample compared to a volume-limited survey, possibly meaning our sample is not a true representation of the sites of (all) massive star-formation. In particular, the Type IIn subclass of SNe (which likely represents a fairly exotic, mass-loss-intense end-phase of stellar evolution itself) and Ibn SNe (analogous to IIn SN but with narrow helium lines) tend to exhibit substantially higher luminosities, causing them to be over-represented. These events make up 21/150 of our sample, so even if they are over-represented somewhat, they are unlikely to exert substantial impact on the properties of the sample. Luminous (ordinary) II SNe could in principle trend towards a different host population than sub-luminous II SNe if, for some reason, the peak luminosity of a SN was a metallicity-dependent quantity. However, \citet{Gutierrez2018} found no significant difference between the properties of CCSN explosions produced in faint, low-mass galaxies and those produced in bright, high-mass galaxies, so a metallicity-dependent effect is probably not significant.

\subsection{SLSNe}

Our SLSN sample is comprised of objects from a variety of surveys. About half our sample of SLSNe were discovered by PTF \citep{2016Perley,Quimby2018}, a devoted transient survey during which substantial effort was placed in securing spectroscopic classifications of as many objects as possible. Even so, only a small fraction of PTF SNe could be followed-up and thus there may be biases associated with this in terms of SN classification. There may also be selection biases (associated with the contrast between the transient and the host galaxy), as explained in more detail by \citet{Frohmaier2017}. However, efforts to identify additional SLSNe in archival PTF data have produced no high-quality candidates other than those mentioned in Table \ref{tab:ptfslsn}, so it is not likely that large numbers of SNe were missed by this effort, but we cannot yet strictly rule this possibility out. 

We constructed the rest of our sample from the literature; since we focus on low-redshift objects, many of these objects were discovered by reanalysis of old data. This SLSN sample may thus potentially be quite heterogeneous. Future surveys with a stronger emphasis on an unbiased selection and follow-up will be needed to ensure this is not the case.

\subsection{LGRBs}
To avoid cosmic evolution effects, we restricted our sample of LGRBs to events closer than $z$ = 0.3, even though these represent a tiny fraction (a few percent) of all LGRBs with known redshifts. However, because most observed LGRBs do not have a successful redshift measurement, it is difficult to know whether the LGRBs that are \emph{known} to be at $z<0.3$ are fully representative of \emph{all} detected LGRBs at $z<0.3$. Low-$z$ LGRBs are often first identified to be nearby on the basis of the appearance of their host galaxies themselves: a catalogued galaxy coincident with an afterglow is a strong motivator for spectroscopic follow-up. This means that, at a fixed redshift, a LGRB host may be more likely to enter our sample if it is luminous than if it is faint. Furthermore, because of the huge pool of LGRBs occurring at higher redshifts, it is quite possible for a LGRB to be \emph{misidentified} as a low-z burst if it happens to align with a lower-redshift galaxy.
Many of these biases are mitigated by requiring a spectroscopically-confirmed supernova in association: not only does this guarantee that the redshift is correct, but the ability to conduct such a search also ensures that the LGRB could be observed readily and excludes a wider pool of events with poor observability which were inferred to be at low-$z$ \emph{only} because of a bright host galaxy. 

Still, some of these SN campaigns may have been conducted only because of the initial detection and redshift measurement of a bright host galaxy in the first place, leaving the possibility of a bias in favour of luminous galaxies and against dim ones in our sample. Whether this is likely to be a significant bias can be investigated case-by-case within our sample. For about half of the events in our sample, the LGRB was either so close that the host galaxy would have been evident almost no matter how luminous or dim it was ($z<0.1$), or the afterglow was so bright that its redshift would have been immediately evident from absorption spectroscopy regardless of its host. About half of our events fall in this category. The remaining events (which may have been missed if their host was fainter or less star-forming) include 031203, 120422A, 150518A, 150818A and perhaps 130702A (on account of its companion). However, omitting these targets would not change our conclusions.

A more delicate issue concerns the use of LGRBs without observed associated SNe, many of which specifically have observations ruling \emph{out} the presence of a SN at or near the luminosity of SN 1998bw. As we have noted, these could in principle represent background objects in dim high-z hosts. They could also represent variants of the short LGRB phenomenon (with T90's at the extreme of the distribution or `extended emission' episodes; e.g., \citealt{Norris+2006,Perley+2009}), or even something else entirely. On the other hand, they could also be genuine LGRBs whose SN was missed, dim, or dust-extinguished, and/or failed entirely \citep{Fynbo+2006}.

The origins of this class are probably heterogeneous: based on examination of individual no-SN events, there is reason to think almost all of the above events are at play. For example, GRB 060614's redshift is unambiguous but a very distinct host galaxy with almost no star-formation may point towards a different progenitor; GRB 051109B had no SN follow-up (despite a massive host, due to poor observability) and may simply be a missed low-z LGRB, though it could also be a background event; the SN in GRB 020903 has been interpreted as having been dust-extinguished \citep{Soderberg+2004}.

Given these uncertainties, we have run our tests both including and excluding the no-SN events; our basic conclusion is unaffected by this choice, although this largely reflects the small sample size of no-SN events. Further work will be needed to securely ascertain whether the no-SN events are associated with a different progenitor.

\subsection{Redshift evolution}

 Our CCSN sample spans redshifts between $0.002<z<0.08$. In contrast, our SLSN and LGRB samples are predominately at redshifts greater than 0.08. While we have used a simple procedure to correct for evolution in star-formation rate, we have not made any correction for stellar mass build-up from $z$ = 0.3 to 0 (as a result of galaxy mergers and from the conversion of gas into new stars)  because this change is very slow across the redshift range of our sample: much less than the differences that we see or the size of our statistical errors. See Fig.~2 of \citet{Furlong2015} who find almost no evolution in the galaxy masses between $z$ = 0.3 to 0, except for in high-mass galaxies. However, since the number of very high-mass ($\gtrsim 10^{11}$ M$_\odot$) galaxies is small (1/29 SLSNe-I, 2/150 CCSNe and no SLSNe-II, LGRBs) we do not attempt to correct for this effect.  
  
Changes in star-formation rate over this redshift range ($z$ = 0--0.3) are present, and while we have corrected for the offset of the main sequence, we have not corrected for possible changes in the distribution of SFR as a function of stellar mass along the main sequence. While these changes are anticipated to be small for most galaxy masses, the number of actively star-forming, very massive galaxies ($\gtrsim 10^{11}$ M$_\odot$) strongly decreases from $z$ = 0.3 to current day due to high-mass galaxy quenching. Thus, high-$z$ transients are more likely to be found in very high-mass galaxies than low-$z$ transients. This is not taken into account by our analysis, but the number of very high-mass galaxies is small in all samples, so its effect is likely to be minor.

\subsection{Extinction effect}

All our SN samples are selected via an optical search and thus are subject to biases if the transient is not easily visible or identifiable due to significant obscuration by dust. This bias is common among all SN searches that discover SN at optical wavelengths. This is especially important if the transient itself is intrinsically low-luminosity or if it is discovered at high redshift (and thus selected in the rest-frame UV). Since all of our samples are exclusively at low-redshift ($z\sim$0--0.3), this effect should be relatively minor and it would affect all three samples in similar ways. 

\subsection{Age effect}

An additional effect we must consider is the results of difference in stellar population ages, associated with different progenitor lifetimes between our samples. Stars that explode as SLSNe and LGRBs are likely $>$12M$_\odot$ and hence have short lifetimes of a few million years. Conversely, our sample of CCSNe is dominated by type II SNe which typically originate from less massive stars of 8--12 M$_\odot$ which may take up to a few tens of millions of years to explode. Therefore, the galaxy populations hosting CCSNe could evolve significantly more that the galaxies hosting SLSN/LGRBs after the actual star-formation episode--such that even if the properties of the galaxies were identical at the time that the SN progenitors were formed, the observable properties may in some cases be different at the time of the SN explosion. 

This effect is crucial when considering precise spatial positions within the galaxy, and to some extent when dealing with emission-line metrics. Fortunately, this issue is lessened in our study because our investigations are limited to quantities derived from the broadband photometry. These wavelengths trace star-formation over a much longer time-scale (10--100 Myr) than other tracers of star formation such as H$\alpha$ which measure the `instantaneous' star-formation (1--10 Myr). Thus, the difference between SLSNe and CCSNe is not likely to be an age effect. If a galaxy is starting from a very small stellar population and then a starburst begins, many more young stars will have formed in the tens of millions of years between the explosions of the first very high-mass stars and the explosions of stars with longer lifetimes. Thus, there may be a mass `build-up' effect seen in CCSNe opposed to SLSNe and LGRBs, leading to systematically higher masses. However, this effect is only important for the very youngest, lowest-mass galaxies, of which we have demonstrated that there are very few.

\subsection{Differences in Photometry Procedure}

We use broad-band photometry and SED fitting using \textsc{Le PHARE} to derive stellar masses and SFRs of each galaxy sample. Thus, the samples are well homogenised since the same modelling technique is used on all samples. The photometry for PTF SLSN hosts and ASAS-SN CCSN hosts are performed using a similar method. There may, however, be differences in background and aperture treatment for photometry measurements we have taken from the literature--but in most cases this photometry is directly validated by comparison to our own measurements, with measurements that disagree to high significance excluded. 

In addition, the  dominant source of uncertainty for the photometry of nearby and massive galaxies in the CCSN sample is from the background subtraction. Therefore the photometric uncertainties may be underestimated for these sources. However, we gather some photometry from the NSA which has a special background optimised subtraction and these host galaxies are not offset from the rest of the sample.

\bsp	
\label{lastpage}
\end{document}